\documentclass[12pt,preprint]{aastex}
\begin{document}
\title{Turbulence in the Inter-galactic Medium: Solenoidal and Dilatational Motions, and the Impact of Numerical Viscosity}
\author{Weishan Zhu$^{1,2}$, Long-long Feng$^{3}$, Yinhua Xia$^{4}$,
Chi-Wang Shu$^{5}$, Qiusheng Gu$^{1,2}$ and Li-Zhi Fang$^{6}$}
\altaffiltext{1}{School of Astronomy and Space Science, Nanjing 
University, Nanjing, 210092, China}
\altaffiltext{2}{Key Laboratory of Modern Astronomy and Astrophysics of the Ministry of Education, Nanjing, 210092, China}
\altaffiltext{3}{Purple Mountain Observatory, Nanjing, 210008, China}
\altaffiltext{4}{School of Mathematical Sciences, University of Science 
and Technology of China, Hefei, Anhui 230026, China}
\altaffiltext{5}{Division of Applied Mathematics, Brown University, 
Providence, RI 02912, USA}
\altaffiltext{6}{Department of Physics, University of Arizona, Tucson, 
AZ 85721, USA}

\begin{abstract}
We use a suite of cosmological hydrodynamic simulations, 
run by two fixed grid codes, to investigate the properties of solenoidal and 
dilatational motions of the intergalactic medium (IGM), and the impact of numerical 
viscosity on turbulence in a LCDM universe. The codes differ only in the spatial difference discretization. 
We find that (1) The vortical motion grows rapidly since $z=2$, 
and reaches $\sim 10 km/s -90 km/s$ at $z=0$. Meanwhile, the small-scale compressive ratio $r_{CS}$ drops 
from $0.84$ to $0.47$, indicating comparable vortical and compressive motions at present. 
(2) Power spectra of the solenoidal velocity possess two regimes, $\propto k^{-0.89}$ and 
$\propto k^{-2.02}$, while the total and dilatational 
velocity follow the scaling $k^{-1.88}$ and $k^{-2.20}$ respectively 
in the turbulent range. The IGM turbulence may contain two distinct phases, the supersonic and 
post-supersonic phases. (3) The non-thermal pressure support, 
measured by the vortical kinetic energy, is comparable with the thermal 
pressure for $\rho_b \simeq 10-100$, or $T <10^{5.5} K$ at $z=0.0$. The 
deviation of the baryon fraction from the cosmic mean shows a preliminary 
positive correlation with the turbulence pressure support. (4) A relatively higher 
numerical viscosity would dissipate both the compressive 
and vortical motions of the IGM into thermal energy more effectively, resulting in less 
developed vorticity, remarkably shortened inertial range, 
and leading to non-negligible uncertainty in the thermal history of gas accretion .  
Shocks in regions outside of clusters are significantly suppressed by numerical viscosity 
since $z=2$, which may directly cause the different levels of 
turbulence between two codes. 

\end{abstract}


\section{Introduction}

The intergalactic medium (IGM) feeds the growth of galaxies and 
receives feedback from star formation 
and even more violent activity like AGN in galaxies. The dynamical and 
thermal states of the intergalactic 
medium reflect this mutual process and are important factors to 
interpret many observations, such 
as the lymann alpha forest and the metal absorption lines in the
spectra of high redshift objects, which could in 
turn support the structure formation theory.  Moreover, more 
information on the state of the IGM would shed 
light on the so-called missing baryon problem at low redshift. 
The observed baryons in galaxies only 
account for one-tenth of the cosmic baryon content that 
given by the high redshift Ly$\alpha$ lines and the cosmic 
microwave background (CMB) studies (Fukugita, Hogan, \& Peebles et 
al. 1998; Bregman 2007 and references therein). 

The thermal state of the IGM has been thoroughly studied by 
cosmological hydrodynamical simulations 
since the late 1990s. Most of those works predicted that the 
majority of baryons rest in the warm
($10^4<T<10^5$), and the warm-hot ($10^5<T<10^7)$ IGM, i.e., WHIM (e,g., 
Cen \& Qstriker 1999; Dave et al. 1999; 
Dave et al. 2001; Cen \& Ostriker 2006). Many efforts have been 
made by observers to detect the WHIM in the low density regions (Bregman 2007). The dynamical state of the 
IGM, however, has not gained much 
attention. Adhesion models have been applied to describe the evolution of the distribution of  baryonic gas in
the mildly non-linear regimes in analytic works(Jones 1999; Matarrese \& Mohayaee 2002), which, however, may only work
 for baryon density contrast $< 5 \sim 10$ and can not capture the complex pattern of baryonic flow in  the highly
non-linear regimes. As an emerging subject, the turbulence in 
the IGM has invoked growing interest in recent years. Observationally, Zheng et al. (2004) fitted the H I and 
He II absorption lines in the high quality spectra of quasar HE 2347-4342 and found comparable velocities of hydrogen 
and He ions, suggesting a turbulence dominated velocity field in the IGM between $z=2.0$ and 
$2.9$.  Estimation of the turbulent velocity in galaxy groups and clusters are available in recent years
via direct and indirect observational means, giving weakly constrained  upper limits of 
hundreds km $s^{-1}$(Churazov et al. 2004; Sanders et al. 2010, 2011,2013;Bulbul et al. 2012) . 

He et al. (2006) studied the intermittency of the velocity field 
of cosmic baryons in cosmological 
simulation, indicating that the motion of baryons might be 
turbulent on large scales.  Ryu et al. (2008) found 
that vorticity could induce turbulent-flow motions in the 
IGM and amplify the intergalactic magnetic 
fields. Oppenheimer \& Dave (2009) used cosmological simulations 
to model the O VI absorbers at $z< 0.5$ 
and required a density-dependent sub-resolution turbulent 
motion in the IGM to match the observation results. Tepper-Garcia et al. (2011) also found the 
necessity of turbulence in the IGM that was not captured in their simulation to mimic the
OVI absorbers at low redshift. In a previous paper Zhu, Feng, \& Fang (2010) (hereafter ZFF2010), 
we studied the evolution of vorticity and its power spectrum in 
the IGM, and showed that the IGM is highly turbulent in the range 
from a couple of Mpc  down to hundreds of kpc at low redshifts. 

An elementary understanding about the turbulence injection in the IGM has 
been established, although many 
details remain unknown. Cosmological hydrodynamic simulations show 
that supersonic gas motions will develop 
during the hierarchical structure formation (Ryu et al. 2003; 
Pfrommer et al. 2006; Skillman et al. 2008;Vazza et al. 2009). Those 
supersonic flow motions, driven by the baryon gas accreting to sheets and 
filaments, and gaseous haloes merging with 
each other, could efficiently induce vorticity in the intergalactic 
velocity field and then trigger turbulence
(Ryu et al. 2008; ZFF2010). In addition to the 
cosmological shocks and vorticity, the shear 
flows produced by the ram pressure stripping of protogalactic and 
collapsed objects moving in the IGM 
will bring turbulent motions too. Also,  galactic outflows driven 
by starburst and AGN might inject 
turbulent kinetic energy into cluster outskirts
(Evoli \& Ferrara 2011).  Iapichinio et al. (2011) used sub-grid 
scale model to include the unresolved small scale turbulence in 
their analysis of turbulent kinetic energy 
evolution as a function of redshifts, and speculated that shock 
interactions and merger induced shear flows 
should be the main source for the production of turbulence in the WHIM 
and ICM, respectively.  

Meanwhile, properties of compressible and supersonic isotropic 
turbulence in the context of fluid 
dynamics and interstellar medium study have been investigated in 
more details in the past two decades
(e.g., Kida \& Orszag 1990, 1992; Porter et al. 1992, 1994, 1998; 
Kritsuk et al. 2007; Federrath et al. 2010).  
Statistics of velocity and its two components, including power 
spectra and structure functions, show that 
both decaying and forced supersonic turbulence behave 
differently from incompressible 
Kolmogorov-type turbulence. As the turbulence in the IGM is likely 
mainly produced by cosmic shocks, its 
scaling law may also deviate from the classical Kolmogorov's. 
Actually, the power spectra of the velocity (Ryu et al. 2008, He et al. 2006) and kinetic energy 
(ZFF2010) of the IGM have already exhibited  non-Kolmogorov features. 

Moreover, the velocity field of the IGM would be dilatationally dominated at 
high redshifts, because the driving force, 
gravity, is curl-free. As vortices emerge at low redshifts, 
the velocity field would transfer into solenoidal 
domination within the turbulence inertial range, driven by 
dilatational forcing. The evolution of vorticity and divergence, 
and their corresponding velocity 
components, solenoidal and dilatational velocity, would be an essential probe to 
the transfer process. More solid 
results concerning those fundamental characteristics can enrich 
and verify our knowledge of turbulence in the IGM. 

On the other hand, the numerical viscosity in cosmological 
hydrodynamic simulation codes may have a significant impact on the development 
of turbulence in the IGM. Because of the relatively low density in comparison with the Earth's gas, 
the IGM is estimated to have very high Reynolds number( e.g. $10^{13}$ in Gregori et al. 2012). 
For the hot intra-cluster medium, however, the physical viscosity may be non-negligible. Fabian et al. 
(2003a,b) gave an estimated Reynolds number of $\sim 1000$ in the central region of the Persues cluster, 
and suggested that the viscosity dissipation would be important in the inner tens of kpc. 
The physical viscosity of the ICM have been considered and  implemented into cluster formation 
simulations(e.g., Sijacki\& Springel 2006; Dong\&Stone et al. 2009; Suzuki et al. 2013). Otherwise, the 
cosmic gas component is usually assumed to be non-viscous and governed by the cosmological version of the Euler 
equation. Despite that many efforts have been made to reduce it, numerical viscosity remains 
non-negligible in almost all of the codes and bring additional numerical dissipation with respect to the physical dissipating 
mechanisms including Jeans dissipation and shock heating. Numerical viscosity will damp the flow instabilities and shock waves, 
and smearing out substructure in cosmological hydrodynamic simulations, which has been confirmed 
by many code comparison works (e.g., Agertz et al. 2007; Vazza et al. 2011; Bauer \& Springel, 
2012). Dolag et al. 2005 found that the 
numerical viscosity in the smoothed particle hydrodynamics (SPH) method can artificially suppress the turbulent 
motions in the intra-cluster medium. Nelson et al. (2013) proposed that the turbulent motions on large scales could be spuriously 
dissipated into thermal energy by numerical viscosity, which would keep the baryonic gas locked in the diffuse, 
hot and low density halos of central galaxies, and consequently slow down the hot gas accretion by an order of magnitude. 
Hence, the various levels of numerical viscosity in different codes are expected to lead to different intensity of turbulence in the IGM. 
To obtain a more concrete understanding of turbulence in the IGM, the impact of 
numerical viscosity needs to be investigated comprehensively.  

Using two kinds of fixed mesh methods with different levels of numerical viscosity,   we performed a suite of high 
resolution cosmological hydrodynamical simulations
to investigate the properties of solenoidal and 
dilatational velocity fields,  and made comparison between two codes 
regarding turbulence and shocks development in the IGM. Differences between samples run 
by these two codes allowed us to assess  
the impact of numerical viscosity, and in turn provided more details on 
the production and evolution of turbulence,  which would help to build a 
more reliable history of gas accreting onto virialized  structures.  

This paper is organized as following.  In \S2, we give a short introduction to the numerical methods
used in this work and then present some simulation 
details. \S 3 investigates the properties of vorticity and 
divergence in the IGM in our simulation samples. 
Statistics of solenoidal and dilatational velocity, and 
non-thermal pressure provided by the solenoidal motion 
are studied in \S4. Then we analyze the distribution of cosmic 
shocks in simulations run by these two different schemes with the same 
resolution in \S 5. Discussion and concluding remarks are summarized  
in \S 6.   The numerical tests for code comparison and resolution convergence 
are presented in the Appendix. 

\section{Numerical Methods}

\subsection{TVD scheme}
In the literature, many modern difference schemes are referred as 
total variation diminishing (TVD) 
method, as long as they fulfill the TVD nonlinear stability condition
for scalar model problems and linear systems, which was first 
suggested by Harten (1983). In this work, we will use Harten's 
original, explicit, second-order TVD scheme, and expand it to three 
dimension by dimension splitting(Strang 1968). This scheme will 
be implemented into the cosmological context in almost 
the same way as in Ryu et al. (1993), except for the treatment of
 internal energy in order to improve 
its performance in high Mach number regions. Usually, non-negligible 
errors may appear in hypersonic regions if the thermal energy is 
calculated by subtracting the kinetic energy from the total energy, 
and sometimes this leads to negative density and pressure. In Ryu et 
al. (1993), a modified entropy method was used to solve the thermal 
energy accurately. In this work, we employ the entropy based dual 
energy method in Feng et al. (2004) to track the thermal 
energy for the TVD scheme. 

\subsection{Positive-Preserving WENO Scheme}

The weighted essentially non-oscillatory (WENO) scheme is a high-order
finite difference scheme first 
developed by Jiang \& Shu (1996). The WENO scheme can simultaneously 
achieve high order accuracy in 
 smooth region, and sharp transition at shock or contact 
discontinuity surface (Shu 1998, 1999). Feng et 
al. (2004) implemented the fifth order finite difference WENO scheme 
into the context of cosmological 
application and demonstrated its capacity of capturing strong 
shocks and complex flow pattern 
emerging in cosmic structure formation.  

Very recently, Zhang \& Shu (2012) designed a positivity-preserving 
strategy combining with high order finite difference WENO 
scheme, which could successfully keep the positivity of density and 
pressure during simulation without losing 
conservation and high order accuracy, and highly improved
the performance of the WENO scheme in solving hypersonic flows. 
Details about the positive-preserving WENO scheme can be found in 
the Appendix. The global Lax-Friedrichs flux splitting method is 
used in this work, which is more robust and compatible with the 
positivity-preserving method than the local Lax-Friedrichs flux 
splitting used in ZFF2010.  This change results in a 
looser restriction on the $cfl$ number and less dispersion over 
spherical shock front, shown by the Sod shock tube, and Sedov blast wave tests
presented in the Appendix, where the performance of the WENO and
TVD schemes are given.  

\subsection{Cosmological Simulation Set Up}

We use the WENO and TVD schemes for ideal fluid dynamics, coupled with the standard 
particle-mesh method for gravity calculation.  In particular, the early WENO scheme 
employed in Feng et al.(2004) is updated with the positivity-preserving version, which maintains the positivity of physical quantities. 
Except for the spatial difference discretizations, all the other modules are the same in both codes, including the gravity solver, time difference 
scheme, cooling and heating processes.  The time integrations are performed using the third 
order low storage Runge-Kutta method. 

All the cosmological simulations are 
performed in a periodic cubic box of size 
$25 \ h^{-1}$ Mpc in a LCDM universe. The cosmological parameters 
are set to the WMAP5 result (Komatsu 
et al. 2009), i.e., $(\Omega_{m}, \Omega_{\Lambda},h,\sigma_{8},
\Omega_{b},n_{s})=
(0.274,0.726,0.705,0.812,0.0456,0.96)$.  We run simulations with equal number of grid cells and dark matter 
particles,  one set with $512^3$ evolved by the WENO and TVD schemes respectively, and another 
with $1024^3$ evolved by the WENO scheme only.  The $512^{3}$ runs have 
a grid resolution of $47.7 \  h^{-1}$ kpc, and  
a mass resolution of $1.04 \times 10^{7} \  M_{\odot}$ for the dark matter 
particle.  The effect of re-ionization 
is included by adding a uniform UV background that is switched on at $z=11.0$, and 
the radiative cooling and heating processes are modeled in the same way as Theuns et al. (1998), 
with a primordial composition $(X=0.76,y=0.24)$. Star formation and 
its feedback are not taken into account.  All the simulations start at redshift $99$ and end at $0$, 
We will refer a specific simulation run in the form of,  e.g., 'WENO-512', where 
the terms before and after the dash 
stand for the hydro scheme and one-dimensional scale, respectively.  

\section{Vorticity and Compressibility in the IGM}

In Ryu et al. (2008) and ZFF2010, the vorticity of the IGM velocity field and its power spectra 
have been demonstrated to be good indicators of turbulence. This section is to investigate 
the behavior of these indicators in our simulation samples.  As a useful probe to the 
growth history of solenoidal motions with respect to dilatational motions under 
a curl-free forcing, the relative importance of vorticity to divergence will also be addressed 
by the small scale compressive ratio.

\subsection{Vorticity Distribution }

The vorticity of the IGM velocity field is defined as 
$\vec{\omega}=\nabla \times {\bf v}$. The dynamical 
equation of the modules of vorticity vector, 
$\omega\equiv |{\vec \omega}|$, in the IGM reads as (ZFF2010)
\begin{equation}
\frac{D \omega}{Dt}\equiv \partial_t {\omega} +\frac{1}{a}{\bf
v}\cdot{\nabla}\omega=\frac{1}{a}\left [\alpha \omega-d\omega +
\frac{1}{\rho^2}\vec{\xi}\cdot (\nabla \rho \times\nabla p )
-\dot{a}\omega \right ],
\end{equation} 
where $\vec{\xi}={\vec\omega}/\omega$, 
$\alpha={\vec\xi}\cdot({\vec\xi}\cdot\nabla){\bf v}$, and 
$d=\partial_iv_i$ is the divergence of the velocity field.

According to eqn.(1), the generation of vorticity in the IGM can be attributed to the 
baroclinity term, $(1/\rho^2)\nabla \rho \times\nabla p$. A non-zero baroclinity  means the 
gradient of pressure and density are not parallel to each other, which occurs in curved 
shocks and complex flow structures.  Stretched by the first term on the right hand side (RHS) of 
Eq(1), the vorticity structures will expand or 
contract along the local baryon gas flow. Meanwhile, the vorticity in 
the IGM will be attenuated by the cosmic expansion. Normalized by the cosmic time $t$, the 
vortical motion of fluid is usually measured by the dimensionless quantity  $\omega t$,  which
actually characterizes the number of turn-overs that vortices can make within the cosmic time. 

\begin{figure}[htbp]
\begin{center}
\vspace{-1.0cm}
\includegraphics[width=7.5cm]{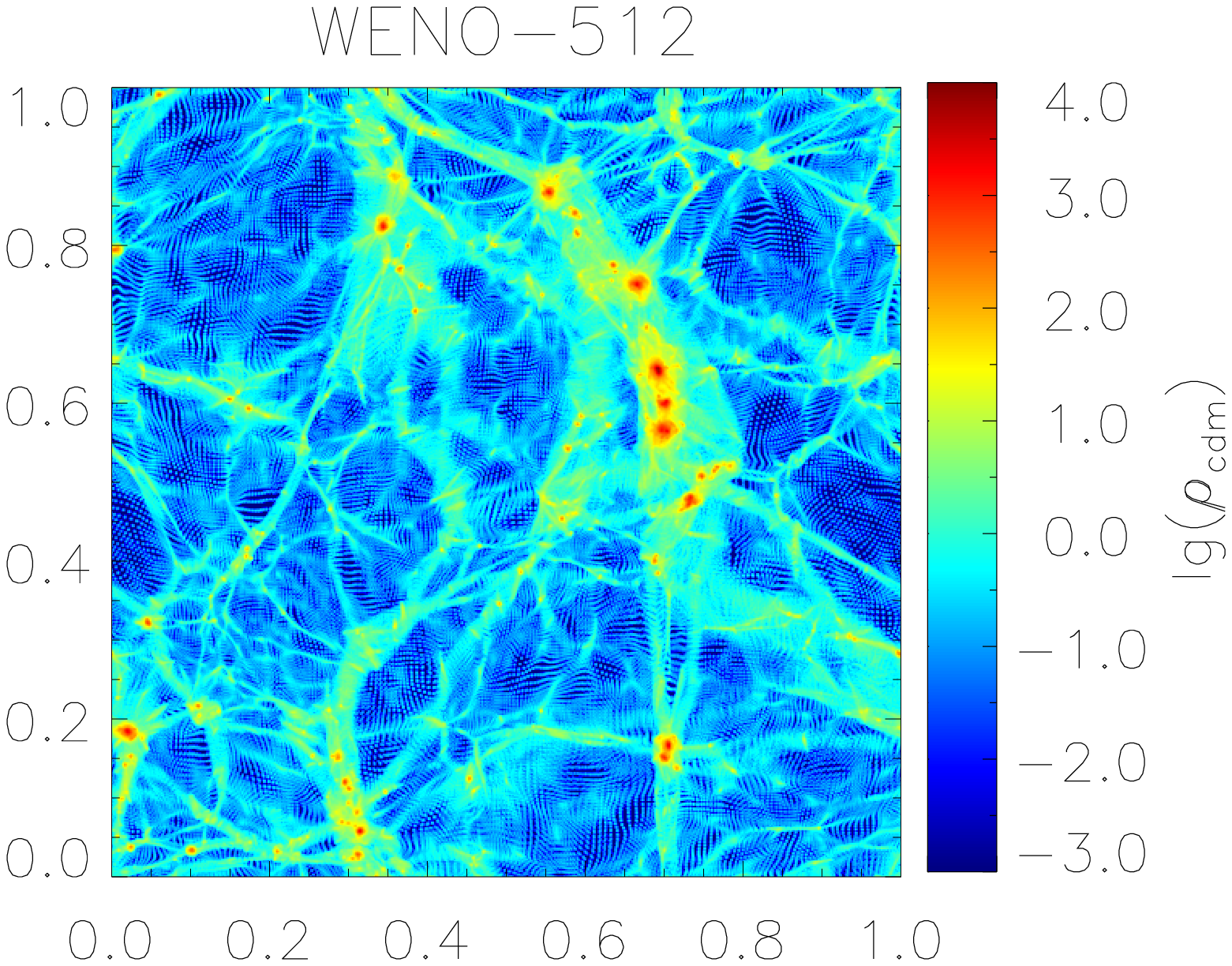}
\includegraphics[width=7.5cm]{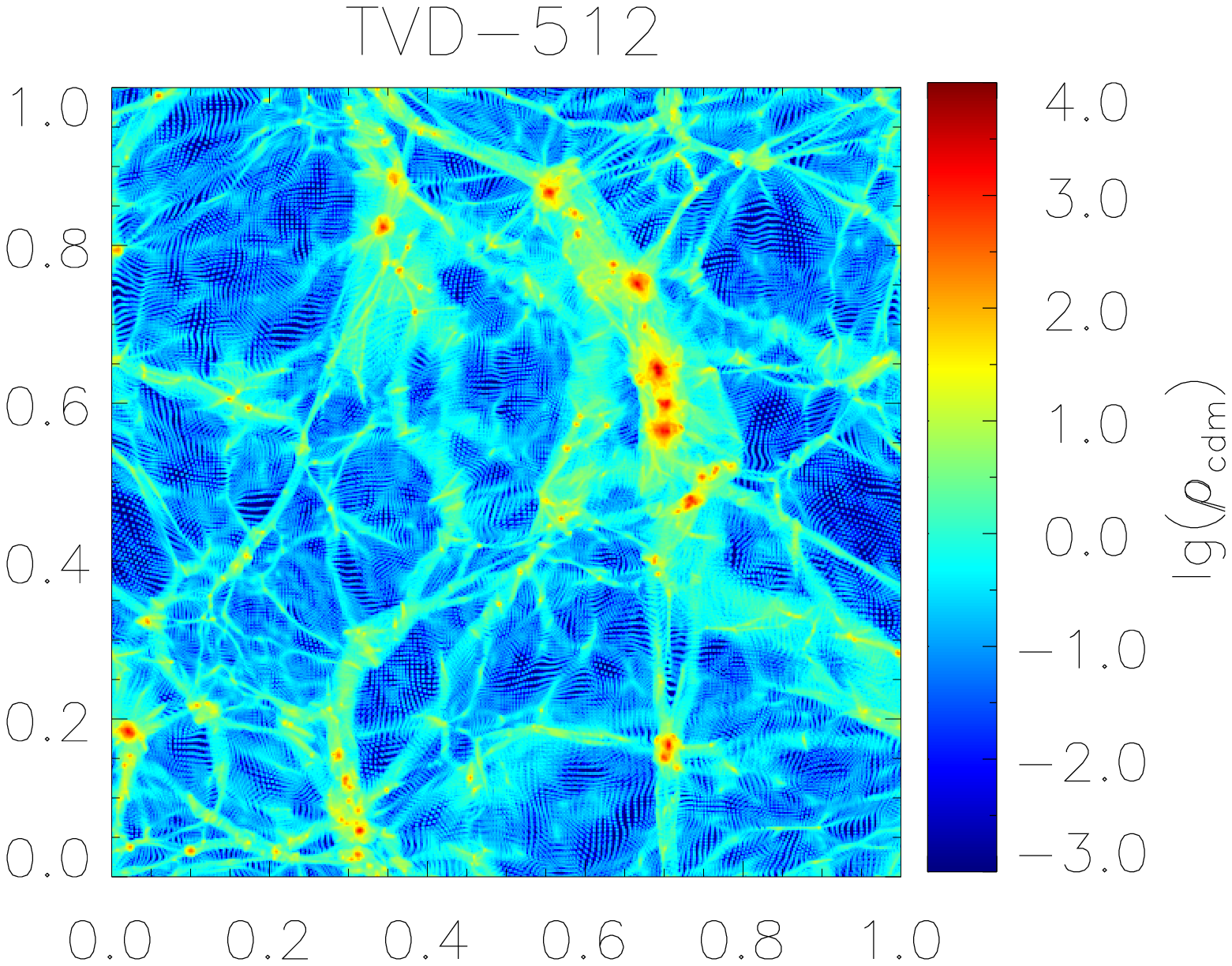}
\vspace{1.0cm}
\includegraphics[width=7.5cm]{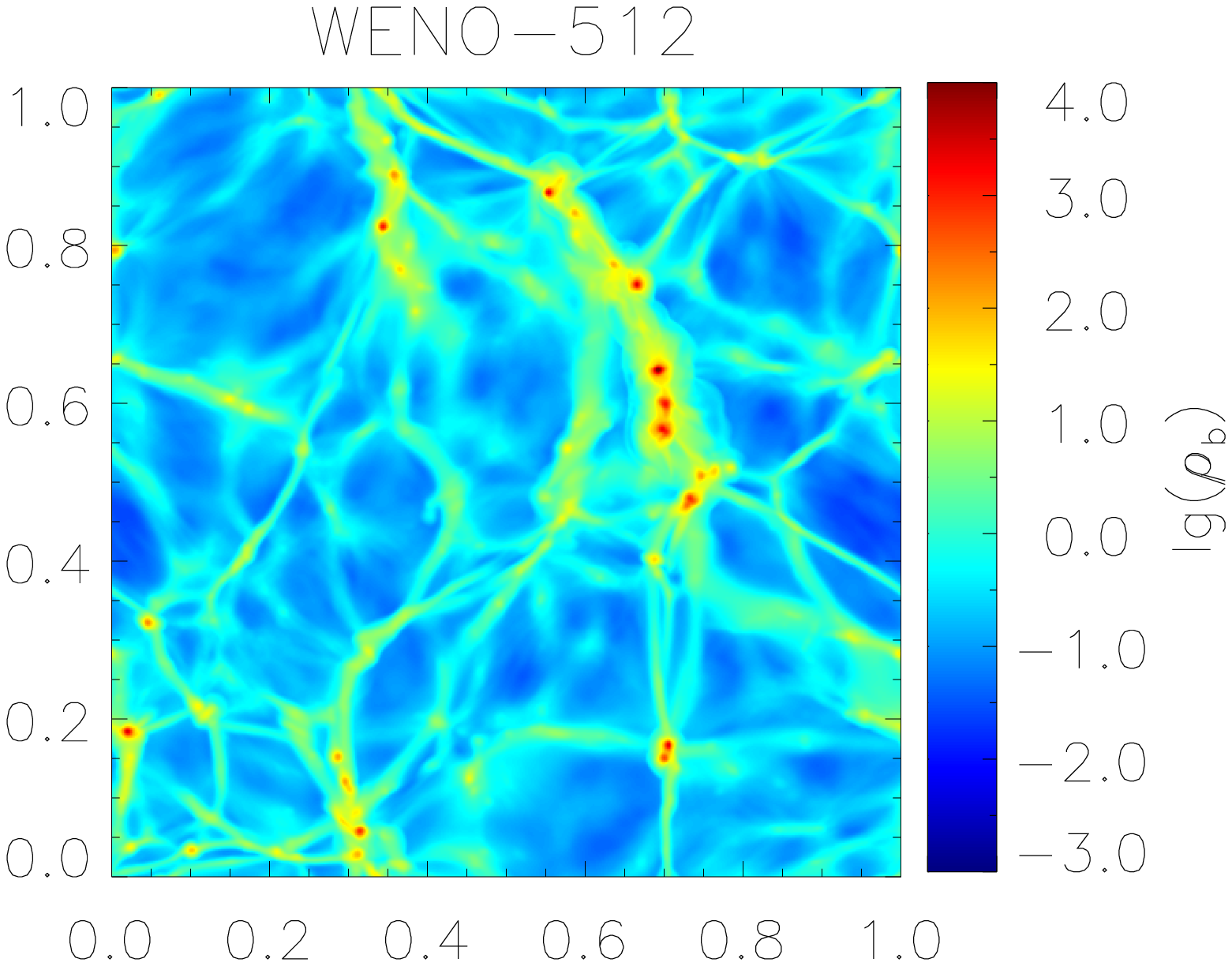}
\includegraphics[width=7.5cm]{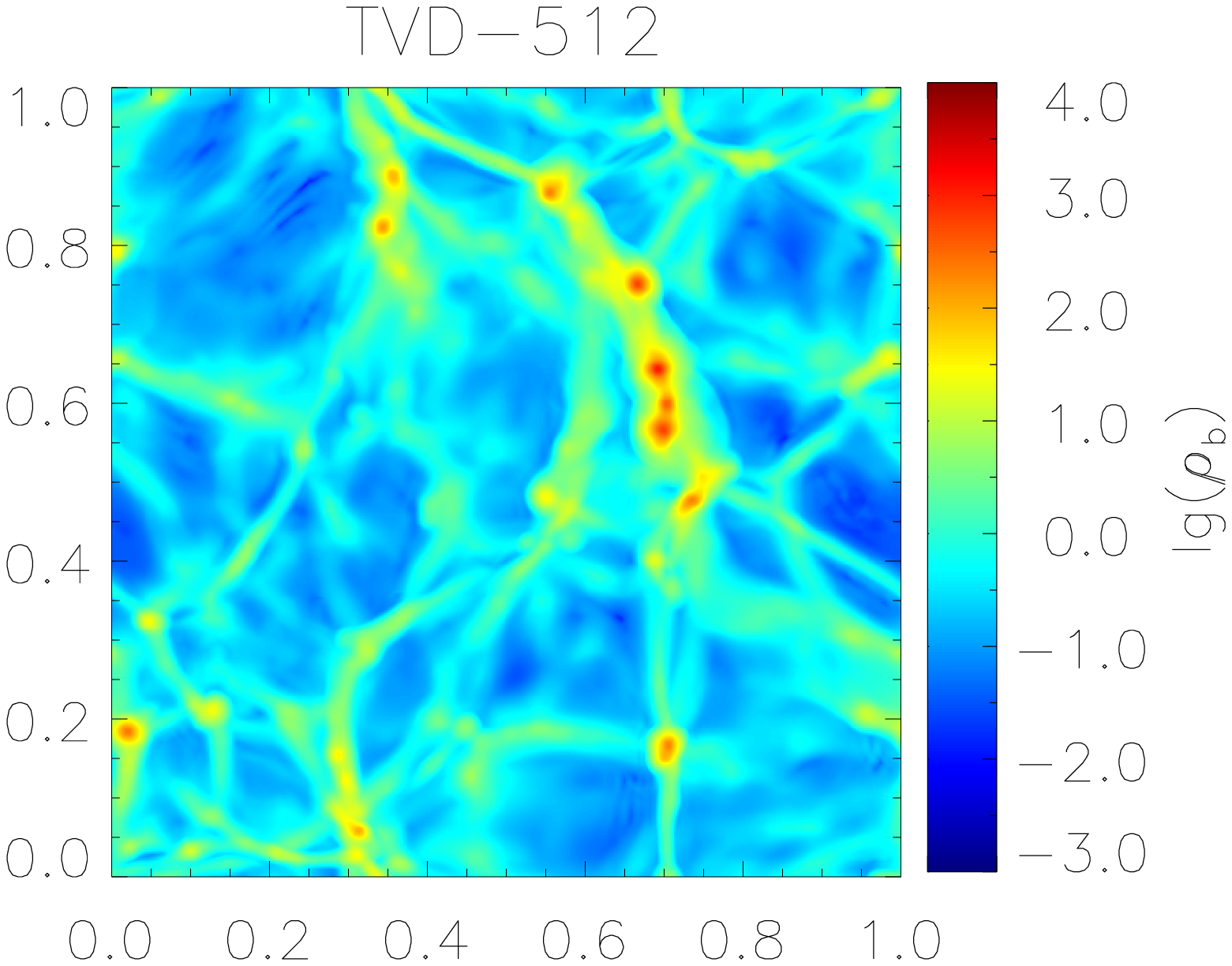}
\vspace{1.0cm}
\includegraphics[width=7.5cm]{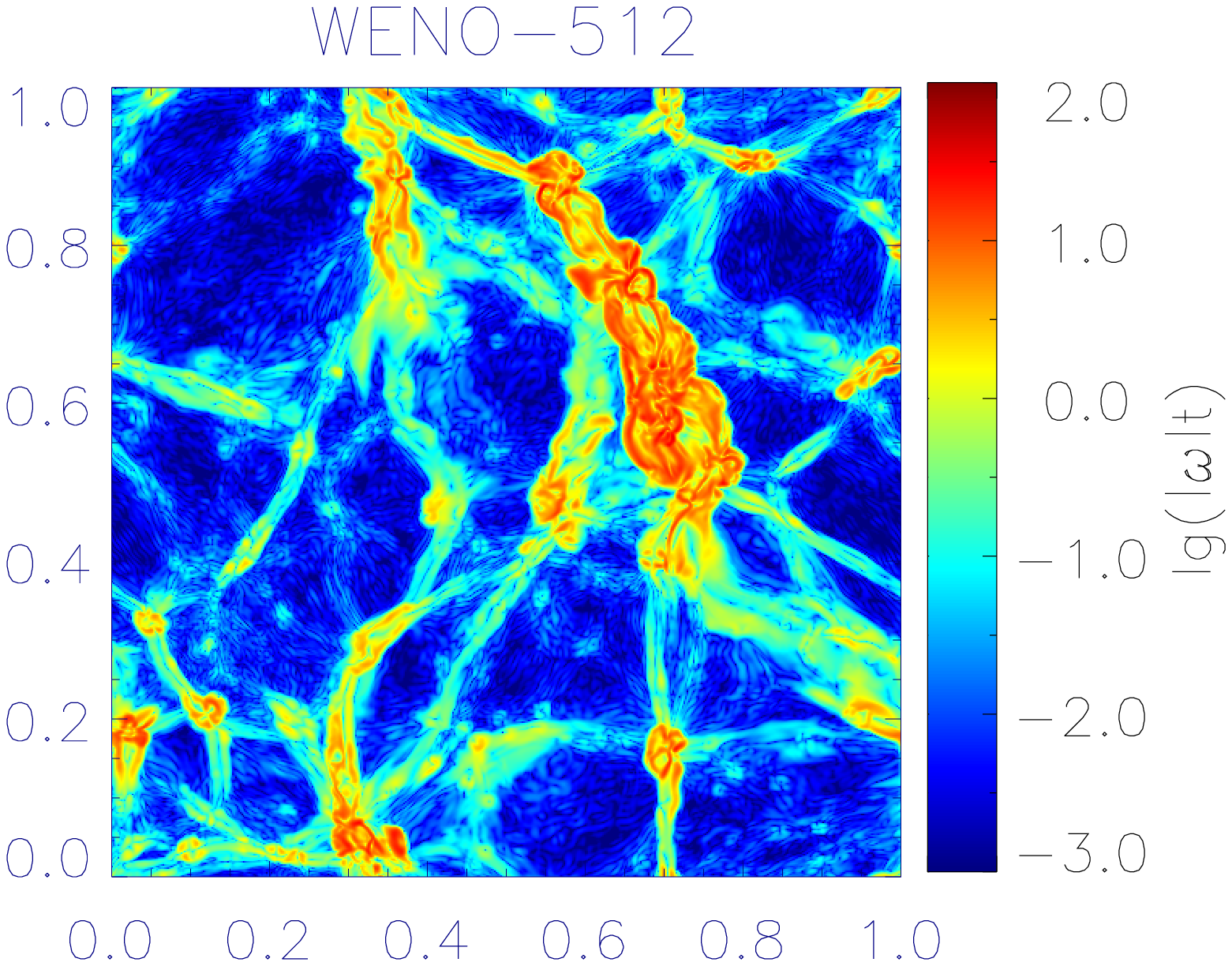}
\includegraphics[width=7.5cm]{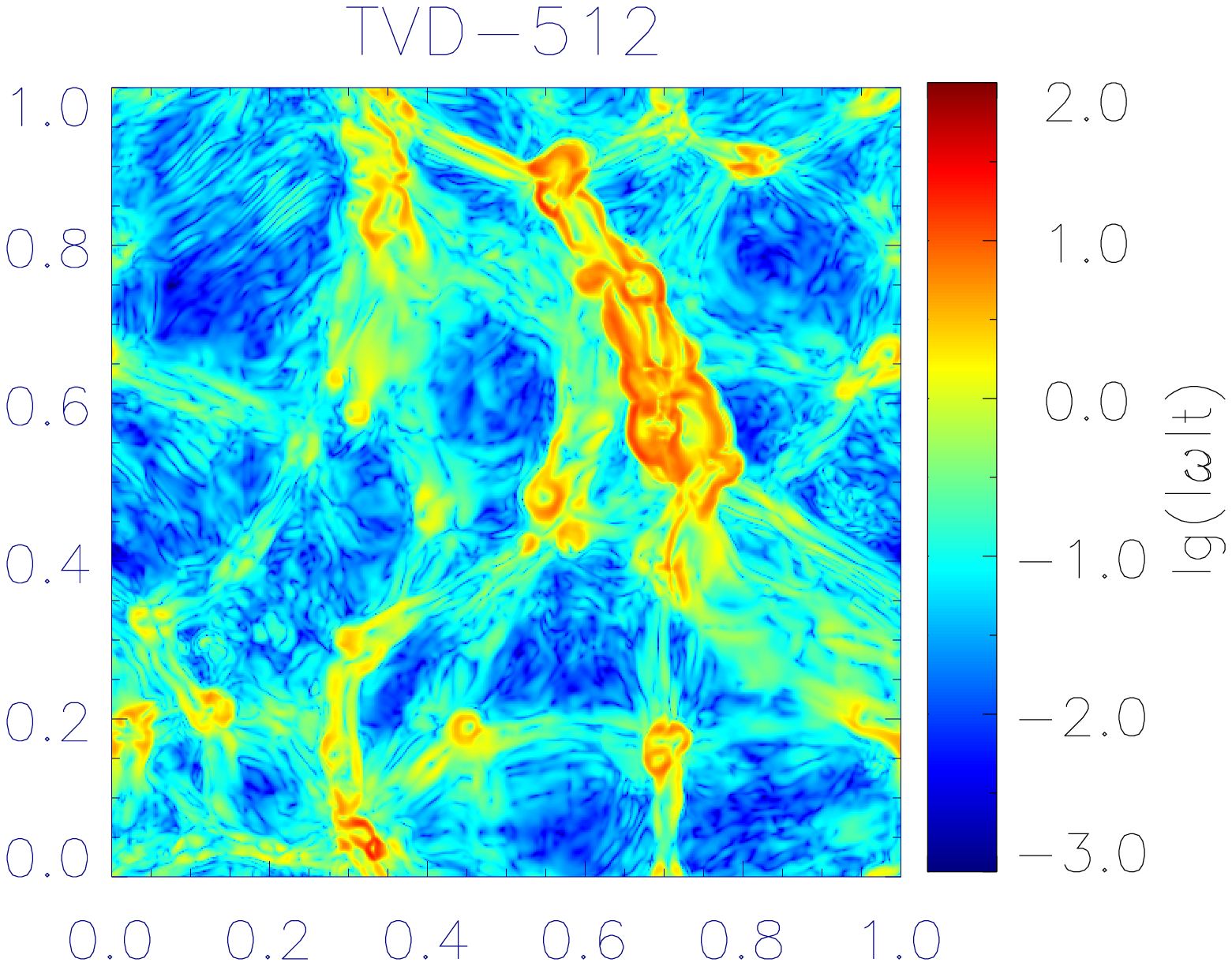}
\end{center}
\caption{Dark and baryonic matter density, in unit of the cosmic mean, 
and vorticity normalized by 
cosmic time in a slice of $25\times 25\times 0.25 \  h^{-3} \ \mbox{Mpc}^3$ in the $512^3$ simulations at $z=0$ .
 The top (middle) row gives dark (baryonic) matter density, and the 
bottom row is vorticity. The left (right) 
column is extracted from the WENO-512 (TVD-512) simulation.}
\end{figure}

Figure 1 presents a straightforward view of the distribution of dark and baryonic 
matter,  and vorticity in a slice of depth $250 h^{-1}$ kpc, which are extracted 
from the WENO-512 and TVD-512 simulation samples respectively. The dark matter density field is 
constructed by partitioning the particle masses onto
the grid cells using the Triangular Shaped Cloud (TSC) mass assignment scheme. The vorticity field is obtained by the fourth order 
finite difference of  $\vec{\omega}=\nabla 
\times {\bf v}$ at grid cells in a post simulation procedure.
 
We first take a look at the density field. While a subtle 
difference can be found in the dark matter distribution between 
 two codes, the difference in the  
baryonic gas distribution is significant.  The gaseous structures 
in the WENO sample are more well developed
than their counterpart in the TVD sample. Since the WENO code attains high order accuracy, it resolves more sub-structures within the high-density regions. For example, we call the reader's attention to the 
gaseous halo located at $\sim(0.7, 0.2)$, 
where two gaseous density peaks are sharply resolved
in the WENO sample, however, only a single clumpy core can be found 
in the TVD sample. 
 
\begin{figure}[htb]
\centering
\includegraphics[width=8.0cm]{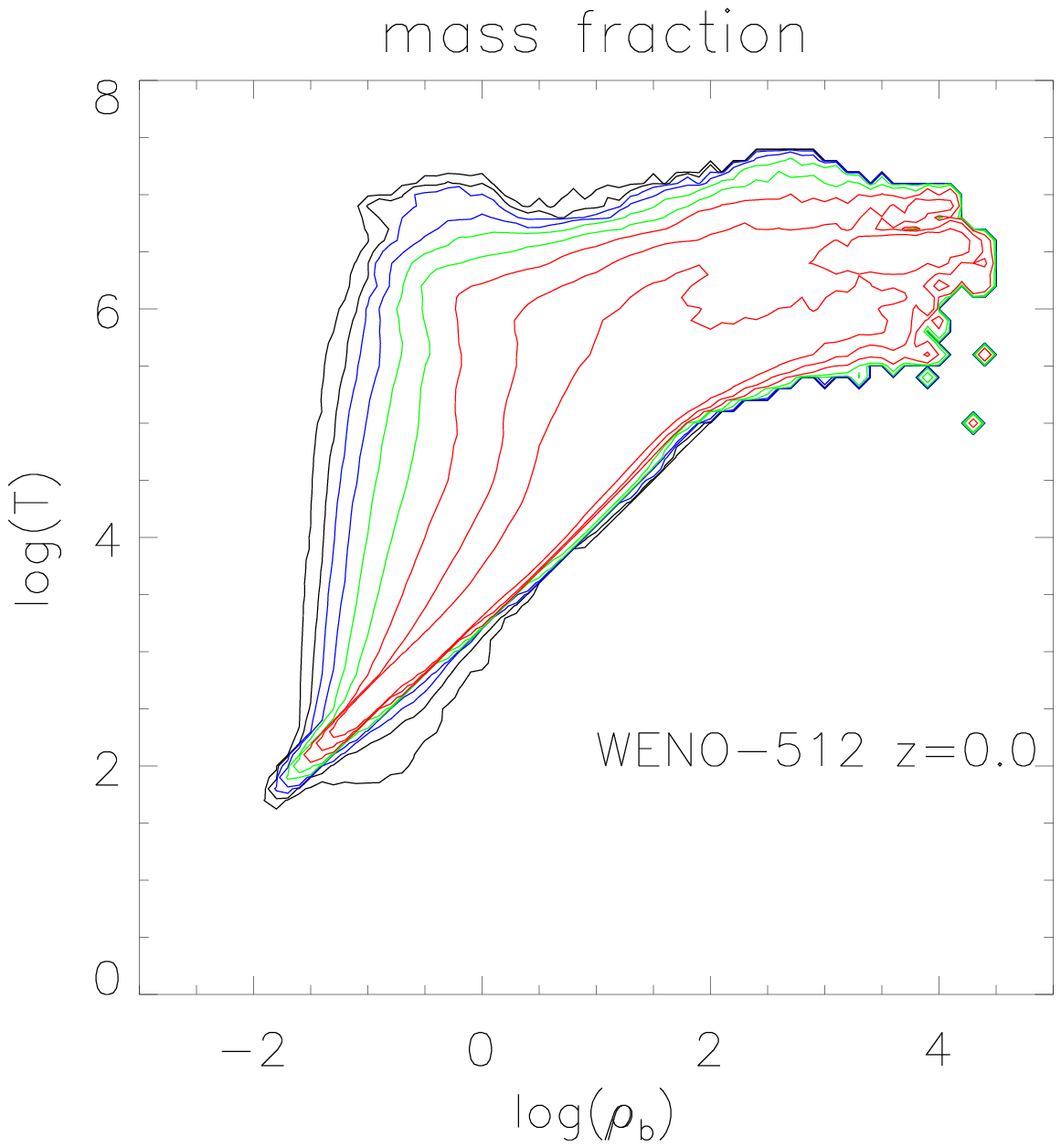}
\includegraphics[width=8.0cm]{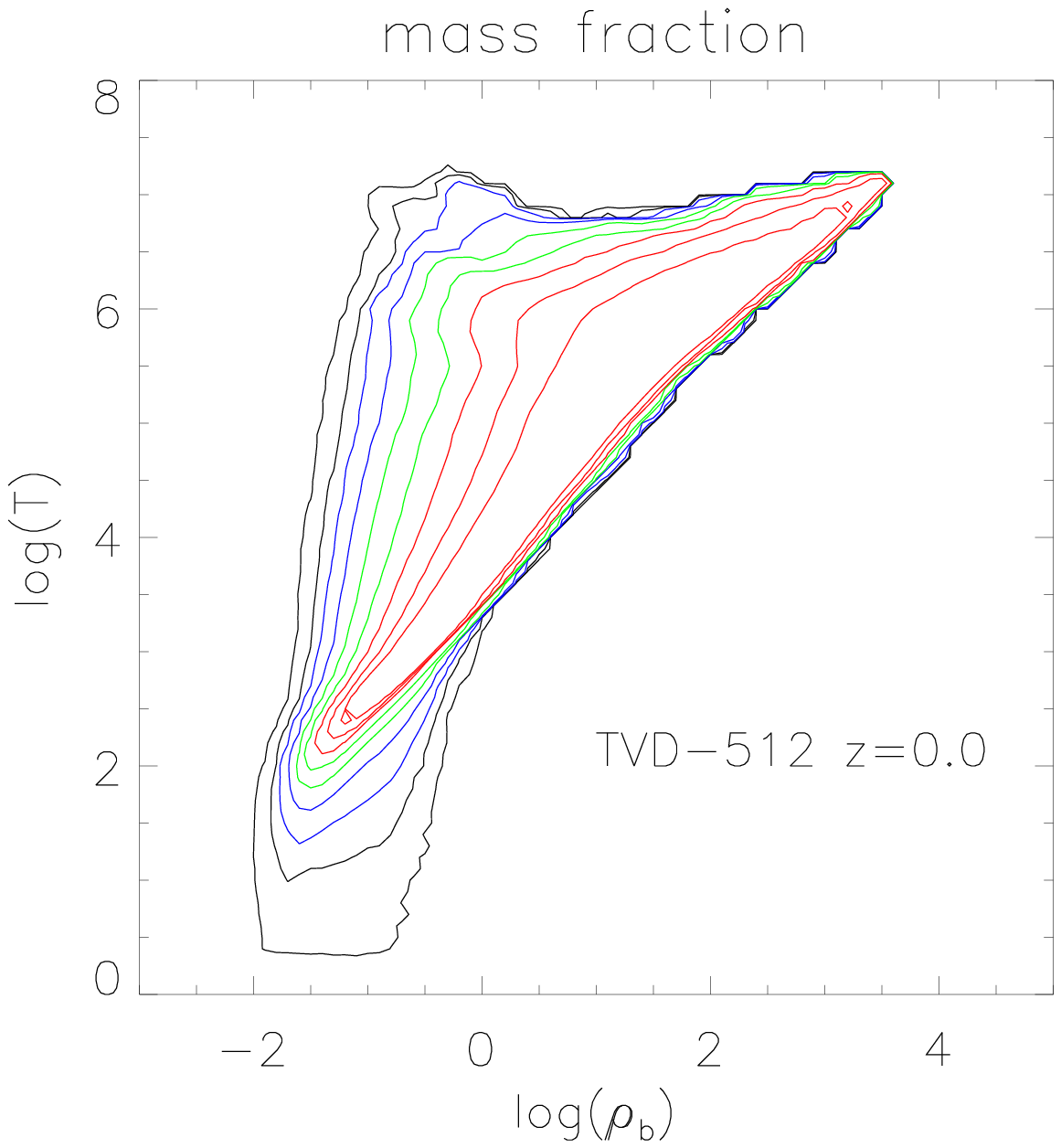}
\caption{Mass distribution as function of gas density and temperature 
in the $512^3$ simulations run by the WENO and TVD codes
at $z= 0.0$. Left: WENO-512. Right: TVD-512.}
\end{figure}
 
This trend is shown more clearly by the gas mass fraction distribution 
in the density-temperature space, 
as displayed in Fig. 2.  Once the normalized over-density exceeds $10$, 
gaseous structures in the WENO-512 run
would go through a relatively more rapid growth in density, and 
slower growth in temperature than the TVD-512 run.  This 
difference may partly result from the different levels of numerical 
viscosity. The relatively higher numerical viscosity in 
the TVD scheme could artificially dissipate the kinetic energy associated with both compressible and incompressible
motions into thermal energy more effectively, and hence 
smear out gaseous density transition in the process of structure formation.  Despite the different resolution and scope, 
this is in analogous to the mechanism that Nelson et al. (2013) described, by which the turbulent motions on large scales in the SPH-based Gadget simulation are spuriously dissipated into thermal energy,  rather than realistically dissipated by cascading to smaller scales in the AREPO. Consequently, part of the baryonic gas component is locked in the diffuse, hot halos of central galaxies in the SPH code, instead of accreting, cooling and entering into more over-dense central region. 

Fig. 1 demonstrates that highly developed vortices also shows filamentary and 
knotted structures. $\omega t$ can be 
as high as $\sim 100$, i.e., the turn-over time is about $~1 Gyr$ at $z=0$. 
The difference in the vorticity field 
between two samples shows a similar trend as in the baryonic matter. 
Vorticity in the WENO-512 simulation has a
higher magnitude and more concentrated configuration. Both the 
vorticity and baryonic density fluctuation variations in the TVD-512 sample lag than those in the WENO-512 sample. 

\begin{figure}[!htb]
\centering
\includegraphics[width=0.55\textwidth]{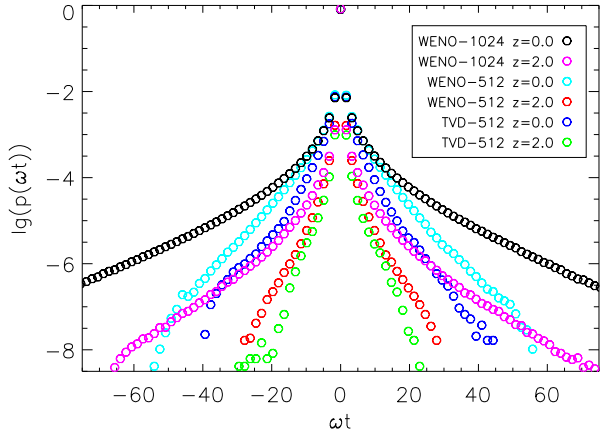}
\caption{Probability distribution function of the normalized 
vorticity $\omega t$ in the $512^3$ and $1024^3$ 
simulations run by the WENO and TVD codes at $z=2.0, 0.0$.}
\end{figure}

The probability distribution functions of vorticity in the
$512^3$ and $1024^3$ simulations are given in Fig. 3.  
Long, extended tails are present in the WENO-1024 since $z=2$, 
confirming the results in ZFF2010. The discrepancy 
between the WENO and TVD schemes is found to appear  earlier than $z=2$ 
and is much more evident at $z=0$.  We note that the vorticity in 
the WENO-512 simulation is under-developed compared to 
 that in ZFF2010, especially for $\omega t>20$, even 
though the resolution is the same. The 
difference should be mainly introduced by the different flux 
splitting methods. The local Lax-Friedrichs method in ZFF2010 bears more dispersion and slightly less dissipation. 

\subsection{Vorticity Power Spectrum and Turbulence-Developed Range}

The power spectra of vorticity can be further used to estimate 
the ranges of spatial scale within which turbulence have been fully developed in the IGM. 
According to the condition suggested by Batchelor (1959),  fully developed turbulence within a given scale range should satisfy the 
equality in wavenumber space, 
\begin{equation}
P_{\omega}(k) \approx k^2P_v(k), 
\end{equation} 
where $k$ is the wavelength, $P_{\omega}(k)$ and $P_v(k)$ are the 
Fourier power spectra of vorticity 
and velocity, respectively.  Namely, in the regions that turbulence 
is highly developed, the fluctuating velocity should be dominated by the curl component. 
In practice, we label the scale range where the 
margin is less than $\sim 0.3$ dex as the turbulent scale range. 

\begin{figure}[htb]
\centering
\includegraphics[width=8.0cm]{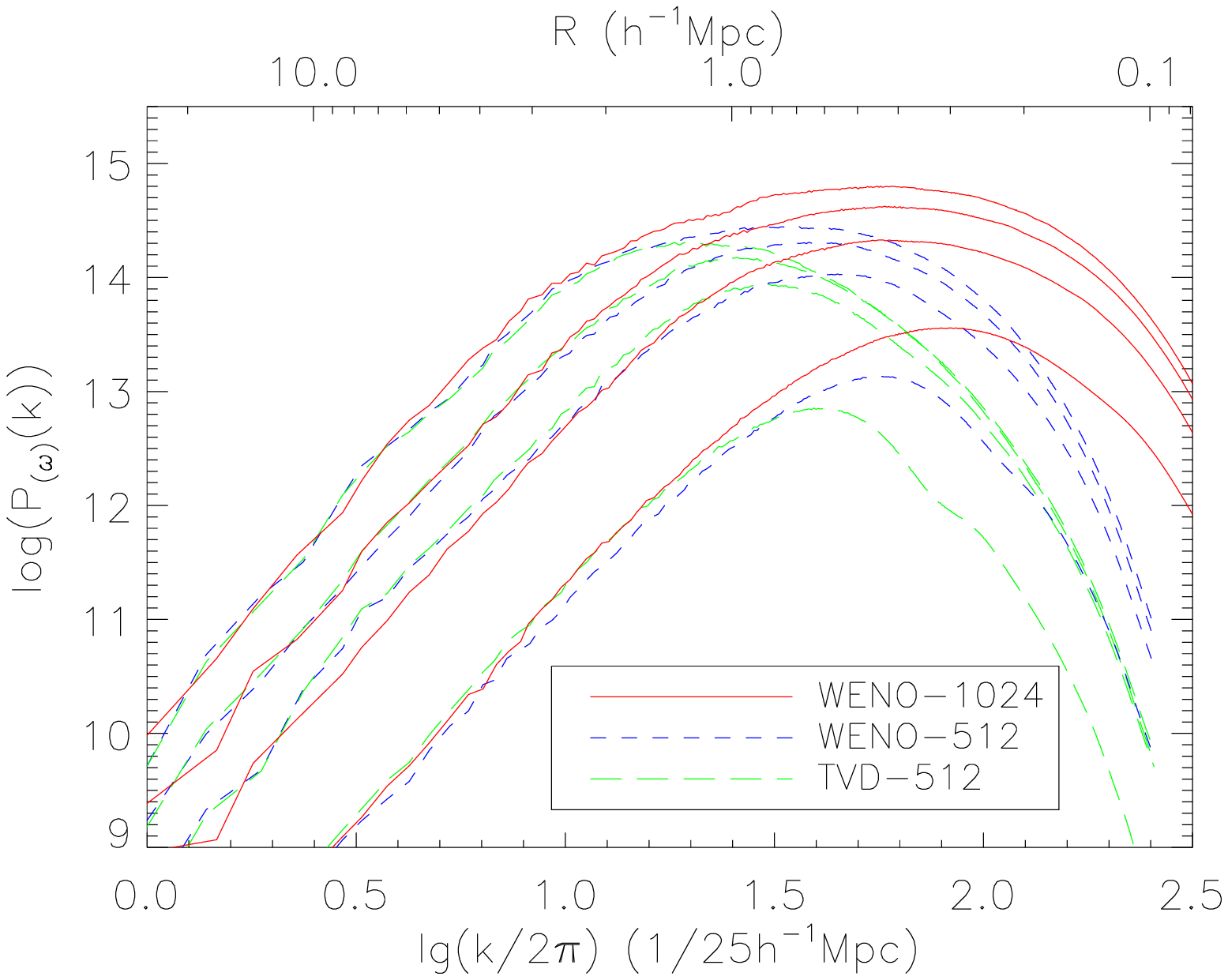}
\includegraphics[width=8.0cm]{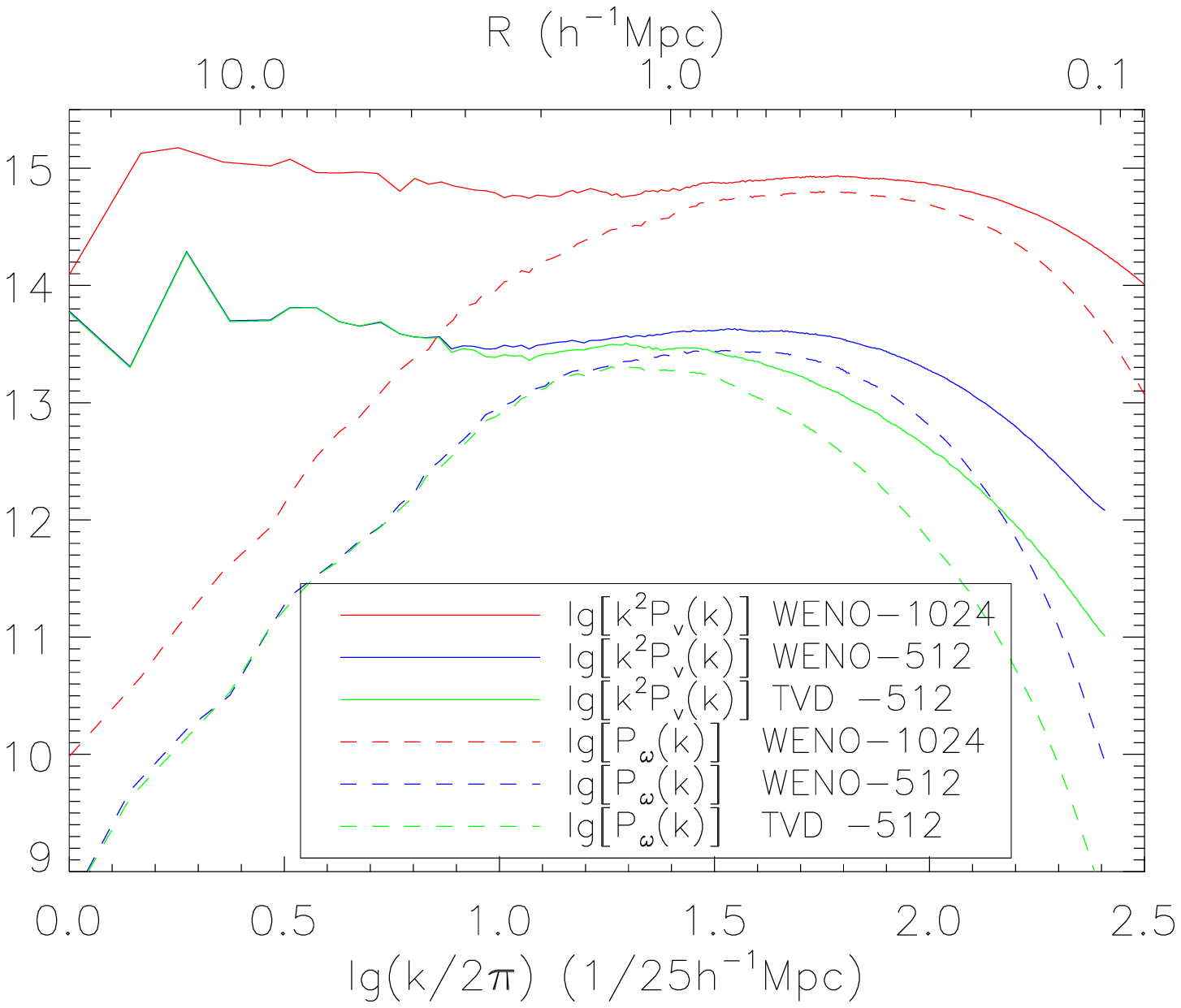}
 \caption{Left: Power spectra of vorticity at $z=2.0, 1.0, 0.5, 0.0
$, arranged in bottom-up order. Solid, dashed 
and long dashed lines represent the WENO-1024, WENO-512 and TVD-512 results.  
Right: Power spectra of 
velocity and vorticity in the $512^3$ and $1024^3$ 
samples at $z=0$. The lines of  the $512^3$ 
samples have been shifted down by an order of magnitude for the sake 
of clarity.}
\end{figure}

The evolution of vorticity power spectra since $z=2$ is given 
in Fig. 4, which suggests that most of the 
power is gained between $z=2.0$ and $z=0.5$, the same epoch in which 
the cosmic structure formation 
activity is the most violent. Both the differences between samples produced by different 
schemes with the same resolution, and by different 
resolution with the WENO scheme are very significant under 
$\sim 1.0 \  h^{-1} \ \mbox{Mpc}$, and can date back to $z=2.0$.  
The former could only result from the effect of dissipation by 
numerical viscosity, and the latter would be 
caused by the gravity force resolution additionally. The margins keep 
growing as the scale decreases 
down to $\sim 200\  h^{-1} \ \mbox{ kpc}$, and reach about an order of magnitude. 
Difference can barely be found at scales larger than $\sim 2 \ h^{-1} \  \mbox{Mpc}$.
 
In Fig. 4, we also compare the power spectra of velocity and vorticity 
at $z=0.0$. The turbulence 
scale range inferred from Eq.(2) are around $300 \ h^{-1} \mbox{kpc} - 2 \ h^{-1} \ \mbox{Mpc} $ and 
$600 \ h^{-1} \  \mbox {kpc} - 2 \ h^{-1} \ \mbox{Mpc}$ in the WENO-512 and 
TVD-512 samples, respectively. The turbulent 
range in the WENO-1024 run is  about $
\sim 200 \  h^{-1} \ \mbox{kpc} -2 \ h^{-1} \ \mbox{Mpc}$ at $z=0$, which is basically consistent with the result in 
ZFF2010, although the higher dispersion introduced by the local flux splitting likely have artificially 
expanded the turbulence range there.  Both the total velocity and its curl part in the IGM would be 
effectively damped if a considerable numerical 
viscosity exists. Actually, the damping effect of numerical 
diffusion on the velocity fluctuation has been 
observed in hydro-only, supersonic turbulence. While the upper end $l_u$ of the IGM turbulent range
 is robust, higher numerical  viscosity would lead to a larger 
dissipation scale, $l_d$, and hence a shorter turbulent range. Results 
of the positivity-preserving WENO schemes 
in this paper show a much weaker so called bottleneck effect
(Kritsuk et al. 2007) near the dissipation scale 
than in ZFF2010. 
 
\subsection{Compressive Ratio}
\begin{figure}[htb]
\centering
\includegraphics[width=8.0cm]{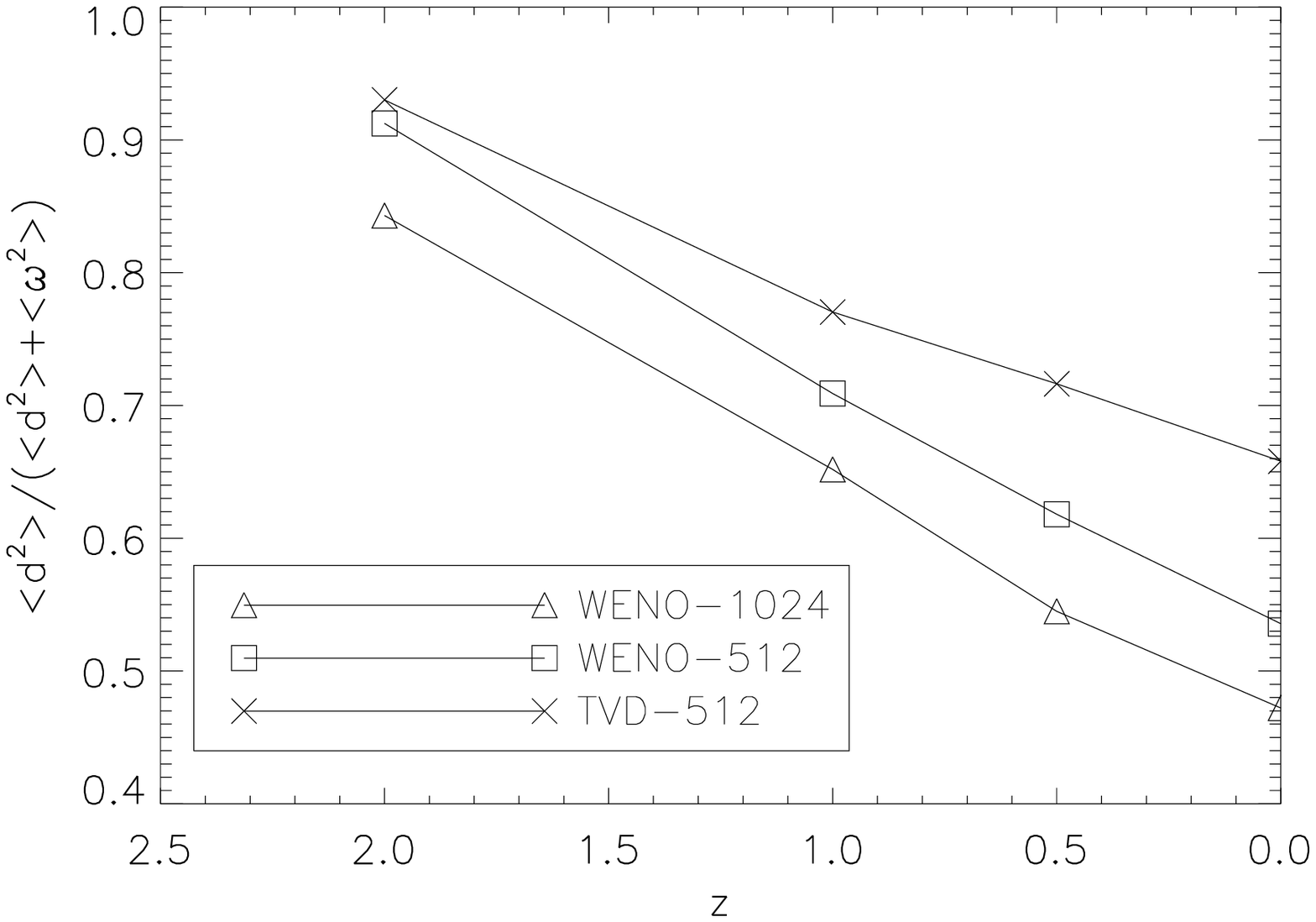}
\includegraphics[width=8.0cm]{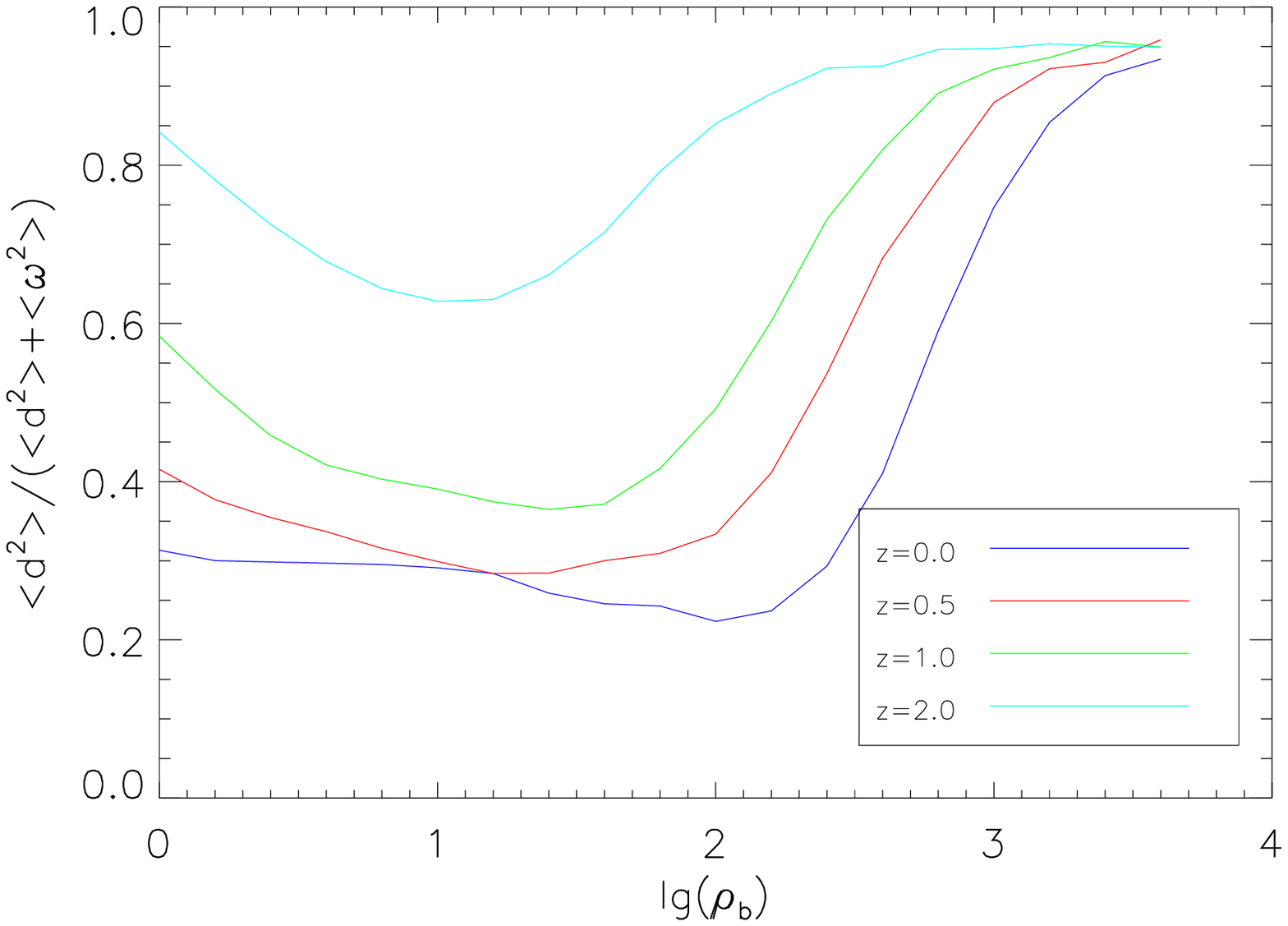}
\caption{Left: Small-scale compressive ratio in the WENO-1024
(triangle), WENO-512 (square) and TVD-512 (cross) 
simulations since $z=2$. Right: Small-scale compressive ratio 
in different baryonic density bins in the 
WENO-1024 simulation.}
\end{figure}

During the hierarchical structure formation, the motions of the 
IGM are driven by the curl-free gravity field, 
and hence are dominated by the compressive mode initially in both the linear and the quasi-linear regimes.
The vortical motion, mostly produced by 
the baroclinity in the post regions of curved shocks, develops rapidly 
since $z \sim 2$ and can overwhelm the 
compressive motion within the turbulence  range.  To characterize this
transition, we investigate the relative 
importance of the vortical over the compressible motions by the
so called small-scale compressive ratio (Kida \& Orszag 1992) 
\begin{equation}
r_{CS}=\frac{<|\nabla \cdot v|^2>}{<|\nabla \cdot v|^2>+
<|\nabla \times v|^2>}=\frac{<d^2>}{<d^2>
+<\omega^2>}
\end{equation}
 
Fig. 5 gives the small-scale compressive ratio $r_{CS}$ in the $512^3$ and 
$1024^3$ simulations. The vortical motion 
undergoes a dramatic growth at $z<2$, and is comparable to the 
compressive motion in the 
$1024^3$ simulation at $z=0.0$,  Consequently $r_{CS}$ drops from $0.84$ to $0.47$ 
during this period. The magnitude of $r_{CS}$ at $z=0$ is close to the compressively driven
isothermal turbulence with root mean square Mach number $\simeq 2.5$ 
in Schmidt et al(2009), and higher than the time-averaged value, $0.28$, for the isothermal turbulence
with  Mach number $M \simeq 6.0$ in Kritsuk et al (2007). Lower resolution and relatively higher 
numerical viscosity attenuate this transition and result in relatively 
higher compressive ratio. $r_{CS}$ is about $0.66$ at $z=0.$ in the TVD-512 sample, suggesting the
motion of the IGM is still dominated by the compressive mode.  
 
We calculate $r_{CS}$ in different gas density bins in the 
$1024^3$ simulation, which is also presented 
in Fig. 5.  In regions with baryonic density $\rho_b \lesssim 10^3$, the vortical motion grows rapidly since $z=2.0$.  
For $\rho_b \lesssim 10^2$, the curl velocity can overwhelm the compressive part as early as $z=1.0$. 
Moreover, $r_{CS}$ can be as low as $0.2$ for $\rho_b \simeq 100$ at $z=0.0$. 
Comparing with the compressive motions, the vortical motion grows more 
rapidly before entering into outskirts of clusters, i.e, $\rho_b \sim 100$,
In the clusters and their outskirts, the vortical motion may dissipate faster than the compressive motions.
The magnitude of $r_{cs}$ here is volume weighted, which would be significantly higher than the corresponding 
mass-weighted. The discrepancy in Schmidt et al.(2009) is around $0.2 \sim 0.3 \ dex$, comparable with the difference between our results and
the mass weighted value in Iapichino et al.(2011).

\section{Properties of Curl and Compressive Velocity}

In this section, we focus on the two components of the velocity 
field of the IGM, $\vec{V}$, which can be 
separated by the Helmholtz-Hodge decomposition (Ryu et al. 2008;
Sagaut \& Cambon et al. 2008). 
\begin{equation}
\vec{V}=\vec{V}_{curl}+\vec{V}_{div}+\vec{V}_{unif}
\end{equation}
where,  $\vec{V}_{curl}$ is the divergence-free curl velocity 
associated with the vorticity,
$\nabla \cdot \vec{V}_{curl}=0$ and $\nabla \times \vec{V}_{curl} 
= \nabla \times \vec{V}$; $\vec{V}_{div}$ is the curl-free component, 
$\nabla \times \vec{V}_{div}=0$ and $\nabla \cdot \vec{V}_{div} = 
\nabla \cdot \vec{V}$. The curl-free component $\vec{V}_{div}$ 
represents the compressive velocity. The 
remaining component $\vec{V}_{unif}$ is both curl-free and 
divergence-free, which is actually negligible in the simulation.

\subsection{Curl Velocity Distribution}
As an example, Fig. 6 shows the velocity field and its curl 
component projected over the gas density in the WENO-512 sample in the same 
slice as Fig.1. The curl velocity is basically associated 
with the high vorticity region, consistent with the definition above. 
The maximum curl velocity in the slice is $~150 km/s$, relative 
to $~200 km/s$ for the total velocity. 

\begin{figure}[htb]
\centering
\includegraphics[width=8.0cm]{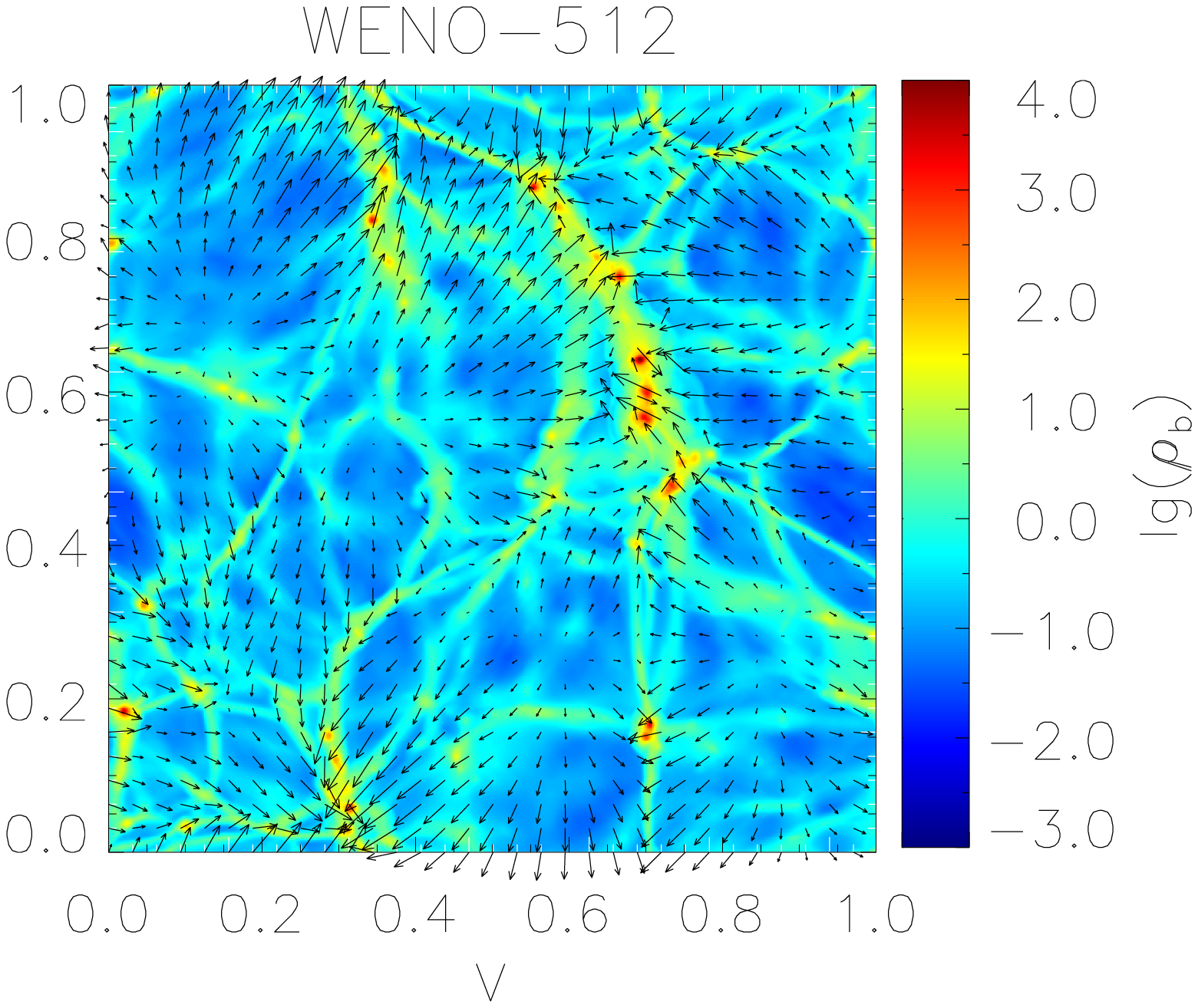}
\includegraphics[width=8.0cm]{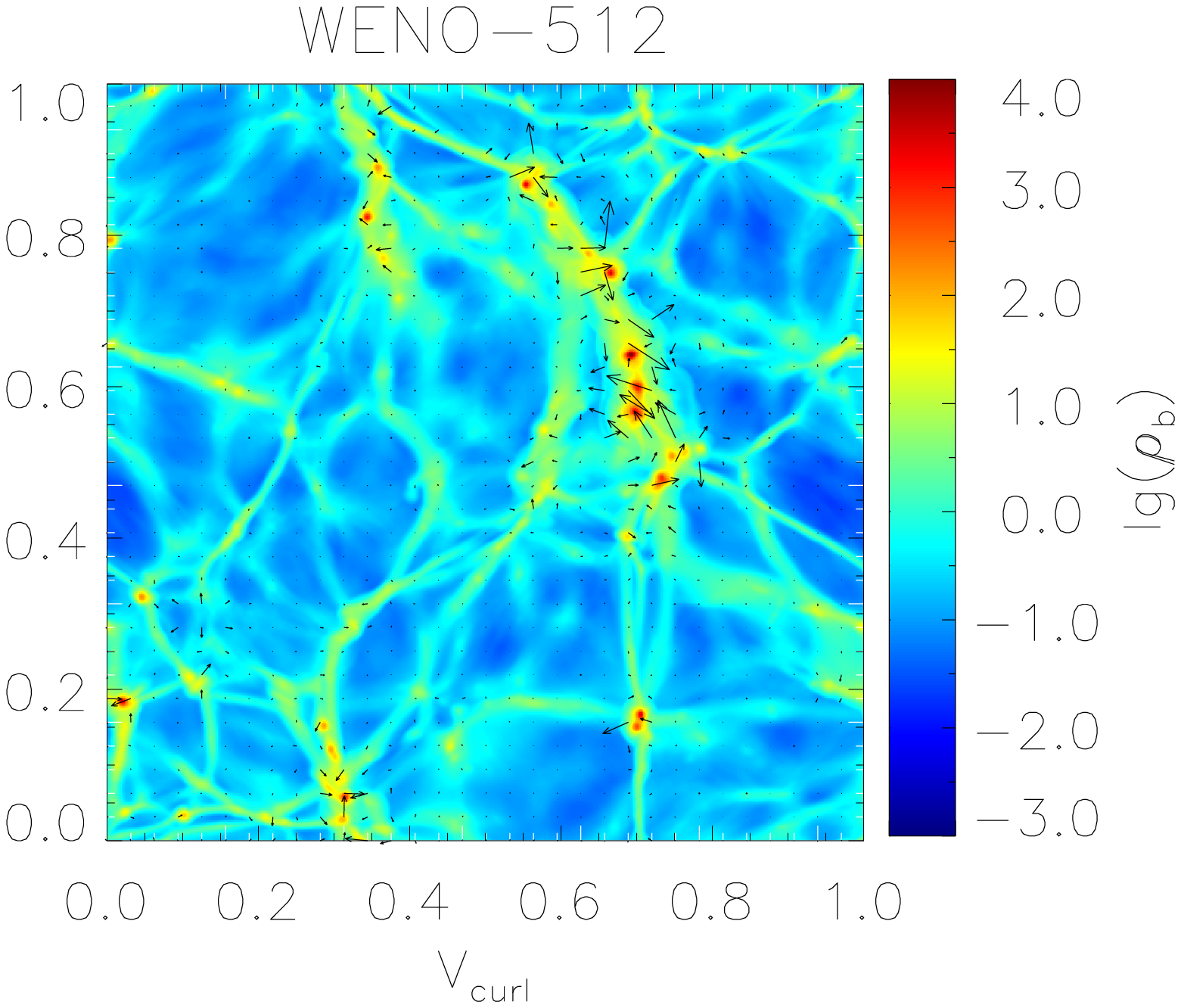}
\caption{Velocity field and its curl component of the IGM projected 
over gas density in the same 
slice as in Fig.1 at z=0, extracted from the WENO-512 sample. The maximum 
velocity and curl velocity are around 
$~200 km/s$ and $~150 km/s$ in this slice.}
\end{figure}

The mean curl velocity as function of gas density, 
$\bar{V}_{curl}(\rho_b)$, and temperature, $\bar{V}_{curl}(T)$ 
from $z=2.0$ to $z=0.0$ are presented in Fig. 7. We focus on 
the region with $\rho_b >1.0$, 
where the vorticity and curl velocity are well developed. 
Correspondingly, the temperature is generally 
higher than $10^4 K$, as shown in Fig. 2. Both the magnitude and 
trends are almost the same in all the 
simulations.  A discrepancy of a few $km/s$ exists all through the 
range. As a very small fraction of cells has gas density 
$ \rho_b >3\times 10^3$ in the TVD samples, comparison is not 
available there. Starting at a level of $10 km/s$ at low density or 
temperature, the magnitude of $\vec{V}_{curl}$ grows rapidly till 
$\rho_b \sim 30$, or $T \sim 10^6 k$, reaches $\simeq 90 km/s$ 
and flattens out henceforth at $z=0.0$. Fig. 7a,  also shows our approximation at z=0 as the black solid line, 
given
\begin{equation}
V_{curl}(\rho_b)=10.0 \times exp\{-[lg(\rho_b/100.0)]^2\}+85.0 \times exp\{-[lg(\rho_b/\rho_c)/2]^2\}  \ km/s ,
\end{equation}
where, $\rho_c=200.0/\Omega_m$. 

The magnitude in highly over-dense 
clusters in our simulation is below the weak constraints placed by direct or indirect 
observation methods, i.e, $\simeq 200 \sim 1000 km/s$(Sanders et al. 2010, 2011,2013).  
The absence of recipes dealing with star formation, magnetic fields and AGN in our simulation may 
underestimate the turbulent motion in clusters.  The level of $ 10 \sim 20 km/s$ at $z=2.0$ 
is also lower than the turbulent line broadening in Zheng et al.(2004) at $z \simeq 2.7$ 
by a factor of  $\simeq 2$. In Oppenheimer \& Dave (2009), the density-dependent sub-resolution 
turbulent velocity added by hand are 
$b_{turb}=13, 22, 40$ and $51 \ km s^{-1}$ for $\rho_b=20, 32, 100$ 
and $320$ at $z=0.25$, in order to match 
the O VI absorbers in simulation with observation. The assumed 
relation between turbulence velocity and 
density is supposed to follow the fitting formula $v_{turb}=13.93 log(n_H)+101.8 km s^{-1}$, for 
$n_H=10^{-4.5}-10^{-3.0} cm^{-3}$.  Their magnitude  
is lower than our results in the $1024^3$ simulation by 
$\sim 15 \  km s^{-1}$. 

\begin{figure}[htb]
\centering
\includegraphics[width=8.0cm]{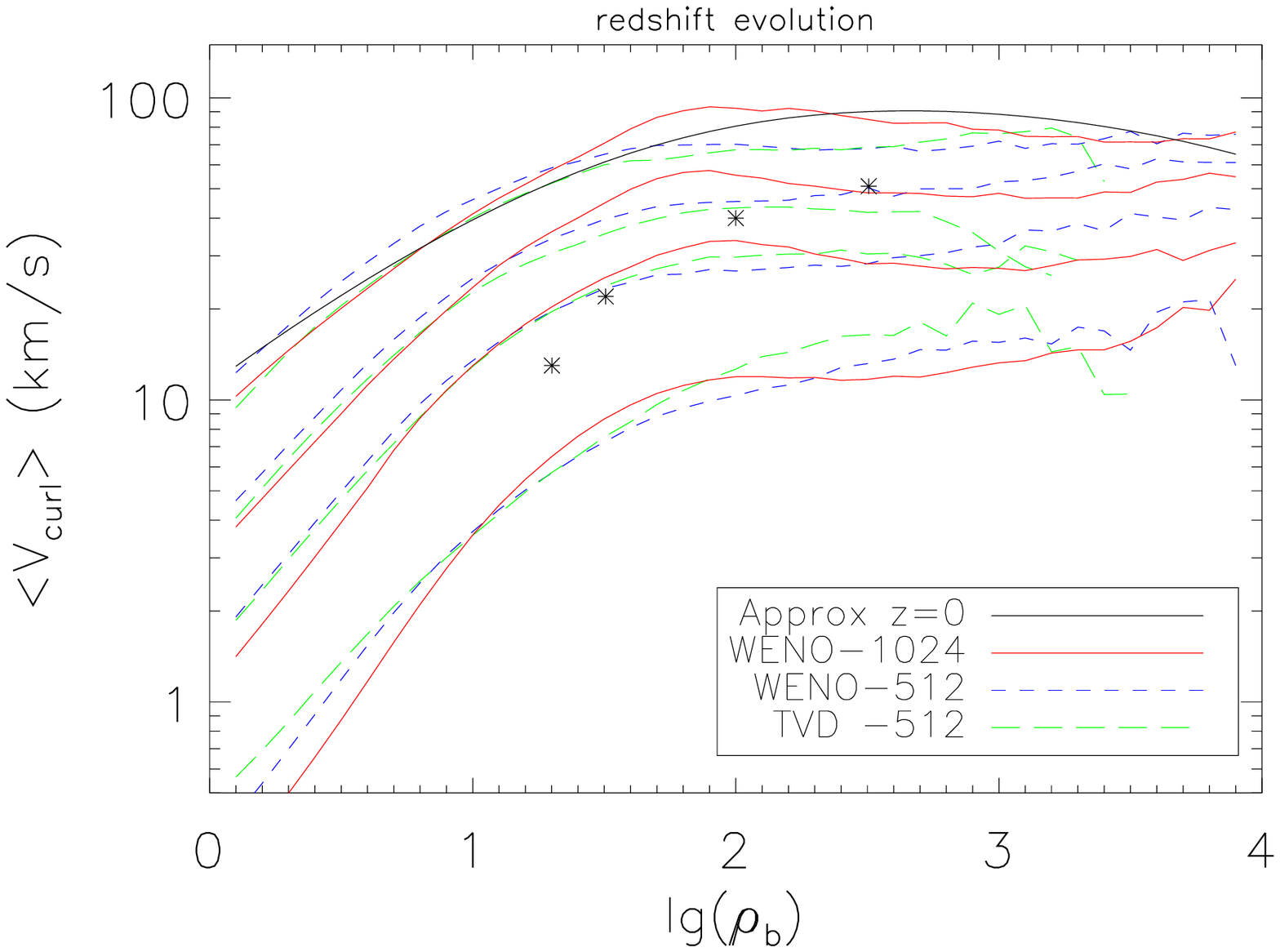}
\includegraphics[width=8.0cm]{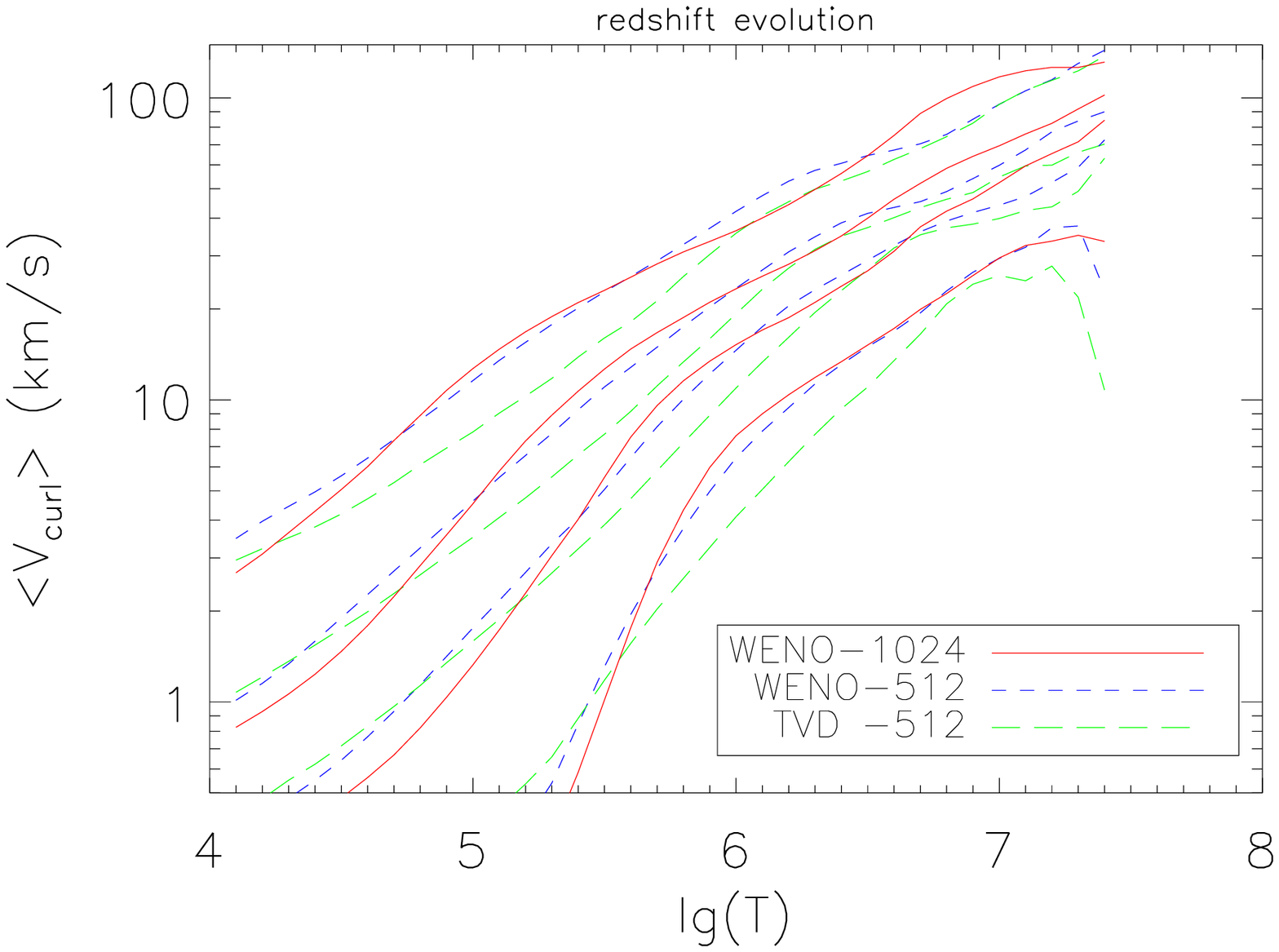}
\caption{The curl velocity as function of gas density (left), and 
temperature (right). Lines of $z=2.0, 1.0, 0.5, 0.0$ are in bottom-up order.  The black solid line in  the left panel indicates the approximation at z=0. 
Asterisks indicate the un-resolved sub-grid turbulent velocity added by hand in OD09 at $z=0.25$.}
\end{figure}

The observed tendency of curl velocity in all the samples would have been affected by 
the UV background, structure 
formation history and  dissipation simultaneously. Re-ionization by 
the UV background will enhance the 
thermal energies in our simulations and heat the gas in the
low density regions and damp the
pressure fluctuations. Once entering into the nonlinear regimes, 
the violent accretion of gas onto sheets 
and filaments will effectively produce shocks and vorticity, and 
hence the curl velocity. Viral relaxation will make  
the vorticity decay in the highly nonlinear regimes and flatten the 
curl velocity function. 

As has been shown in Fig.2, there is a notably larger fraction of 
gases located in the high density regions in the 
WENO-512 samples. Difference in the total kinetic turbulent energy 
between two codes is         
considerable, as $\bar{V}_{curl}(\rho_b)$ is about the same. Rather 
than cascading into small scales, a 
substantial part of the kinetic turbulent energy in the TVD-512 
samples may have been dissipated into 
thermal energy by numerical viscosity during cosmic 
structure formation. 

\subsection{Curl and Compressive Velocity Power Spectra}

For the fully developed turbulence in incompressible fluids, the velocity power spectra 
follow the well known Kolmogorov law as 
$P_v(k) \sim k^{-5/3}$ in the inertial range. Since the early 1990s, 
two- and three-dimensional high resolution 
simulations of both forced and decaying supersonic turbulence in 
the fields of hydrodynamic and interstellar 
medium, however, show that the power index of the velocity power 
spectra deviates from $-5/3$.  A $k^{-2}$ power law is obtained 
in both mildly and highly compressible turbulence
(Porter et al. 1992, 1998; 
Kritsuk et al. 2007, Federrath et al. 2010), and is close to 
the Burgers scaling. 

\begin{figure}[htb]
\centering
\includegraphics[width=8.0cm]{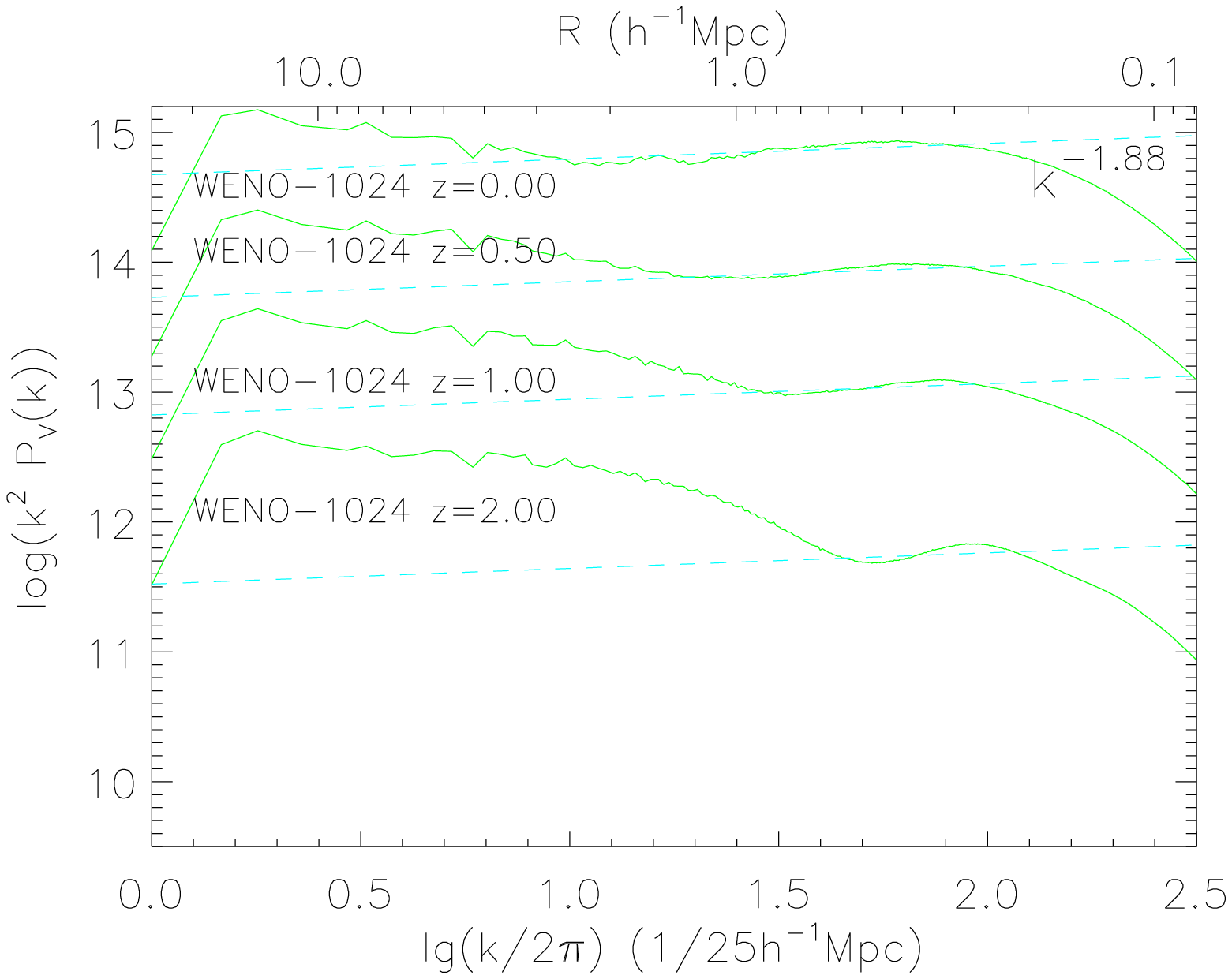}
\includegraphics[width=8.0cm]{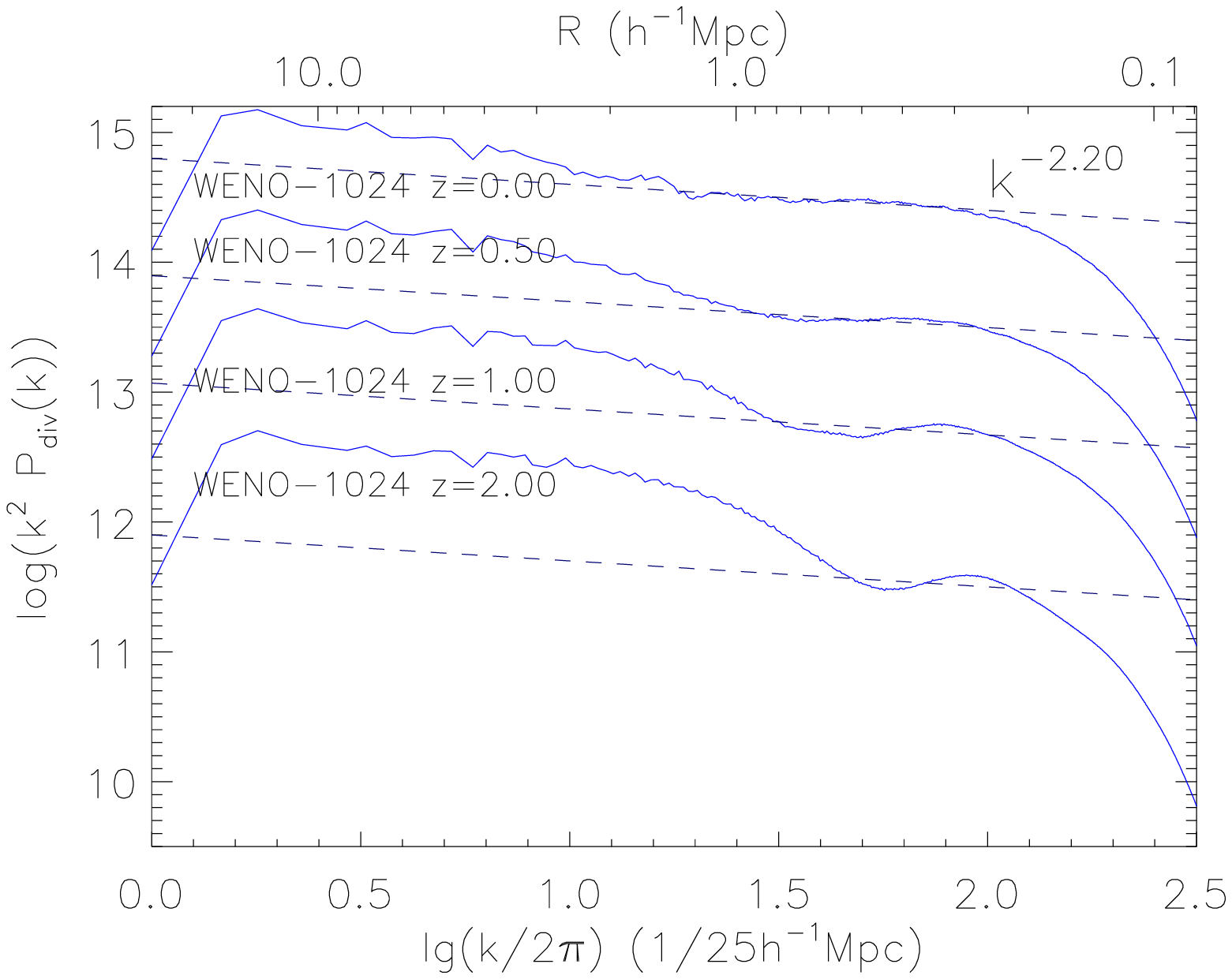}
\includegraphics[width=8.0cm]{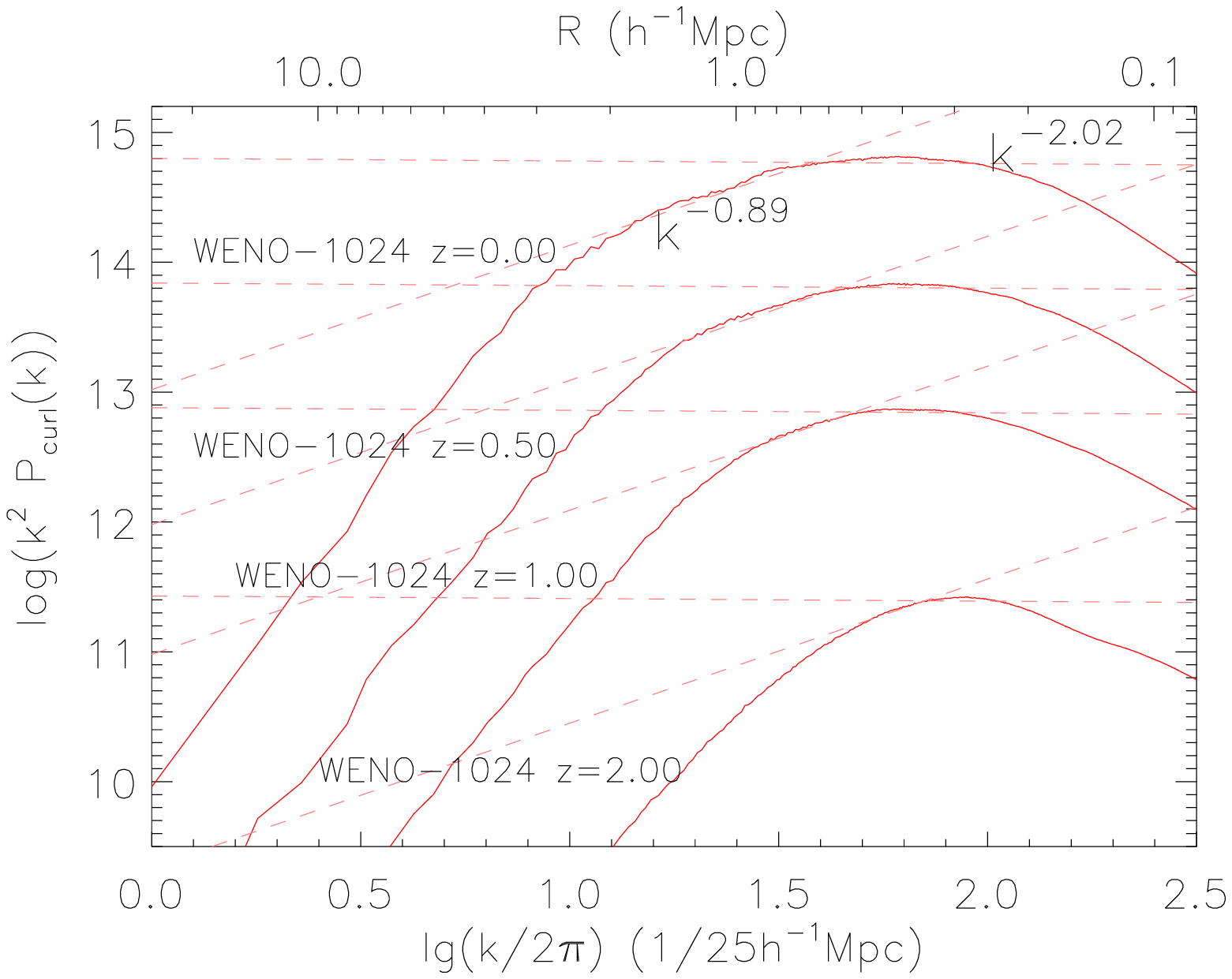}
\includegraphics[width=8.0cm]{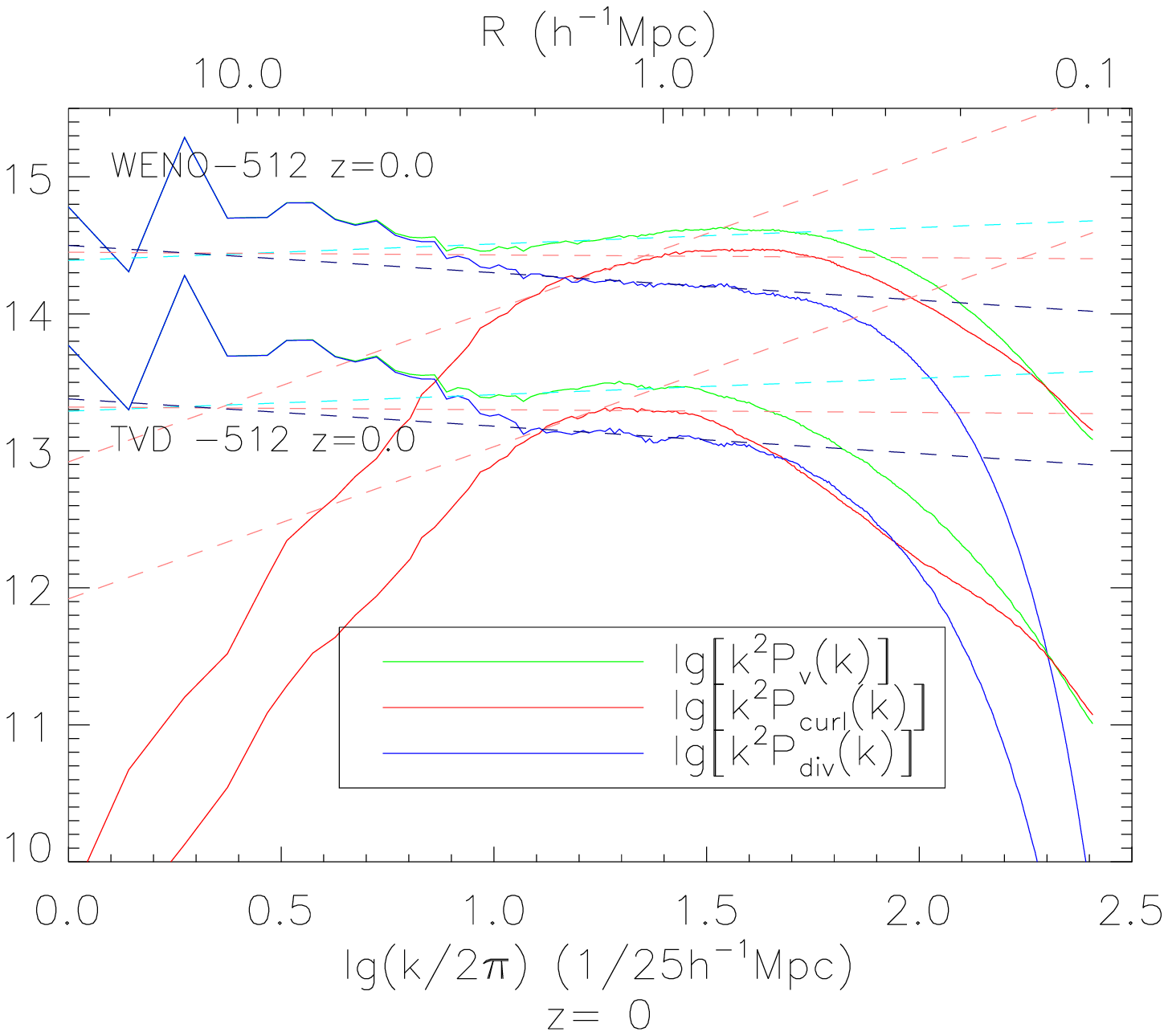}
\caption{Compensated power spectra of velocity (top left), and its 
compressible (top right) and curl
(bottom left) components in the WENO-1024 simulation at $z=2.0, 1.0,
0.5, 0.0$, arranged in bottom-up order in each panel. Lines of $z=0.5, 1.0, 2.0$ have been shifted down by 
$0.5, 1.0, 1.5$ dex respectively for the sake of clarity. Bottom right: Results of the WENO-512 
and TVD-512 samples at $z=0$. Dashed straight lines are the fitted power law.}
\end{figure}

Fig.8 displays the compensated power spectra of flow velocity, 
$k^2P_v(k)$(upper left), and the curl, $k^2P_{v,curl}(k)$ (lower left)
as well as  the compressive components $k^2P_{v,div}(k)$ (upper right) measured in our highest resolution run, 
WENO-1024. Clearly, 
$k^2P_{v,curl}(k)$ overwhelms $k^2P_{v,div}(k)$ in 
the inertial range of fully developed turbulence identified by Eqn.(2).  At $z=0$,
the velocity power spectra is found to be well fitted by $k^{-1.88}$ within the turbulence range. 
Actually, this scaling behavior was established 
as early as $z=1.0$. Similarily, the power spectra of the 
compressible velocity follows a scaling law of $k^{-2.20}$ in the 
same range. While for the curl velocity,  the scaling behavior shows 
a transition from $k^{-0.89}$ to 
$k^{-2.02}$.  The transition scale $l_t$ grows along with the development of turbulence, namely, 
grows with the cosmic time as the upper-end scale, $l_u$, does. For instance, for the WENO-1024 simulation, 
$l_t$ is about $0.32h^{-1}Mpc$ at $z=2.$ and goes up to $0.68h^{-1}Mpc$ at $z=0$.  Meanwhile, the range for the fully developed turbulence 
is enlarged from  $0.2 \sim 0.6 h^{-1}Mpc$ at $z=2$ to $0.2 \sim 2.0 h^{-1}Mpc$ at $z=0$. 
The same scaling law is also justified in the $512^3$ samples , but the disspation scale, $l_d$, is found to be a little 
bit larger than the WENO-1024 samples due to the numerical viscosity.  For the WENO-512, it is about $0.3h^{-1}Mpc$.
The $512^3$ samples follow the same scaling law in their shortened 
turbulence inertial range, with $l_t$ is enlarged as $l_d$ does due to numerical viscosity. 

The scaling law of the velocity power spectrum in our samples 
indicates that the turbulence in the IGM is 
more likely following the Burgers turbulence scaling between 
$l_t$ and $l_d$, consistent with results in 
previous works of supersonic isotropic turbulence
(Porter et al. 1992, 1998; Kritsuk et al. 2007, Federrath et 
al. 2010). This range would correspond to the supersonic phase, as
proposed by Porter et al. (1992, 1994), 
where both the curl and compressive velocity power spectra follow 
$\sim k^{-2}$.  During this phase, the 
vorticity can be effectively produced by the baroclinity term, as 
shocks form and interact with each other. 
The error introduced by the bottleneck effect, of which more 
details can be found in Kritsuk et al. 
2007, is about $0.05$ in the power index of the total velocity spectrum.  

On the other hand, the velocity field in the turbulence range between the transition scale,
$l_t$, and the upper-end scale, $l_u$, exhibits a similar scaling as the 
post-supersonic phase in Porter et al. (1992, 1994), where the 
compressive velocity component still follows a $\sim k^{-2}$ law 
while the curl velocity follows a $\sim k^{-1}$ law.  The vorticity 
evolution in this phase is subject to subsonic 
vorticity dynamics, dominated by vortex interaction and decay, 
although shocks are still active. In summary, turbulence in the
IGM may possess two different phases simultaneously. 

\subsection{Turbulent Pressure Support and Baryon Fraction}

\begin{figure}[htb]
\centering
\includegraphics[width=8.0cm]{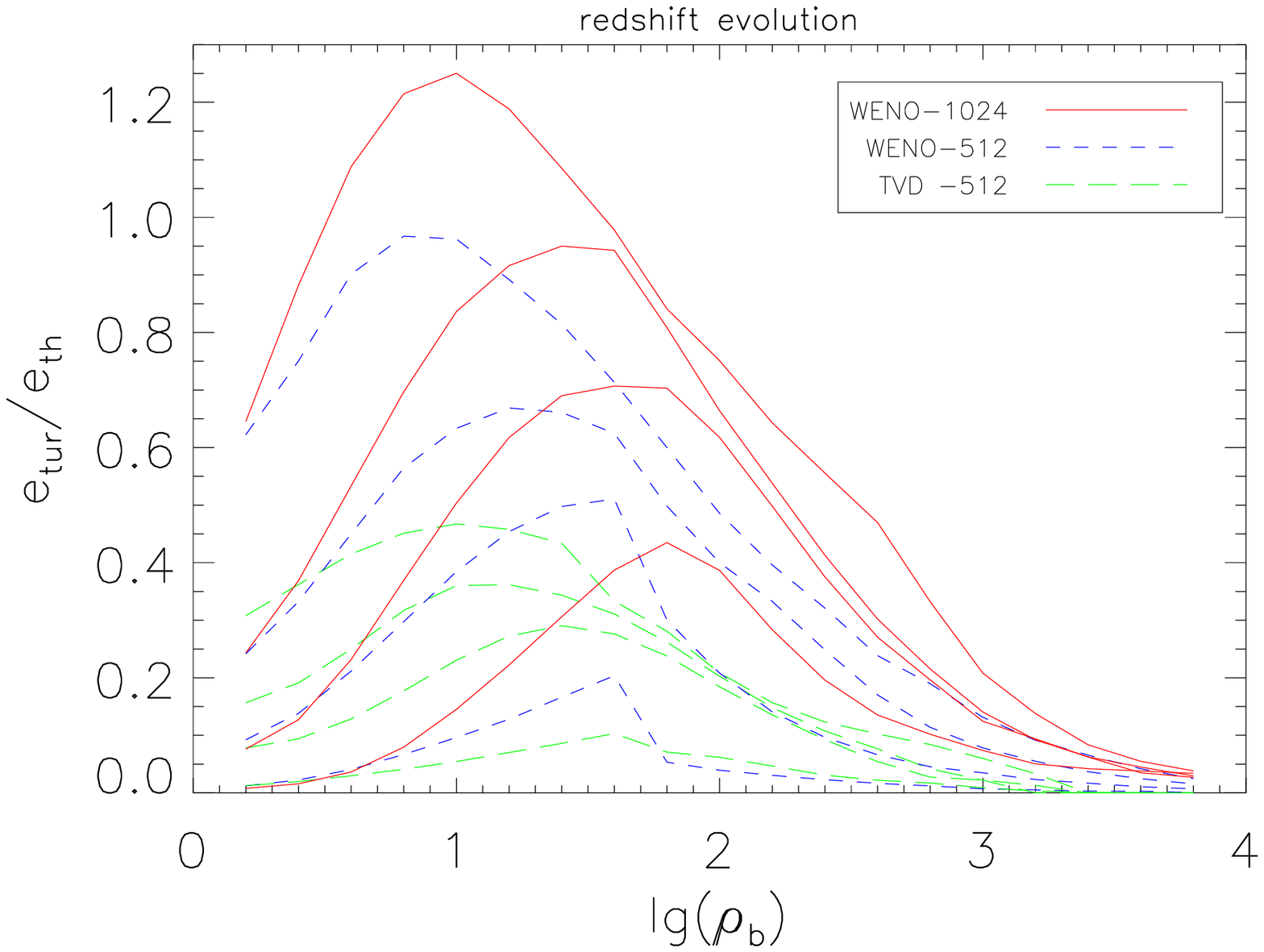}
\includegraphics[width=8.0cm]{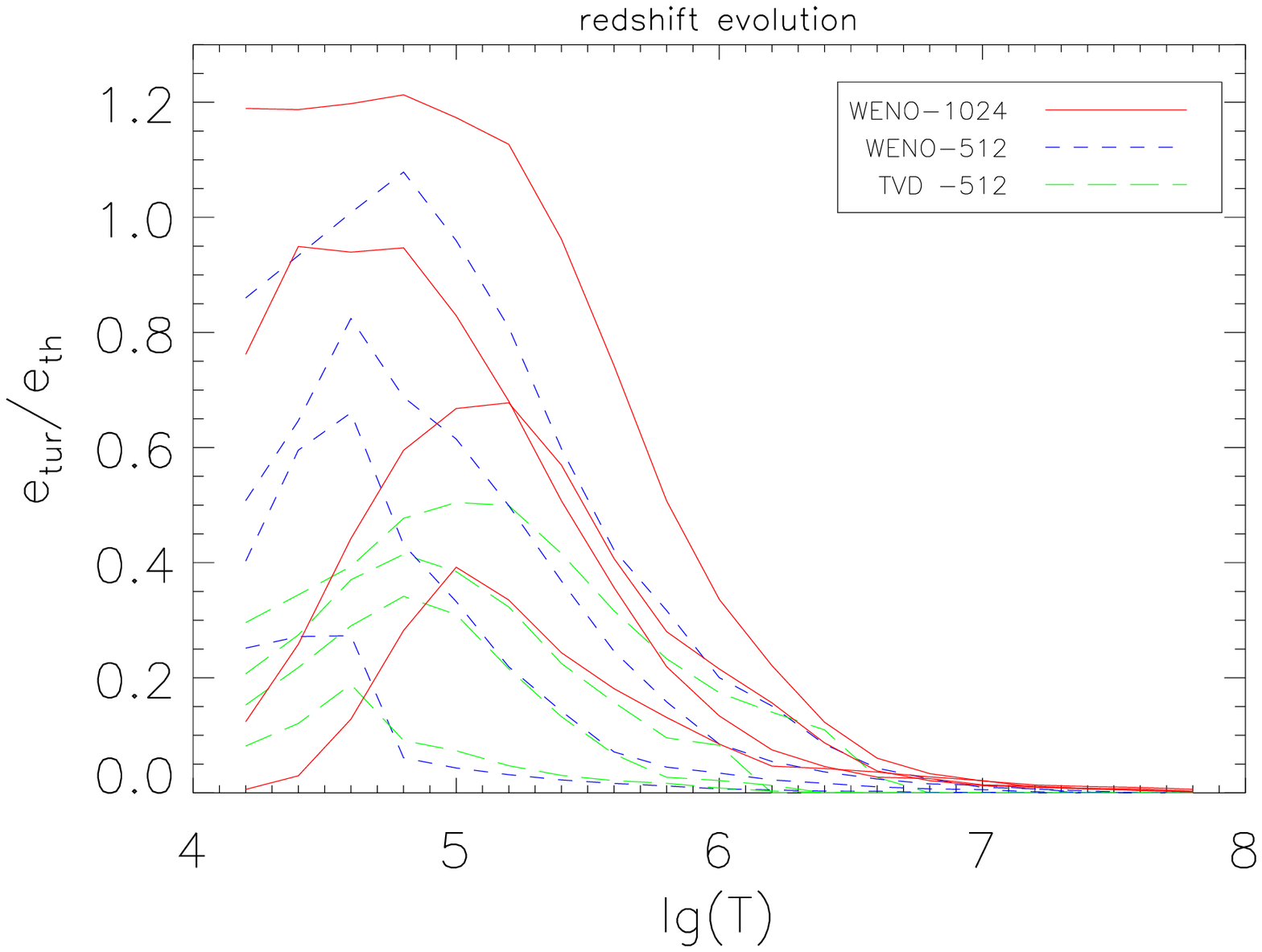}
\caption{The ratio of kinetic turbulent energy to thermal energy as 
 function of gas density (left) and 
temperature (right). Lines of $z=2.0, 1.0 , 0.5, 0.0$ are in bottom-up order.}
\end{figure}

The vortical motions would provide a non-thermal turbulent
pressure during gravitational collapse
(Chandrasekhar 1951a, 1951b).  Ryu et al. (2008) used the kinetic 
energy associated with vortices, $ \rho 
V^2_{curl}/2$, to estimate the kinetic turbulent energy in the IGM. 
The turbulence pressure support  is 
then measured by the ratio of  $\varepsilon_{tur} 
\sim \rho V^2_{curl}/2$ to the internal  thermal energy, 
$ \varepsilon_{int}$.  Fig. 9 shows the mass weighted mean 
non-thermal support ratio as function of gas 
density, $<\varepsilon_{tur}/\varepsilon_{int} (\rho_b)>$, and 
temperature, $<\varepsilon_{tur}/
\varepsilon_{int} (T)>$ at different redshifts in our simulations. The trends with
increasing temperature and density, and the 
evolution with decreasing redshift are almost the same in 
different samples.                

For the same resolution, the ratio in the WENO-512 sample is significantly 
higher than the TVD-512 sample, and is close to $1.0$ for $\rho_b 
\simeq 10$, or $ T \simeq 10^5 T $ at $z=0$.  This result is actually in agreement 
with the curl velocity distribution $\bar
{V}_{curl}(\rho_b)$ and $\bar{V}_{curl}(T)$.  As shown in Fig.2, the 
baryonic mass fraction in the $\rho_b-T$ phase diagram varies much between the WENO and TVD simulations.  
For overdense regions $\rho_b>1.0$, the mass weighted mean 
temperature, $\bar{T}(\rho_b)$ in the WENO-512 sample is lower, namely,   
the thermal energy density $
\varepsilon_{int} \sim \rho \bar{T}(\rho_b)$ is also lower.  
Since $\varepsilon_{tur}\sim \rho_b \bar{V}_{curl}^2(\rho_b)$ are comparable in two codes, consequently, the ratio 
of $\varepsilon_{int}/\varepsilon_{tur}$ in the WENO-512 sample is getting higher.    

The mean ratio as a function of temperature
 is more complicated. In the range of $10^4-10^5 K$, the 
mass distribution at a given temperature, $f(\rho_b, T)$, 
does not vary much between two codes. Higher $<
\varepsilon_{tur}/\varepsilon_{int} (T)>$  in the WENO-512 results 
from  higher $<\varepsilon_{tur}/\varepsilon_
{int} (\rho_b)>$  for $\rho_b=1.0-10^4$.  For $T >10^5K$, 
the mean density, $\bar{\rho_b}(T)$,  
in the WENO-512 is shifted to higher density region comparing to the TVD-512, and is 
generally larger than tens of the cosmic mean. On 
the other hand, $\varepsilon_{tur}/\varepsilon_{int}$ decreases
 as the gas density increases for $
\rho_b > 10$ in both codes. Consequently, the margins in 
$<\varepsilon_{tur}/\varepsilon_{int}(T)>$ between 
the WENO-512 and TVD-512 simulations peak at around $10^5 k$, depending on 
the redshift, and would decrease toward 
higher temperature. 

In the reference run WENO-1024, the non-thermal pressure support 
is substantially stronger than 
the WENO-512 sample, and is comparable to the thermal pressure for 
$\rho_b \simeq 10 \sim 100$, or $T <10^{5.5} K$ at $z=0.0$. 
The kinetic turbulent energy ratios as a function of temperature 
obtained here are different from Ryu et al.
(2008) at $T<10^{4.5} K$. More specifically, the ratio grows rapidly as the temperature decrease at $T<10^{4.5} K$ 
in Ryu et al.(2008). Meanwhile,the magnitude there is larger than ours by about $0.2$ dex. 
These differences should mainly result from the UV background, which is not included in Ryu et al. (2008). 
The UV background would enhance the temperature of gas and might smear out pressure 
fluctuations and hence the baroclinity outside of clusters. In addition, the magnetic field included in Ryu et al.(2008) might 
amplify the turbulent motion in the IGM. 

\begin{figure}[htb]
\centering
\hspace{-10mm}
\includegraphics[width=8.0cm]{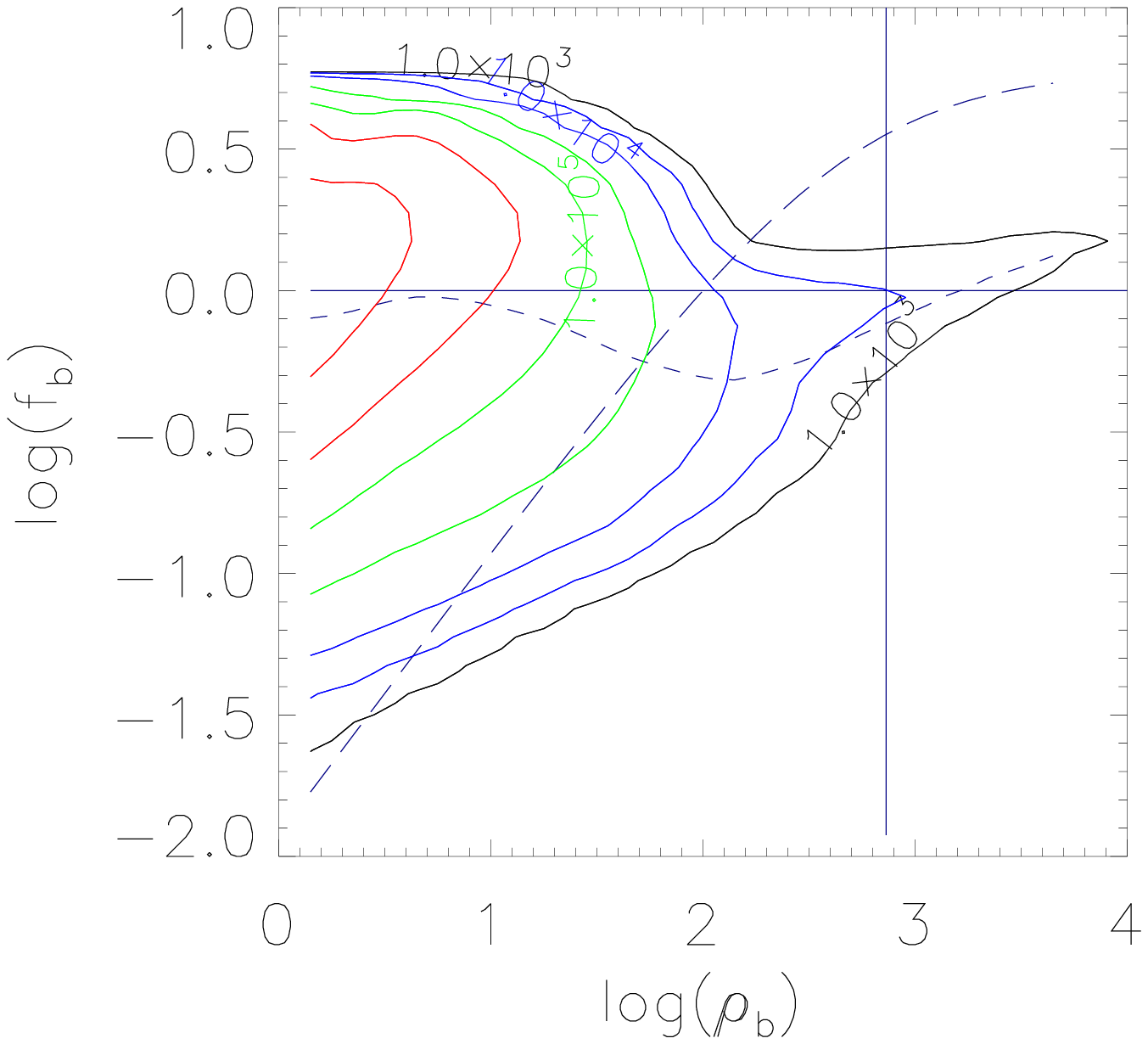}
\hspace{-10mm}
\includegraphics[width=8.0cm]{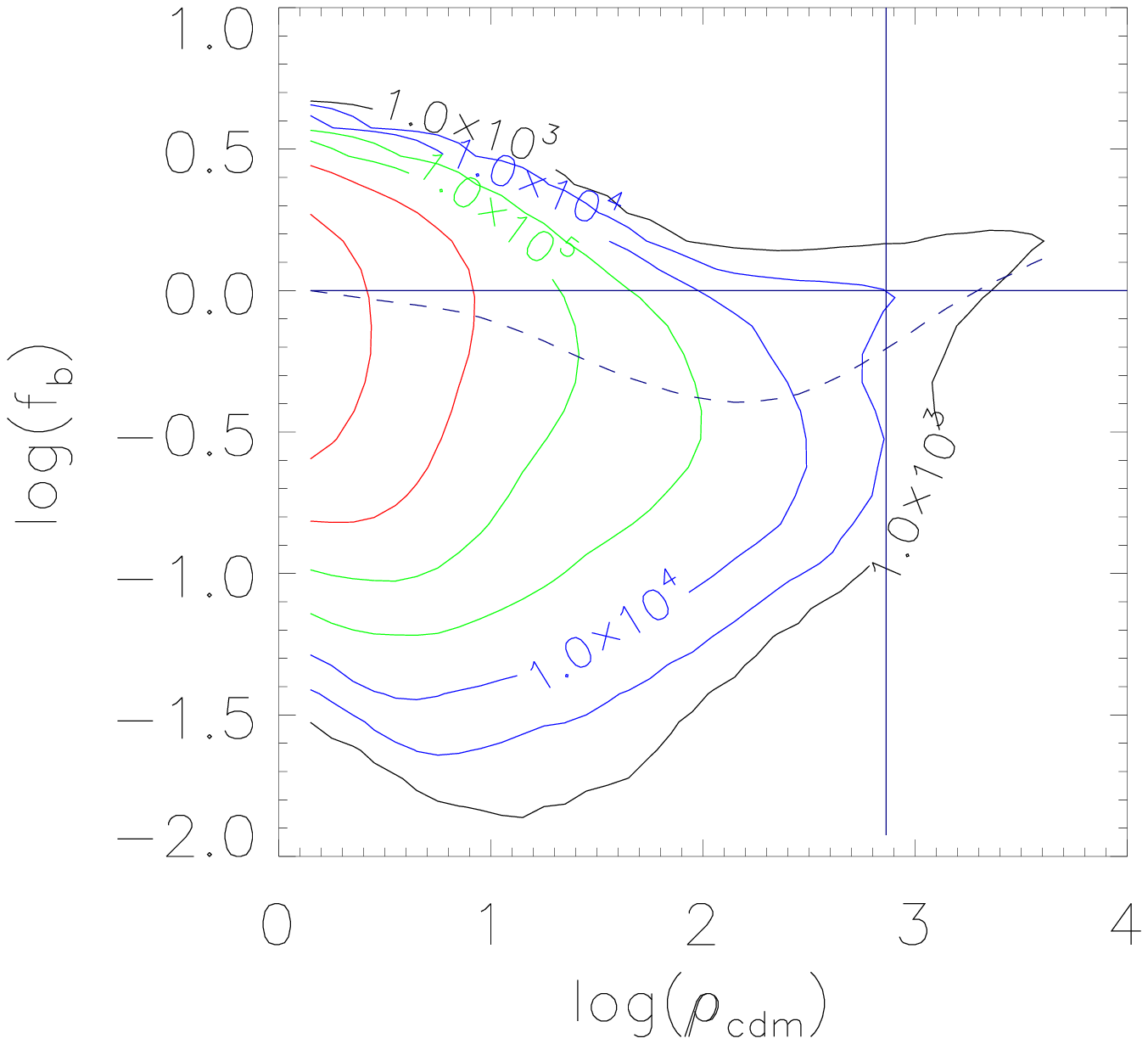}
\hspace{-10mm}
\caption{Left (right) panel shows contours of cell-counting 
in the phase space of baryonic (dark) matter density and normalized baryon 
fraction in the WENO-1024 sample at $z=0.0$. Short 
dashed lines are the 
averaged baryon fraction $\bar{f}_{b}(\rho_{b})$($\bar{f}_{b}(\rho_{cdm})$) in different 
density bins. Long dashed lines in the left 
panel indicate a constant dark matter density of $100.0$.}
\end{figure}

Turbulence pressure of the IGM may cause both the density and velocity 
fields of baryonic gas to be different 
from those of underlying collisionless dark matter, result in 
the baryon fraction deviating from the cosmic 
mean $F_b^{\rm cosmic}$ (Zhu, Feng, Fang 2011).  Fig.10 shows the 
contours of cell numbers in the space of normalized baryon 
fraction $f_b$ and gas density drawn from the $1024^3$ simulation 
at $z=0.0$.  Both the gas and dark matter density at grid cells 
are smoothed with a radius of one grid cell, and then used to calculate 
the baryon fraction. A considerable volume of the cells does 
deviate from the cosmic mean, and shows significant scatter.   
  
\begin{figure}[htb]
\centering
\includegraphics[width=14.0cm]{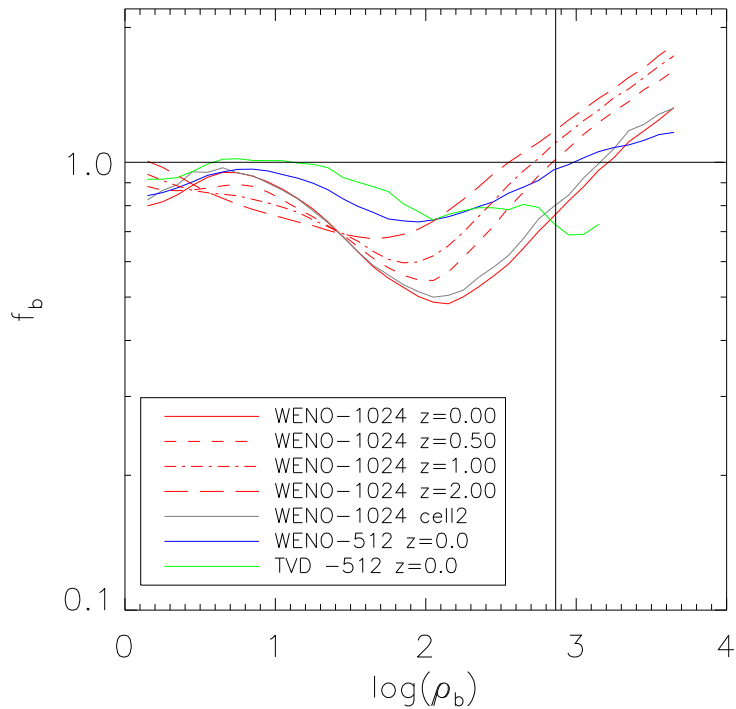}
\caption{Averaged baryon fraction $\bar{f}_{b}(\rho_b)$ as a function 
of baryonic matter density in the $1024^3$ and $512^3$ simulations. The cell2 represents
averaged baryon fraction in cubic boxes of side length of two grid cells. }
\end{figure}

The baryon fraction in different gas density bins, shown as the 
short dash line in the left panel of Fig.10, suggests that the 
maximum deviation lies in $\rho_b \sim 100$, coincidentally overlapping
with the range that the maximum turbulence pressure support and lowest 
compressive ratio takes place. Contours in the dark matter density 
space are also presented in Fig. 10.  The deviation is much more significant
and no longer follow a simple linear function of $f_b(\rho_b)$, which can be 
illustrated by the long dashed line that represents a constant dark 
matter density $100$ in the left panel.  In the outskirts of clusters, 
and halos with size close to one grid cell, the baryon fraction is 
around $0.4-0.55$.  However, in the central regions of clusters with 
smoothed density $>10^3$, the baryon fraction is larger than the 
cosmic mean, suggesting feedback from star formation may 
be needed.  For halos whose sizes are larger than a couple of grid cells, 
the total baryon fraction would be some value between those in the central
and outer regions.   

Fig.11 gives the evolution of the baryon fraction in different 
gas density bins since $z=2$. While the deviation in the range of 
$10<\rho_b<10^3$  is amplified with time, the discrepancy in the 
central region is gradually reconciled.  The deviation in the 
$1024^3$ sample, calculated in cubic boxes of side length of two grid
cells, i.e., equal to the size of one grid cell in the $512^3$ samples,  
overwhelms results in the $512^3$ simulation at $z=0$, except for the 
cluster region in the TVD run. A possible explanation is that 
the compressive motions in the TVD simulation have been effectively 
dissipated by the numerical viscosity, rather than cascading into 
small scales and accreting on to higher density structures. 

In summary, the baryon fraction in the IGM significantly 
deviates from the cosmic mean in our simulations. The deviation 
shows a preliminary positive correlation with the turbulence pressure support.  
The case in virialized halos is much more complicated, and demands more careful investigation.
However, small halos are more likely to suffer from baryon deficit. Its link with the 
baryon missing in dwarf galaxies may require more extensive 
investigation, especially need to consider the impact of radiative, thermal and kinetic feedback. 
 
\section{Properties of Shocks}

The shocks formed during the hierarchical structure formation 
can effectively produce vorticity via the 
baroclinity term (Kida \& Orzsag 1990; Ryu et al. 2008; ZFF2010), 
and hence play an important role in injecting
turbulence into the IGM. Relatively higher numerical viscosity 
would smear out shocks more significantly, and 
consequently may result in less intensive turbulence. In this 
section, we investigate the properties of shocks 
in our simulation samples of $512^3$ produced by the WENO and TVD schemes. Results 
from the $1024^3$ simulation are not included as 
we intend to study the isolated effect of numerical viscosity.  
The impact of grid cell size and force resolution on 
the shocks can be found in Vazza et al. (2009, 2011). 

We detect shocks in our samples in a post-simulation way based 
on the conventional method in Miniati et 
al. (2000) and Ryu et al. (2003),  in which three requirements should be 
fulfilled to tag a grid cell as shocked region, \\
(1) $\nabla P \cdot \nabla S > 0$ \\
(2) $\nabla \cdot V < 0$ \\
(3) $|\Delta log(T)|>0.11$

A shock is typically resolved by two or three grid cells and its 
center is defined as the cell that has the 
minimum $\nabla \cdot V$.  The mach number of a shock $M$ is 
obtained according to 
\begin{equation}
\frac{\rho_2}{\rho_1}=\frac{4M^2}{M^2+3}
\end{equation}
\begin{equation}
\frac{T_2}{T_1}=\frac{(5M^2-1)(M^2+3)}{16M^2}
\end{equation}
for gas with $\gamma=5/3$,  and the subscripts $1$ and $2$
denote for the pre-shock and post-shock regions, respectively
(Landau \& Lifshitz 1959). 

The third condition in shock detection, $|\Delta log(T)|>0.11$, 
corresponds to a Mach number of $1.3$.  
We add one more condition, the density jump over a shock with 
Mach number $1.3$, i.e., $\rho_2/\rho_1 
> 1.4$,  to prevent overestimating the number of low Mach number shocks
by the temperature conditions (Vazza 
et al. 2009), and exclude pressure jumps that might be caused 
by errors in calculating internal energy. The shock detection procedure is performed in a 
one-dimensional way according to the above criterion along three different Cartesian axes.  
As Skillman et al. (2008) had shown that, this one-dimensional shock-finding algorithm 
would over-estimate the number of shocks weaker than $M \sim 10$ by a factor of $\sim 3$. 
The number of low Mach number shocks found in the two samples, WENO-512 and TVD-512, will both suffer from 
this limitation, while the discrepancies between the samples are expected to be systematic. 
Fig. 12 illustrates the shock fronts identified in the same slice shown 
in Fig. 1.  The skeleton of distributions is almost the 
same in two samples, but the WENO-512 run produces more shocks 
surrounding the sheets, filaments and 
compact gas halos. We also present the baroclinity term within this
slice. The spacial pattern of the baroclinity term is well 
confined by shocks and resembles the vorticity shown in Fig.1. 

\begin{figure}[htb]
\centering
\vspace{-0.5cm}
\includegraphics[width=8.0cm]{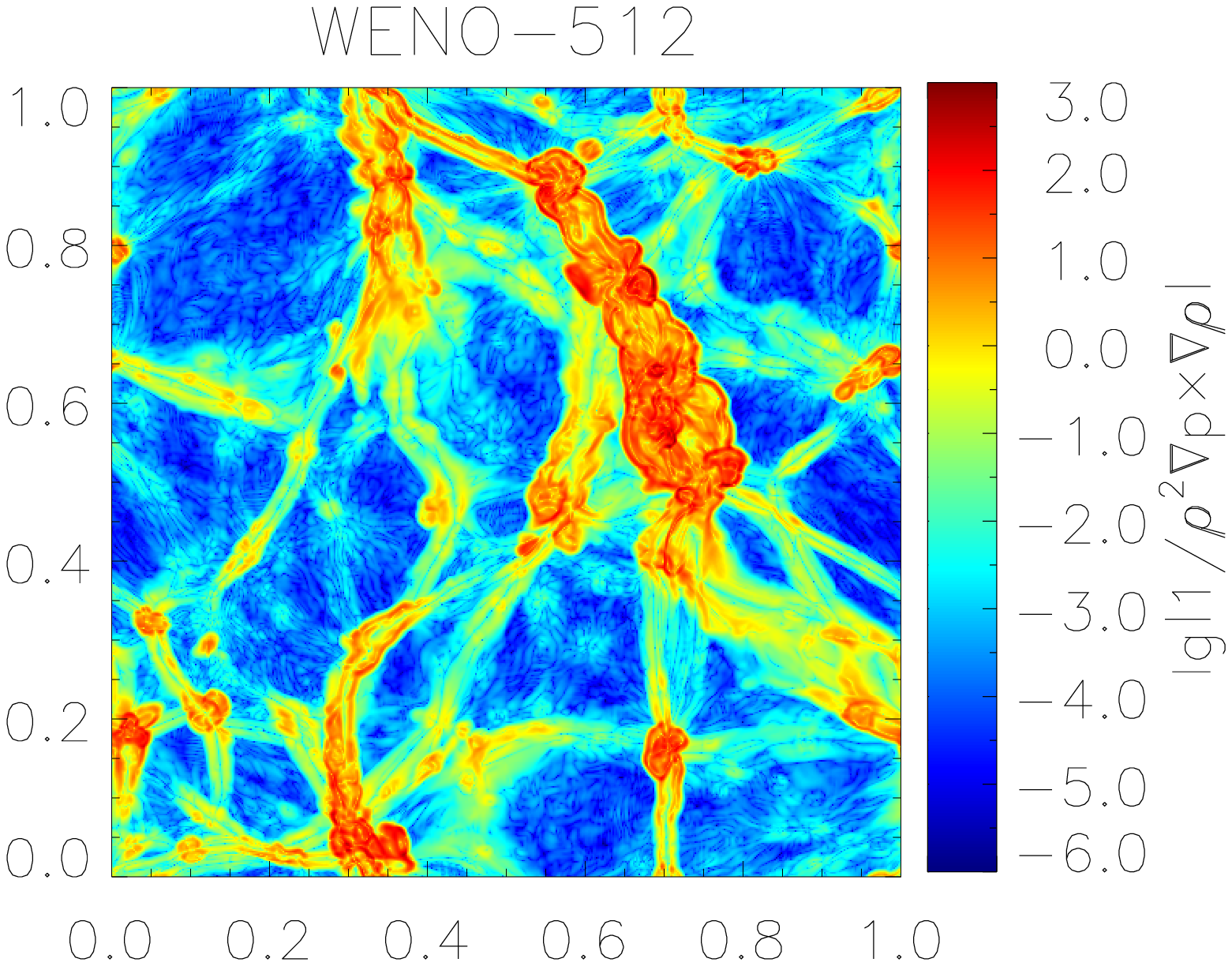}
\includegraphics[width=8.0cm]{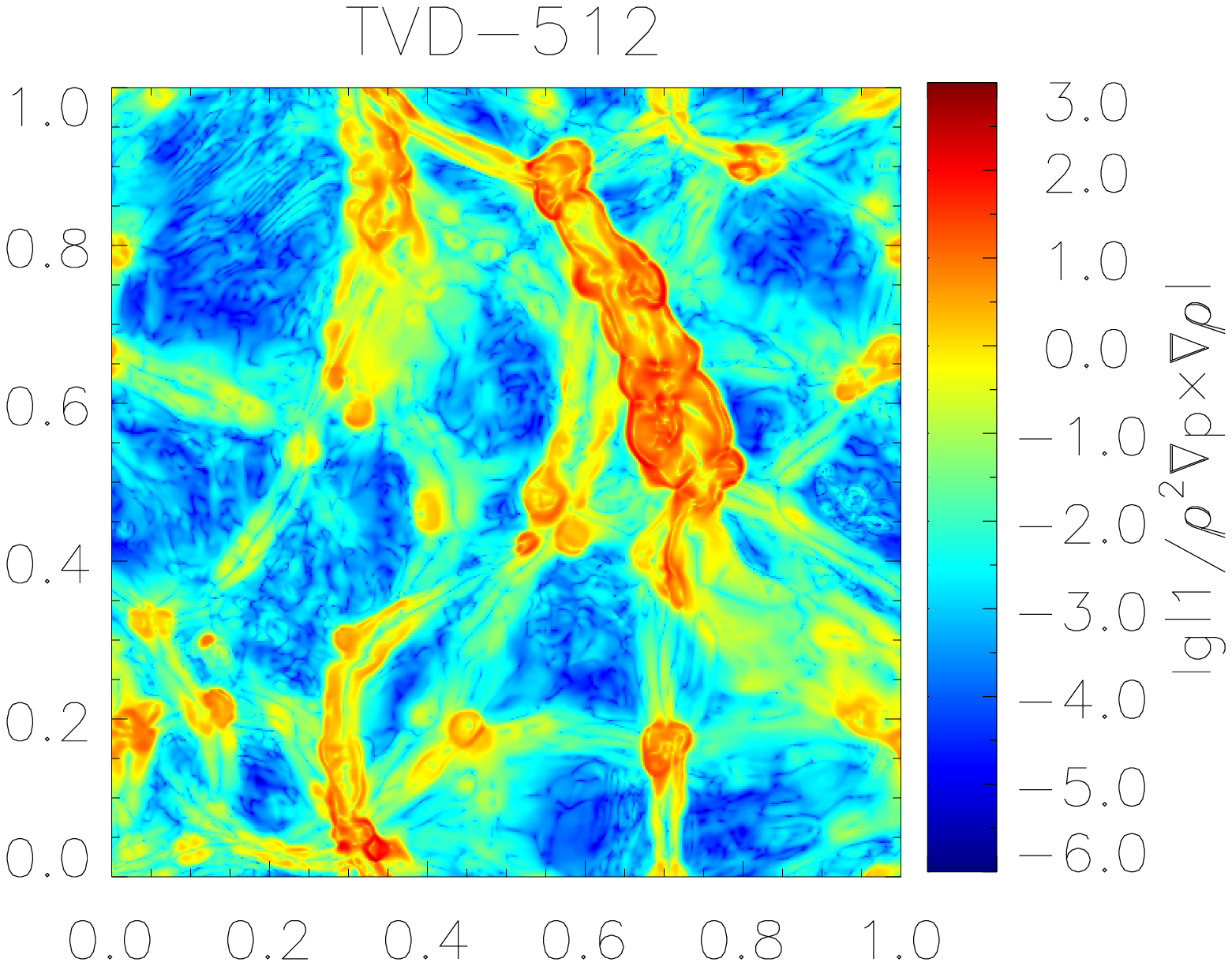}
\vspace{0.5cm}
\includegraphics[width=8.0cm]{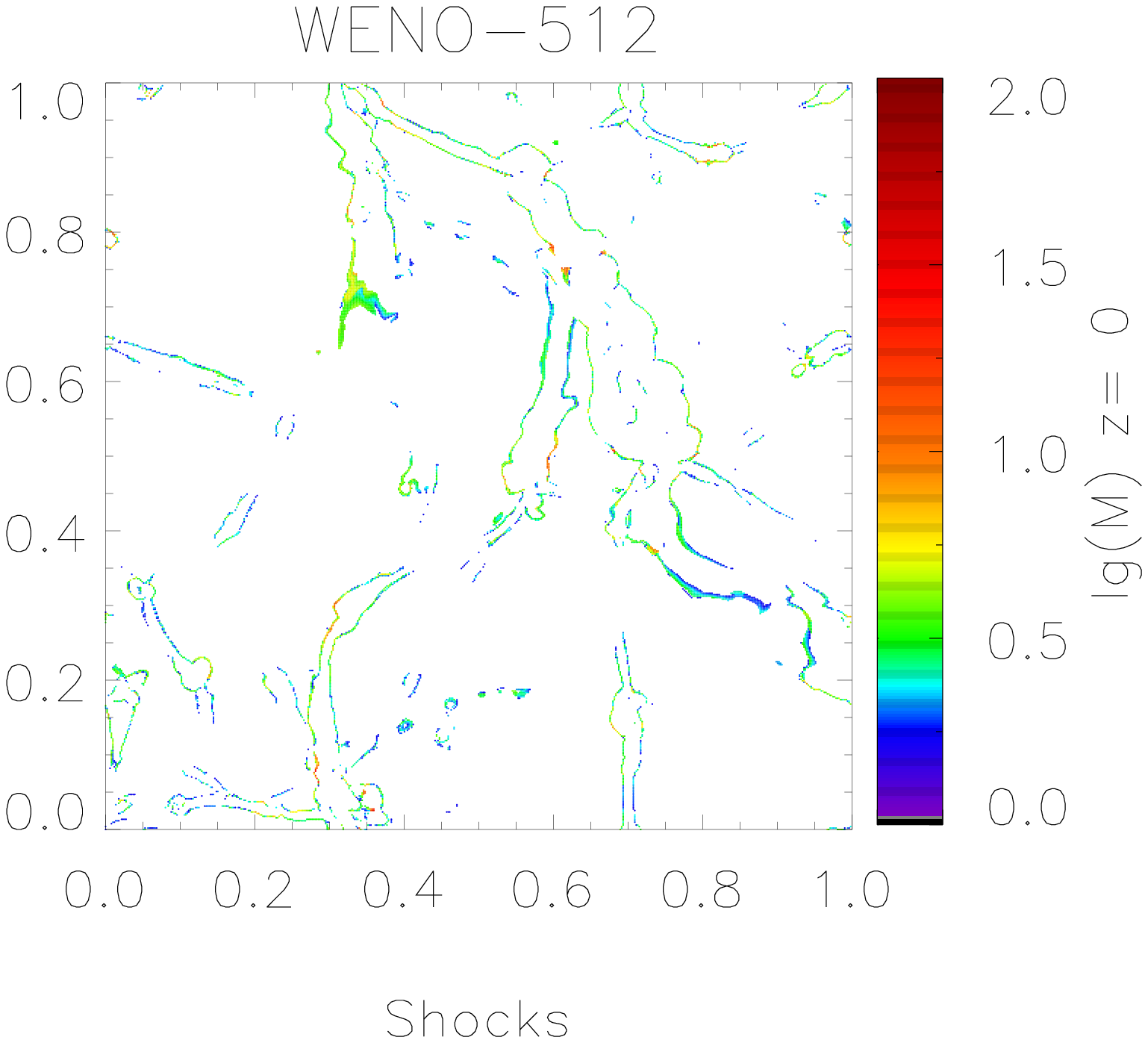}
\includegraphics[width=8.0cm]{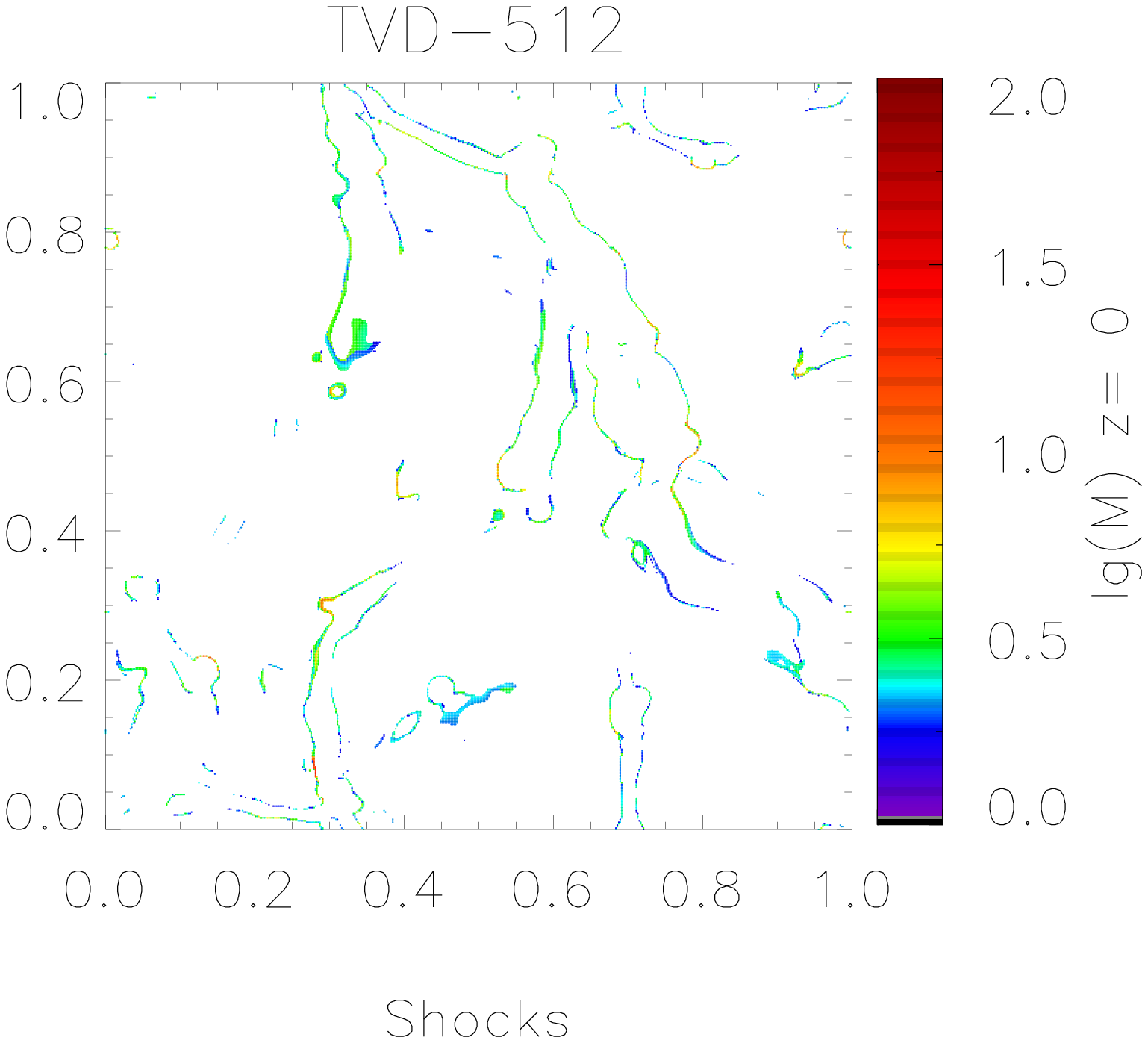}
\caption{Top (bottom) row: Baroclinity (shocks) in the same slice 
as in Fig. 1.  Left (right) column: WENO-512
(TVD-512).}
\end{figure}

Cosmic shocks are driven by the accretion of gas into sheets, filaments and 
clusters, and merging between 
gaseous haloes. As shocks are the sources of vorticity and turbulent 
velocity, the distribution of shocks in 
different structure patterns may give us more information about how 
the vorticity and turbulent motion are 
damped by numerical viscosity. We report the details of shock 
distribution in the following subsections. 

\subsection{External and Internal Shocks}

Depending on whether the pre-shock gas is previously shocked or not, 
shocks in the simulations can be divided 
into two categories, the external and internal shocks. In 
practice, shocks with $T_1<10^4 K$ are 
called the external shocks and vice versa (Miniati et al. 2001; Ryu 
et al. 2003; Kang et al. 2007). External shocks 
are formed when unshocked, low-density gas accretes onto 
nonlinear structures, such as sheets, 
filaments and knots. Internal shocks are distributed within the 
regions bounded by external shocks. The 
internal shocks are produced by: (1) gas infalling from sheets 
to filaments and knots, and from filaments to 
knots, (2) gaseous halos merging with each other. 

\begin{figure}[htb]
\centering
\includegraphics[width=8.0cm]{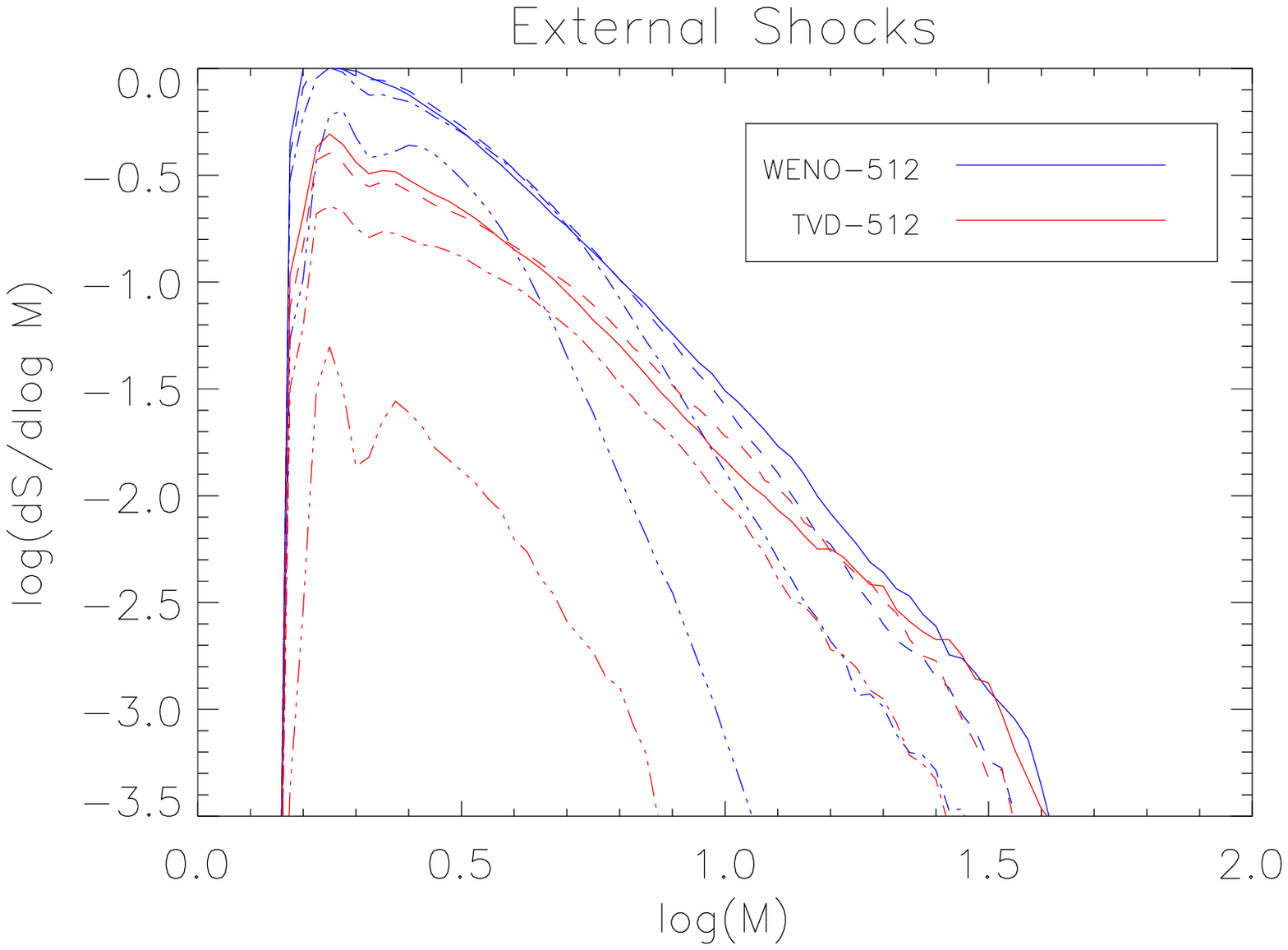}
\includegraphics[width=8.0cm]{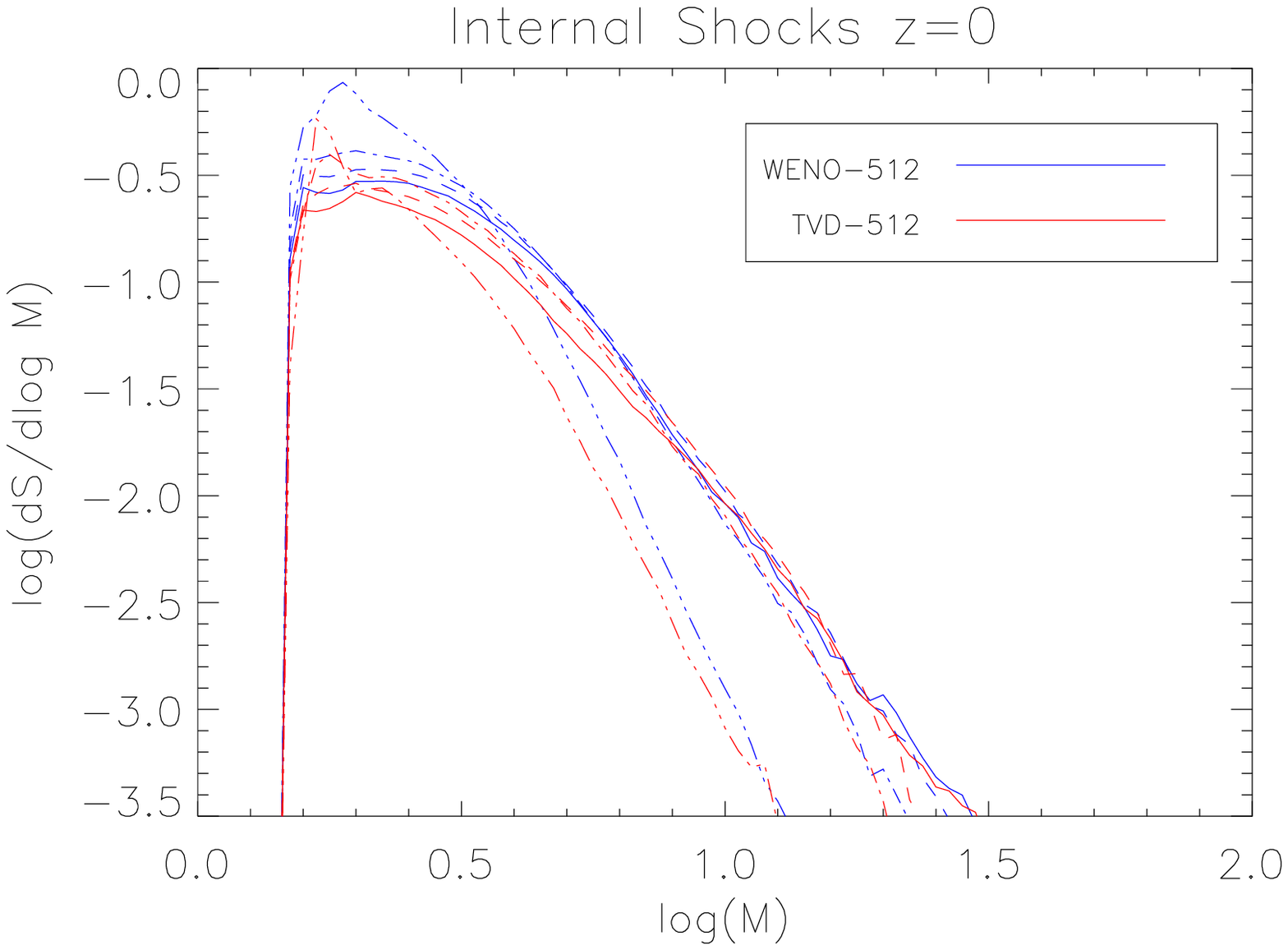}
\caption{Surface area of external and internal shocks in $512^3$ simulations. 
Blue lines: WENO-512. Red lines: 
TVD-512. Solid (dash, dash-dotted, dash-triple-dotted) lines are 
the results at different redshifts $z=0,0.5,1.0,2.0$.}
\end{figure}

Fig.13 compares the frequency of external and internal shocks between
 two codes,  measured by the surface area 
of shocks in logarithmic Mach number bin, $dS(M,z)/d log M$
which was firstly proposed in Miniati et al.
(2000) and indicates the inverse of distance between shocks.  
The most striking feature in Fig. 13, is the 
significant difference in the frequency of external shocks between
two codes. Many more external shocks are 
developed in the WENO-512 simulation from $z=2$ to $z=0$, while moderately 
more internal shocks are formed 
only at $z=2$.  Numerical viscosity has smeared out many shocks in the
low density regions in the TVD codes. 

Comparing with the result in Ryu et al. (2003), the distribution of 
external shocks here have shallower 
tails over $M>10$, while the distribution of internal shocks is 
more extended. The former should be mainly 
due to the UV background and the latter results from the 
improvement of simulation resolution, $47.7 h^
{-1} kpc$ here versus $97.7 h^{-1} kpc$ in Ryu et al.(2003). The number of
external and internal shocks weaker than $M \sim 10$ in our TVD-512 run are consistent with
Ryu et al.(2003), which should be significantly over-predicted in comparison with results by the
coordinate unsplit algorithm proposed in Skillman et al.(2008)(see their Fig. 4). 

\subsection{Shocks in Various Cosmic Structures}

The shock distribution in various cosmic environments can be 
addressed more explicitly according to either the 
gas or the total matter density at shock centers. Skillman et al.(2008) and Vazza et al.(2009)
studied the shock distributions for varying pre-shock gas density or total matter density, 
and their redshift evolution. Here, similar to Vazza et al. 
(2009), we divide the structures that shocks 
are embedded into four categories according to the total density,\\
(1) $\rho_t=\rho_b+\rho_{cdm}<6\rho_{crit}$: voids and under-dense region\\
(2) $6<\rho_t/\rho_{crit}<36$: sheets and filaments \\
(3) $36<\rho_t/\rho_{crit}<200$: outskirts of clusters\\
(4) $\rho_t/\rho_{crit}>200$: virialized clusters 

\noindent
where, $\rho_t$, $\rho_{cdm}$, and $\rho_{crit}$ are the total, 
dark matter and critical density respectively. The 
boundaries, $6, 36, 200$, are slightly different from those
in Vazza et al. (2009), and are identical to the mean density 
of sheets, filaments and halos predicted in Shen et al. (2006).

\begin{figure}[htb]
\centering
\includegraphics[width=7.0cm]{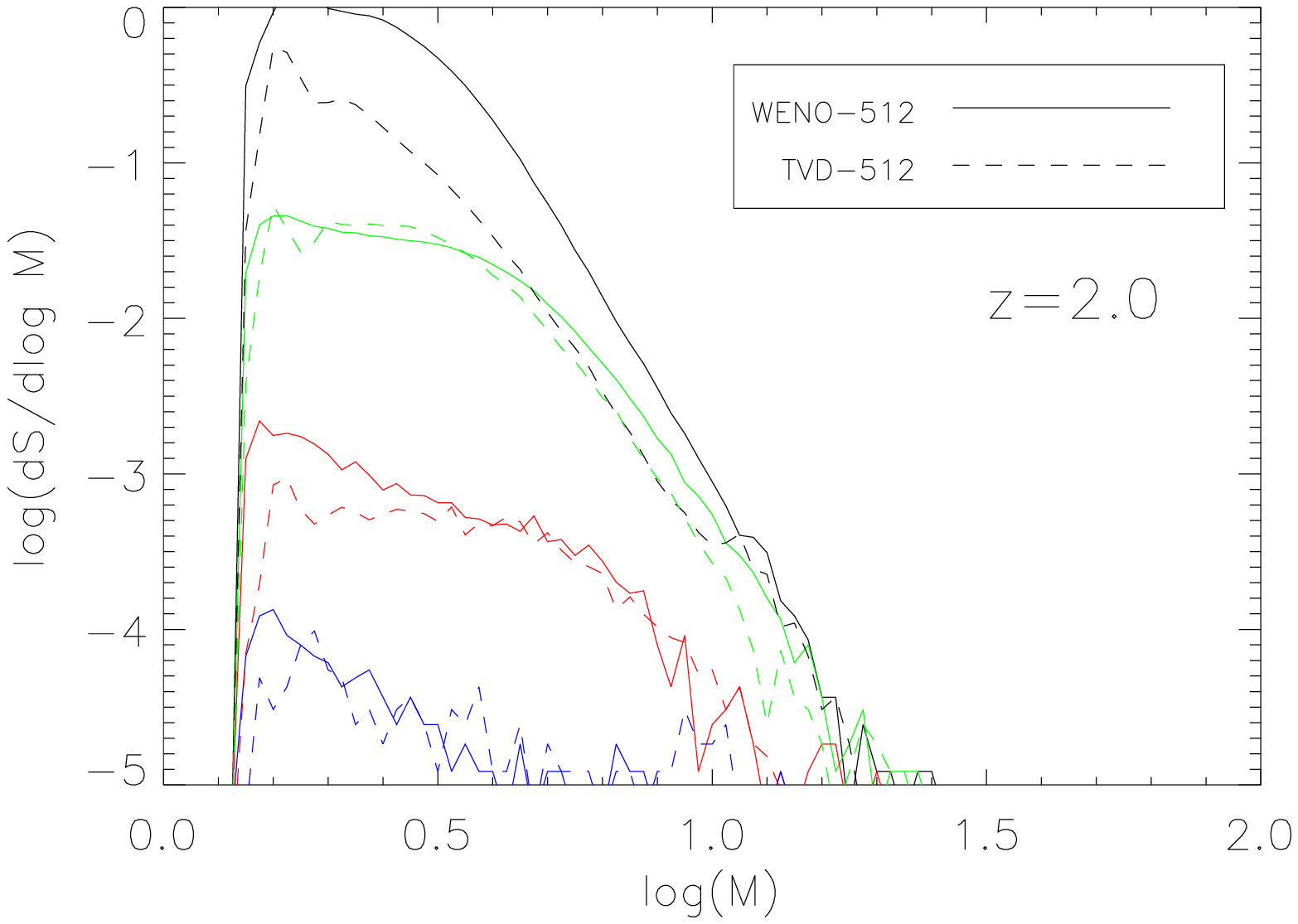}
\includegraphics[width=7.0cm]{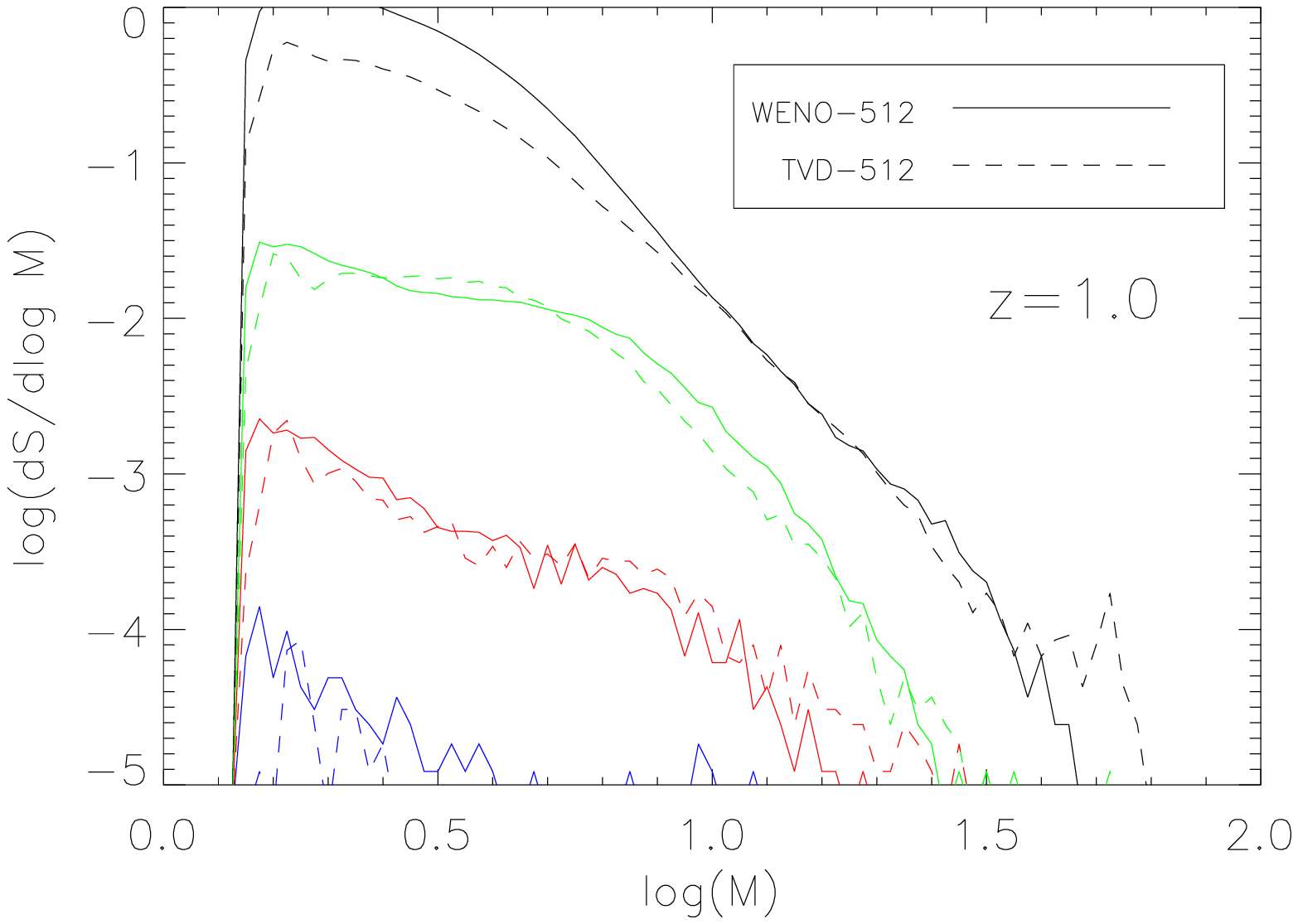}
\includegraphics[width=7.0cm]{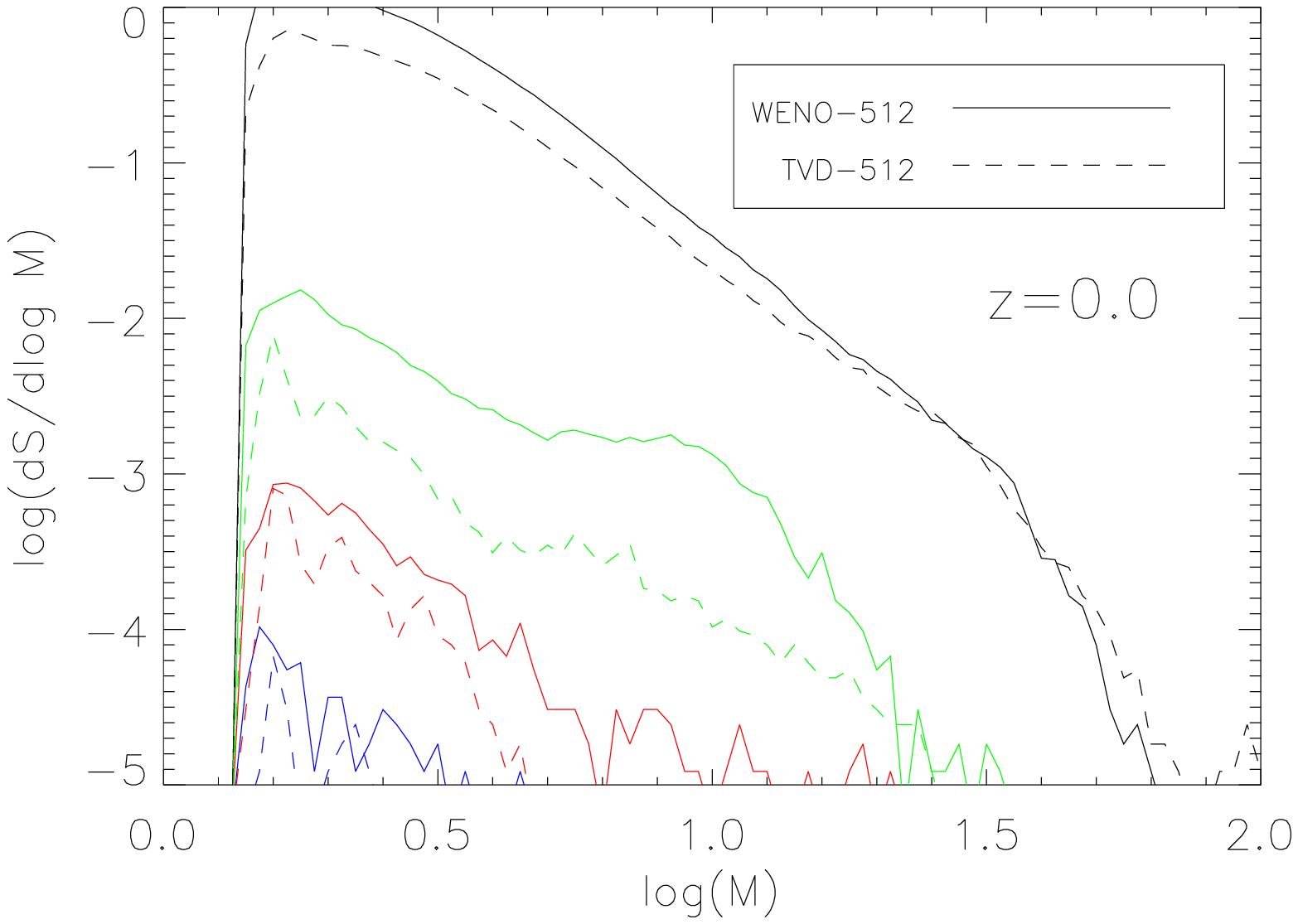}
\includegraphics[width=7.0cm]{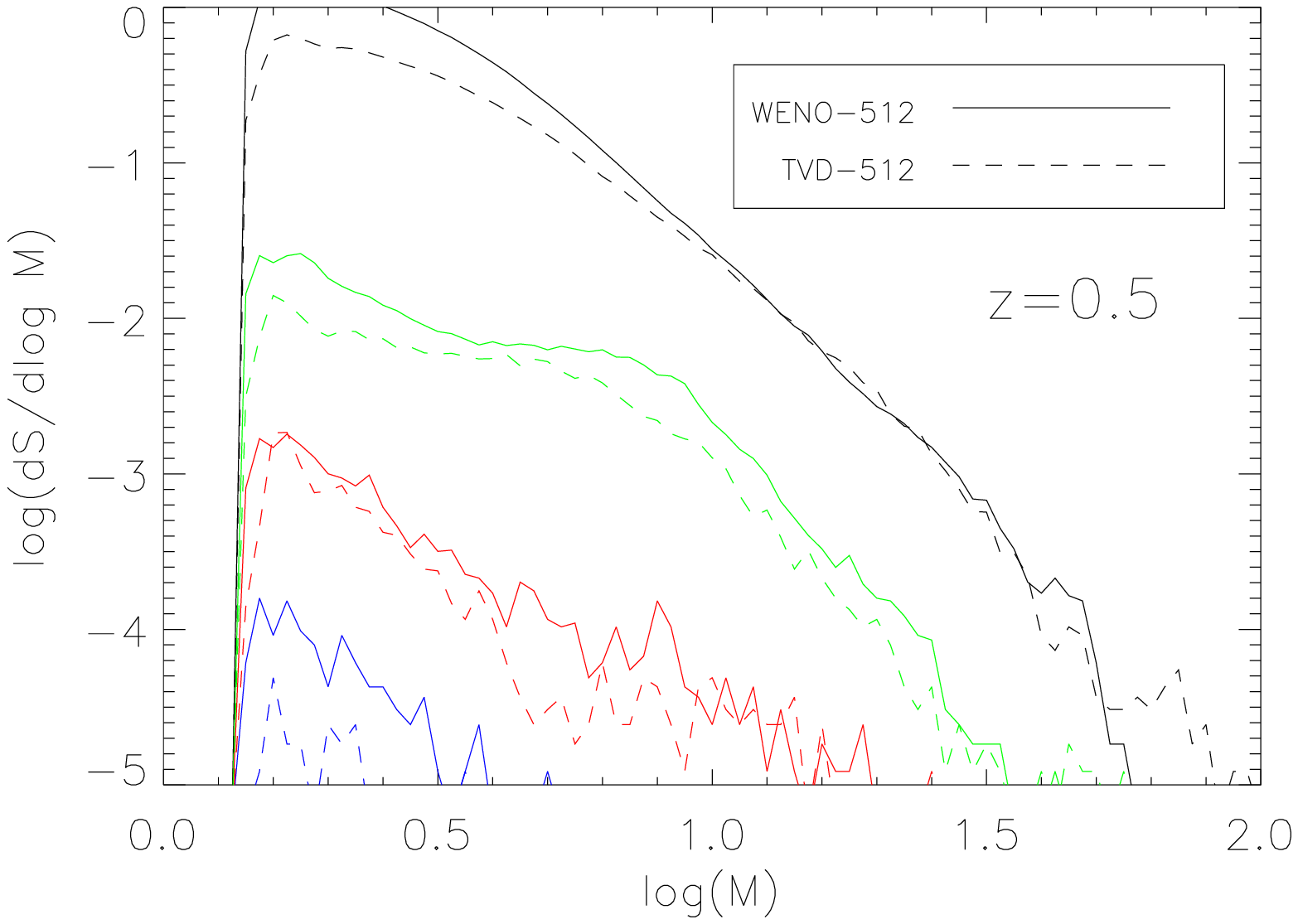}
\caption{Surface area of shocks in various structures in the simulated samples at $z=2.0,1.0,0.5,0.0$ (Arranged in 
clockwise direction). Solid (dotted) lines indicate the WENO-512 (TVD-512) result. Pairs of lines 
indicate shocks in voids, filaments, outskirts of clusters and clusters from top to 
bottom. }
\end{figure}

Fig.14 plots the distribution of shocks in the four types of 
cosmic structures from $z=2.0$ to $z=0.0$. The 
general trends with decreasing redshifts are identical in
two codes: more shocks keep appearing in the 
under-dense region where the gas keeps accreting to nonlinear 
structures; more high Mach number and 
less low Mach number shocks are being developed within the 
sheets and filaments; the frequency of 
shocks in clusters and their outskirts, however, is not changed much.  

On the other hand, the difference between the WENO-512 and TVD-512 simulations takes place 
in different structures at different 
redshifts. At the redshift $z=2$, the shocks in the under-dense 
region are less developed due to the damping of 
velocity and density fluctuations by the higher numerical viscosity in 
the TVD code. Notable difference remains in the 
under-dense region at $z=1$.  Moderate discrepancies exist in all 
the structures at $z=0.5$.  More shocks are 
formed within the sheets, filaments and outskirts of clusters in the 
WENO-512 run at $z=0$. 

The majority of shocks in our samples are hosted in low-density regions including 
voids and filaments, which is in general agreement with Skillman et al.(2008) and Vazza et al(2009).  
Again, the frequencies of low Mach number shocks are much higher than Skillman et al(2008), 
by a factor of $3 \sim 10$, in all kinds of structures, which may mainly result from the different 
shock detecting methods. The number distribution of shocks as a function of Mach number
became shallower when moving towards dense environment for $M>3$ in our simulations, 
which is contradict with Vazza et al.(2009). The evolution in Skillman et al.(2008), however, is much more complicated. 
To answer weather this discrepancy is merely brought by the selection of shock finding algorithm or is a 
complicate process involving both the physical modules and the numerical algorithm in different work, 
much more carefully investigation is needed, which however, is out the scope of this work.

Combining with the result in the last subsection, our simulations 
suggest that the numerical viscosity smears 
out the shocks more seriously in regions outside of clusters than 
within clusters. As shocks are the sources of 
vorticity, numerical viscosity will significantly damp the 
vortices in regions outside of clusters, which may, 
to a great extent, be responsible for the differences regarding 
turbulence development in the IGM between two codes. 

\section{Discussions and Concluding Remarks}

The evolution and properties of curl and compressive velocities, 
corresponding to the vorticity and divergence respectively, 
reveal the development history of turbulence in 
the IGM, and reflect the dynamical state of gas accreting onto different structures.
The numerical viscosity embedded in numerical schemes 
may have significant impact on the vorticity and 
turbulent motions in the IGM in simulations. In this paper, 
we have addressed these two issues using a suite of simulations performed by two fixed 
mesh based codes that differ only in the spatial difference 
discretization,  the second-order TVD scheme and 
the fifth-order positivity-preserving WENO scheme. 

The development of turbulence in the IGM may undergo two phases. 
During the first supersonic phase, vortical motion is triggered by the
baroclinity over strong curved external shocks.  The scaling law in 
this phase follows Burgers's $k^{-2}$ law, which actually 
characterize turbulence consisting of strong shocks. In the 
second post-supersonic phase, vortices interact with each other 
and decay, although shocks are still formed. The governing dynamics 
for the vortices is subsonic. In the context of hierarchical 
structure formation history, the first phase may correspond
to the stage that low density gas accretes onto sheets, filaments and clusters, while the second phase 
represents the interaction between gaseous protogalactic structures, 
e.g.  merging processes. The growing history of the transit scale 
$l_t$ and the upper end of the turbulence scale $l_u$ 
as visualized in our simulation since $z=2$ may be a natural 
consequence of cosmic structure formation.

The turbulence development in the IGM may be affected by the 
numerical viscosity in two ways. Large 
scale velocity and density fluctuations in the baryonic matter 
would be damped, and hence the 
production of shocks and vortices in regions outside of clusters 
is suppressed. Namely, the turbulence 
injection rate is reduced by numerical viscosity.
The kinetic turbulent energy transfer rate will also be 
reduced. On the other hand, the kinetic energy of
large size vortices would be dissipated into thermal energy much 
more rapidly due to numerical viscosity, rather than cascading into small 
size eddies. In other words, the decaying rate 
is enhanced.  The simulated dissipation scale, $\sim 600  h^{-1}  \mbox{kpc}$, in the TVD-512
run is significantly larger than $\sim 300  h^{-1} \mbox{kpc}$ in the WENO-512 run, indicating
that the numerical dissipation might be comparable to the physical dissipation effects like 
shock heating and Jeans dissipation under corresponding scale in the TVD-512 code. 
As a byproduct, the thermal and density evolution of gas accreting
onto highly over-dense structures would also be changed by the numerical viscosity
(e.g. Nelson et al. 2013).   

The non-thermal pressure support, provided by the turbulent motions 
in the IGM, might slow down the 
accretion of baryon into dark matter structures during structure 
formation. Consequently, the baryon fraction in the highly turbulent 
regions would deviate from the cosmic mean. However, the situation 
would become more complicate in the presence of high level numerical 
viscosity. Both the compressible and vortical 
components of gas motions would be dissipated into thermal energy 
more rapidly. The lag in gas density fluctuation development with 
respect to dark matter may be caused by the non-thermal pressure 
support and the damping of compressible motions simultaneously. 
  
The properties of turbulence in the IGM may be changed by many
ingredients that are not considered in our simulations. The most important 
one may be the physical viscosity of the IGM. The baryonic gas is 
assumed to be non-viscous in our simulations, which also holds 
for almost all of the cosmological hydrodynamics 
simulations in the literature. The physical viscosity of the IGM 
is poorly known so far, although it is generally believed that the IGM should have very high Reynolds number(e.g. Gregori et al. 2012). 
However, the properties of turbulence in the IGM may change a lot from the simulation 
if the physical viscosity is non-negligible in the inertial range of this work, i.e. $\sim 2  h^{-1}  \mbox{Mpc}-200  h^{-1}  \mbox{kpc}$. 
Simulation works shows that including the physical viscosity of the ICM would modify the pattern of gas flow and 
the evolution of gaseous sub-structures in the simulated clusters (Sijacki \& Springel 2006; Dong \& Stone et al. 2009; 
Suzuki et al. 2013). Tentative study suggests that the physical viscosity might be important on scales around tens of 
$kpc$ (Brunetti \& Lazarian 2007). Analysis of the morphology of filaments
and bubbles in the Perseus cluster revealed by deep Chandra X-ray 
observation in Fabian et al.(2003a, 2003b) indicated that the hot ICM 
may have non-negligible viscosity and affects the cooling within the inner $50$ kpc of the cluster.  
The expected dissipation scale due to physical viscosity is likely about tens of $kpc$ and below the simulated dissipation 
scale $\sim 200 h^{-1} kpc$ in the WENO-1024 simulation. Future cosmological hydrodynamic simulations that 
implemented the physical viscosity properly would reveal more solid results on the development of turbulence in the IGM.
In addition, including feedback from star formation, magnetic fields and AGN activity in the galaxies may 
trigger more intense turbulence below the dissipation scale.  On the other hand, 
more precise constraints on the viscous properties and the turbulent motion in the IGM and ICM from observation
are needed to verify the simulated results. 
 
Finally, we summarize our results as follows,

(1) The small-scale compressive ratio $r_{CS}$ drops dramatically, from $0.84$ at $z=2.0$ 
to $0.47$ at $z=0.0$, as the vorticity is effectively produced 
since $z=2$, and can be comparable to the divergence at low 
redshifts.  The curl velocity associated with the vorticity increases
from $\sim 10 \ km/s$ at $\rho_b \sim 1$ to $\sim 90 \ km/s$ at 
$\rho_b \sim 100$ and is saturated henceforth for more dense region 
at $z=0.0$, which is  $\sim 15  \ km/s$ higher than the value that Oppenheimer \& Dave 
(2009) used as the sub-resolution turbulence term to match the 
observation, and lower than observational upper limits by a factor of $2 \sim 3$. 
The estimated turbulence scale is $2 \mbox{Mpc}$ down to $200 \mbox{kpc}$, 
which is consistent with ZFF2010.

(2) The power spectrum of the velocity field of the IGM shows 
two different phases within the turbulence scale range simultaneously, i.e., 
the supersonic phase and the post-supersonic phase. In the former 
regime, corresponding to $\sim 0.2 - 0.68 \ h^{-1} \  \mbox{Mpc}$ at $z=0$, 
the scaling law follows $k^{-1.88}$, $k^{-2.20}$,  $k^{-2.02}$ for the total, compressive and 
curl velocity respectively, and is close to a Burgers turbulence.  
In the latter regime, i.e., $\sim 0.68 - 2.0 \  h^{-1} \  \mbox{Mpc}$, the power spectrum of the curl velocity 
transits to $ k^{-0.89}$, while the other two are unchanged.  The transition scale grows from $\sim 0.32 \  h^{-1} \ \mbox{Mpc}$ at $z=2.0$ 
to $\sim 0.68 \ h^{-1} \mbox{Mpc}$ at $z=0.0$. 

(3) The non-thermal pressure support, measured by the ratio of 
kinetic turbulent energy to internal energy, is comparable to the thermal pressure 
for $\rho_b \simeq 10-100$, or $T < 10^{5.5} K$ at $z=0.0$. The mean baryon 
fraction in different density bins of the IGM significantly 
deviates from the cosmic mean, and can be as low as $0.4$ in regions that have the
highest turbulent pressure support and the lowest compressive ratio. 

(4) Relatively higher numerical viscosity would artificially 
dissipate both the compressive and vortical motions in the IGM 
more strongly. Consequently, the density fluctuation and vorticity 
are significantly damped in the TVD samples. The turbulence dissipation scale is
remarkably shifted toward larger scale by a factor of $2$, and hence 
shortens the turbulent inertial range. Numerical viscosity could 
lead to non-negligible uncertainty in the simulated thermal history of gas accretion
during structure formation. 

(5) The shocks in the regions outside clusters are significantly
suppressed by relatively higher numerical viscosity since $z=2$ in the TVD samples.  
As cosmic shocks are the sources of vorticity, the different 
levels of turbulence in the IGM between two codes may be directly 
caused by the un-resolved shocks in the regions outside clusters
 in the TVD code. 


\begin{acknowledgements}
Acknowledgements: The authors thank the anonymous referee
for very useful comments. The simulations presented here were run on SGI Altix 4700 at Supercomputing 
Center of Chinese Academy of Sciences.  WSZ thanks the support from China Postdoctoral Science Foundation and 
the Fundamental Research Funds for the Central Universities. This work is supported under the National Natural Science
Foundation of China under grant 11203012, 11133001 and 10878010. Feng is supported by NSFC grant 11273060 
and 91230115; Shu is supported by ARO grant W911NF-11-1-0091 and NSF Grant DMS-1112700.  

\end{acknowledgements}

\setcounter{figure}{0} 
\renewcommand{\thefigure}{A\arabic{figure}}
\appendix

\section{Grid Difference Schemes}

We first give a brief description of the positivity-preserving 
finite difference WENO scheme and then show its performance 
on two classical numerical tests, the shock-tube test and the 
Sedov blast wave test, comparing to the TVD 
scheme. We refer the interested reader to the references 
in this subsection for more details about the schemes. 

\subsection{Positivity-Preserving WENO Scheme}

In Zhang \& Shu (2012), a simple limiting
strategy is designed for high order finite difference WENO schemes, which
can guarantee the positivity of density and pressure if they
are positive initially, while also maintain the original high
order accuracy of the scheme.  The procedure is a generalization
of an earlier strategy in Zhang \& Shu (2010) for finite volume
WENO schemes and discontinuous Galerkin methods.  We will only
give a brief description of this positivity-preserving
procedure in this section, and
refer the readers to Zhang \& Shu (2012) for more details.

We first look at the one-dimensional Euler equation
$$
w_t + f(w)_x =0
$$
with $w=(\rho, \rho v, E)^T$ and $f(w)=(\rho v, \rho v^2 +p, v(E+p)$,
where $\rho$ is density, $v$ is velocity, $E$ is total energy, and
$p$ is pressure, which satisfies $p=(\gamma -1)(E-\frac 12 \rho v^2)$
for a gamma-law gas.  For a finite difference scheme, given the point 
values $w^n_i$ at time level $n$, we let $A_{i+\frac12}$
denote the Roe average matrix of the two states $w^n_{i}$
and $w^n_{i+1}$, and let $L_{i+\frac12}$ and $R_{i+\frac12}$ denote
the left and right eigenvector matrices of $A_{i+\frac12}$ 
respectively, i.e.,
$A=R\Lambda L$ where $\Lambda$ is the diagonal matrix with 
eigenvalues of $A$ on
its diagonal. Let $\alpha=\max (|v|+c)$ where $c$ is the sound
speed, where the maximum is
taken either globally for $w^n_j$ over all $j$ or locally for $j$ over the
the WENO reconstruction stencil.
We use the Lax-Friedrichs flux splitting, namely
  \begin{equation}
   \label{zad4}
w^{\pm}=\frac12\left(w\pm\frac{f(w)}{\alpha}\right).
  \end{equation}
At each fixed $x_{i+\frac12}$, the algorithm proceeds as follows, where
we omit the subscripts $i+\frac12$ for
notational simplicity:
 \begin{itemize}
  \item[1.] Let $w^{\pm} (x) = \frac{1}{\Delta x} 
\int_{x- \Delta x/2}^{x+\Delta x/2} h^{\pm} (\xi) d\xi$, then
we are given the cell averages of $h^{\pm}(x)$ by
$(\bar{h}^{\pm})^n_j=(w^{\pm})^n_j$.
  \item[2.] Transform all the cell averages $(\bar{h}^{\pm})^n_j$
for $j$ in a neighborhood of $i$ within the WENO reconstruction
stencil to the local characteristic fields by setting
$(\bar{u}^{\pm})^n_j=L (\bar{h}^{\pm})^n_j$. 
  \item[3.] Perform the WENO reconstruction for each component of
$(\bar{u}^{\pm})^n_j$ to obtain approximations of
  the point values of the functions $u^{\pm}=Lh^{\pm}$ at the point
$x_{i+\frac12}$.
Notice that the stencil used for the WENO reconstruction
for $\bar{u}^{+}$ is $\{ i-2, i-1, i, i+1, i+2 \}$, and the one
used for the  WENO reconstruction
for $\bar{u}^{-}$ is $\{ i-1, i, i+1, i+2, i+3 \}$, due to
upwind-bias for stability. 
  \item[4.] Transform back into physical space by
$h^{\pm}_{i+\frac12}=Ru^{\pm}$.  A simple scaling limiter is then
applied to $h^{+}_{i-\frac12}$, $h^{-}_{i+\frac12}$ and
an intermediate quantity to ensure positivity of density and
pressure, see Zhang \& Shu (2012) for the details.
Then we form the numerical
flux by $\hat{f}_{i+\frac12} = \alpha (h^+_{i+\frac12}  + h^-_{i+\frac12})$.
The final scheme is
\begin{equation}
\label{zad5}
w^{n+1}_i=w^{n}_i-\lambda\left(\hat{f}_{i+\frac12}-\hat{f}_{i-\frac12}
\right) .
 \end{equation}
\end{itemize}
Combined with the third order TVD Runge-Kutta method, the scheme
is guaranteed to maintain positivity of density and pressure.
Generalization to multi-dimensions is straightforward in a
dimension-by-dimension fashion.

\subsection{Test Examples}

(1) Shock tube test

\begin{figure}[htb]
\centering
\includegraphics[width=8.0cm]{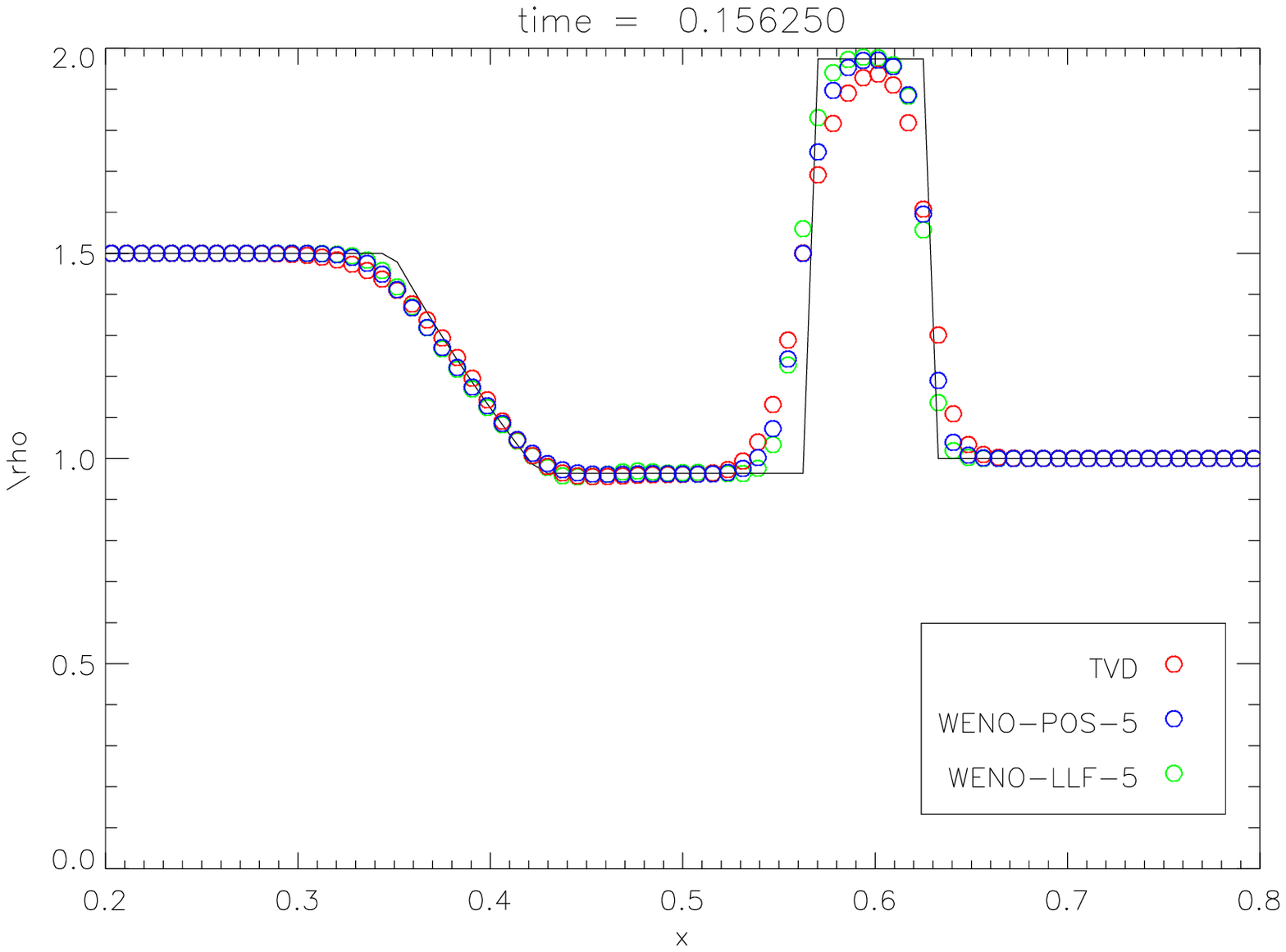}
\includegraphics[width=8.0cm]{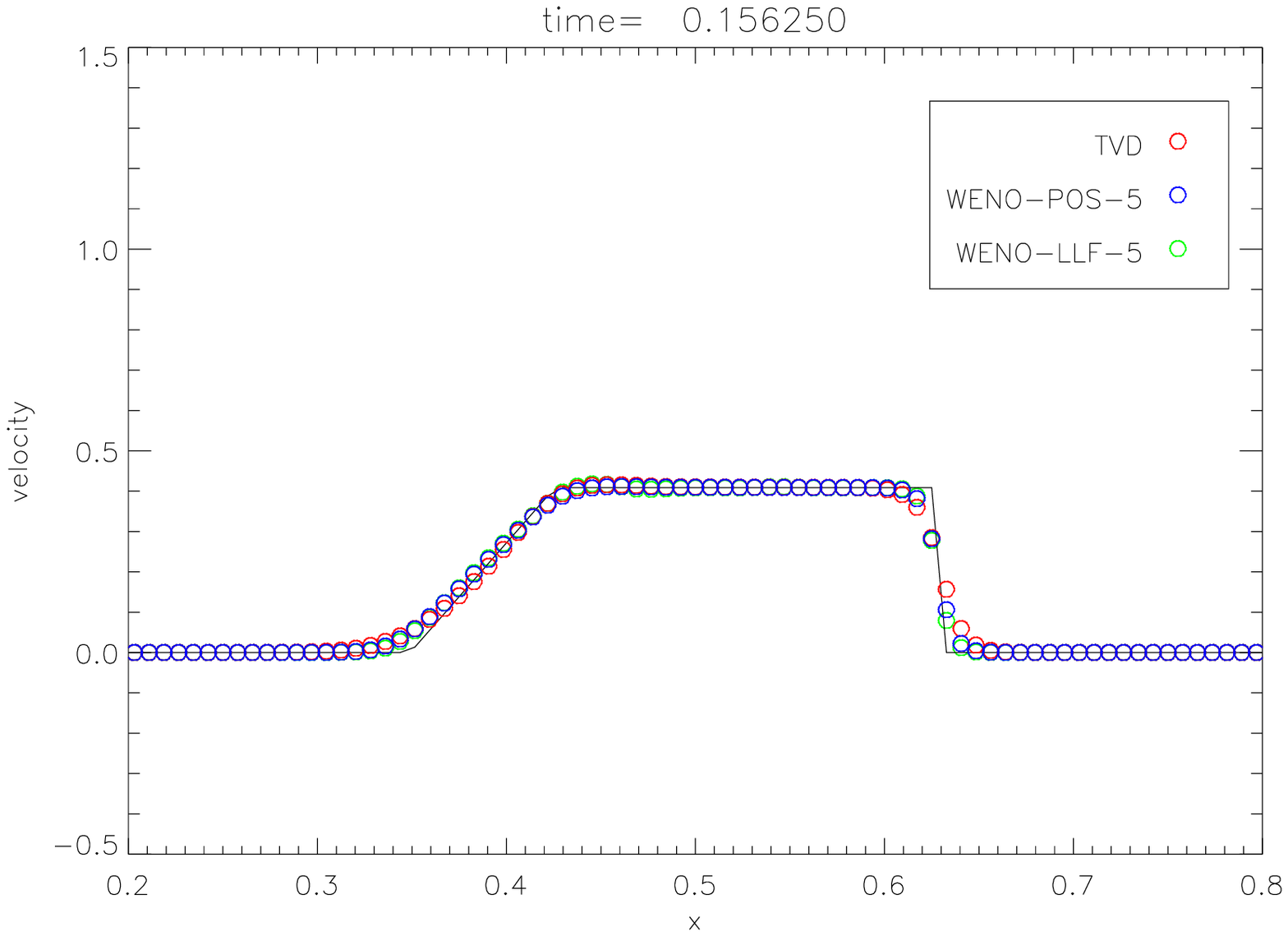}
\caption{One dimensional shock tube test of the TVD and WENO schemes with 
resolution 128.}
\end{figure}

We present the results of the TVD and WENO schemes in solving the one-dimensional 
Sod shock tube problem with 
resolution $128$ in Fig. A1.  The initial state of the left and 
right region are given as ($\rho_L=1.5$, $v_1=0$,
$p_L=1.0$) and ($\rho_R=1.0$, $v_2=0$, $p_R=0.2$) respectively. 
The polytropic index is 1.4.  The CFL 
number is set to the same value $0.32$. Both schemes can well 
capture the shock, although the TVD scheme 
shows more damping over the shock and the discontinuity surfaces. 
This result is not out of expectation, as 
the accuracy of the TVD scheme will degenerate over shocks more 
seriously than the WENO. The $L^1$ errors for a smooth solution 
indicate that the TVD scheme achieve the order of $2.0$ and the 
WENO-POS is $5.1$. Also shown is the 
result run by the original fifth WENO scheme with the
local Lax-Friedrichs flux splitting (WENO-LLF), which was 
used in ZFF2010 to study the turbulence in the IGM. 
The WENO-LLF code uses a CFL 
number $0.2$ in order to keep it stable.  Slightly improved performance 
is achieved by the WENO-LLF 
scheme at the cost of around $50 \%$ more cpu hours. 

(2) Sedov blast wave test

\begin{figure}[htb]
\centering
\includegraphics[width=7.0cm]{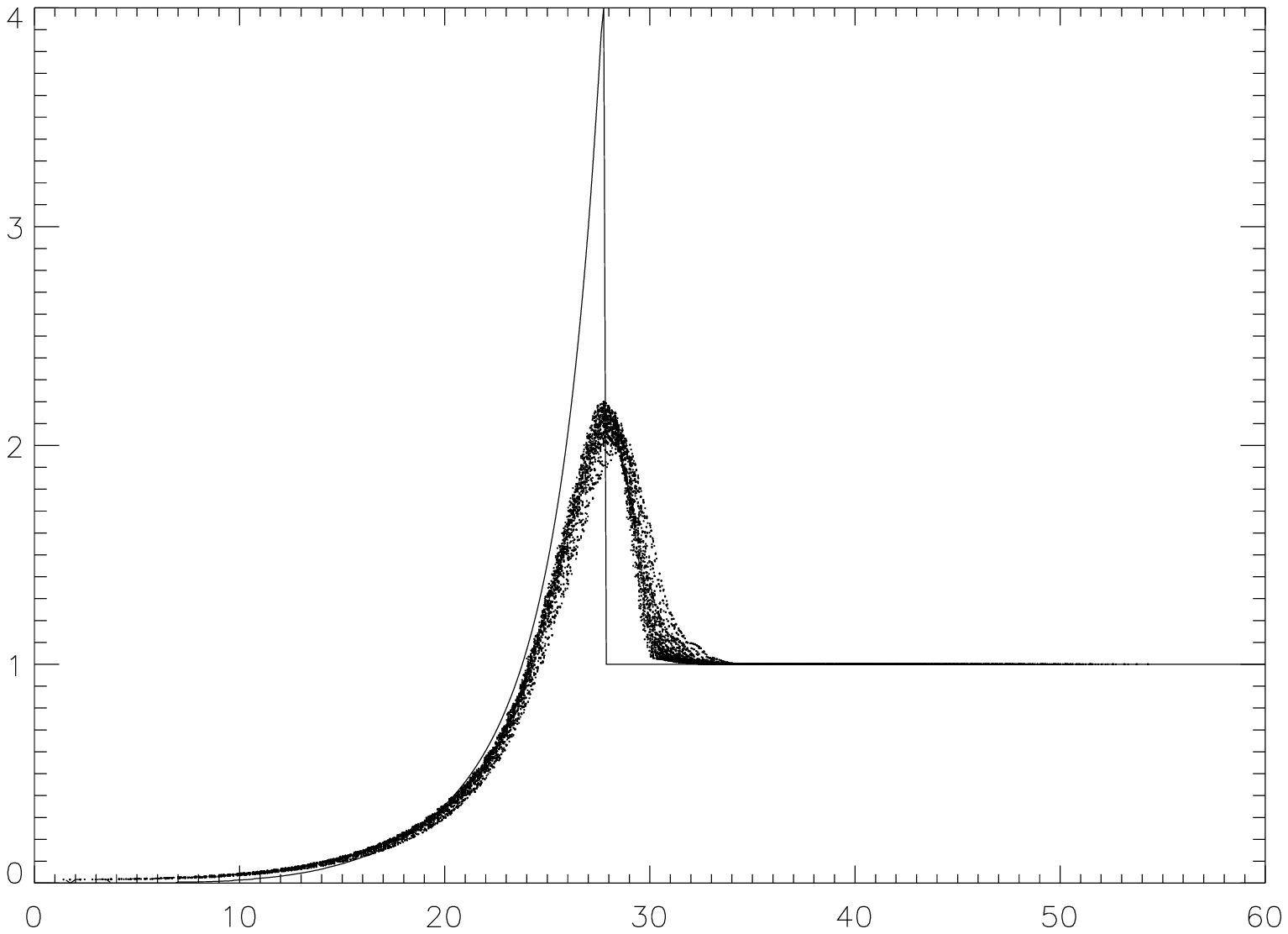}
\includegraphics[width=7.0cm]{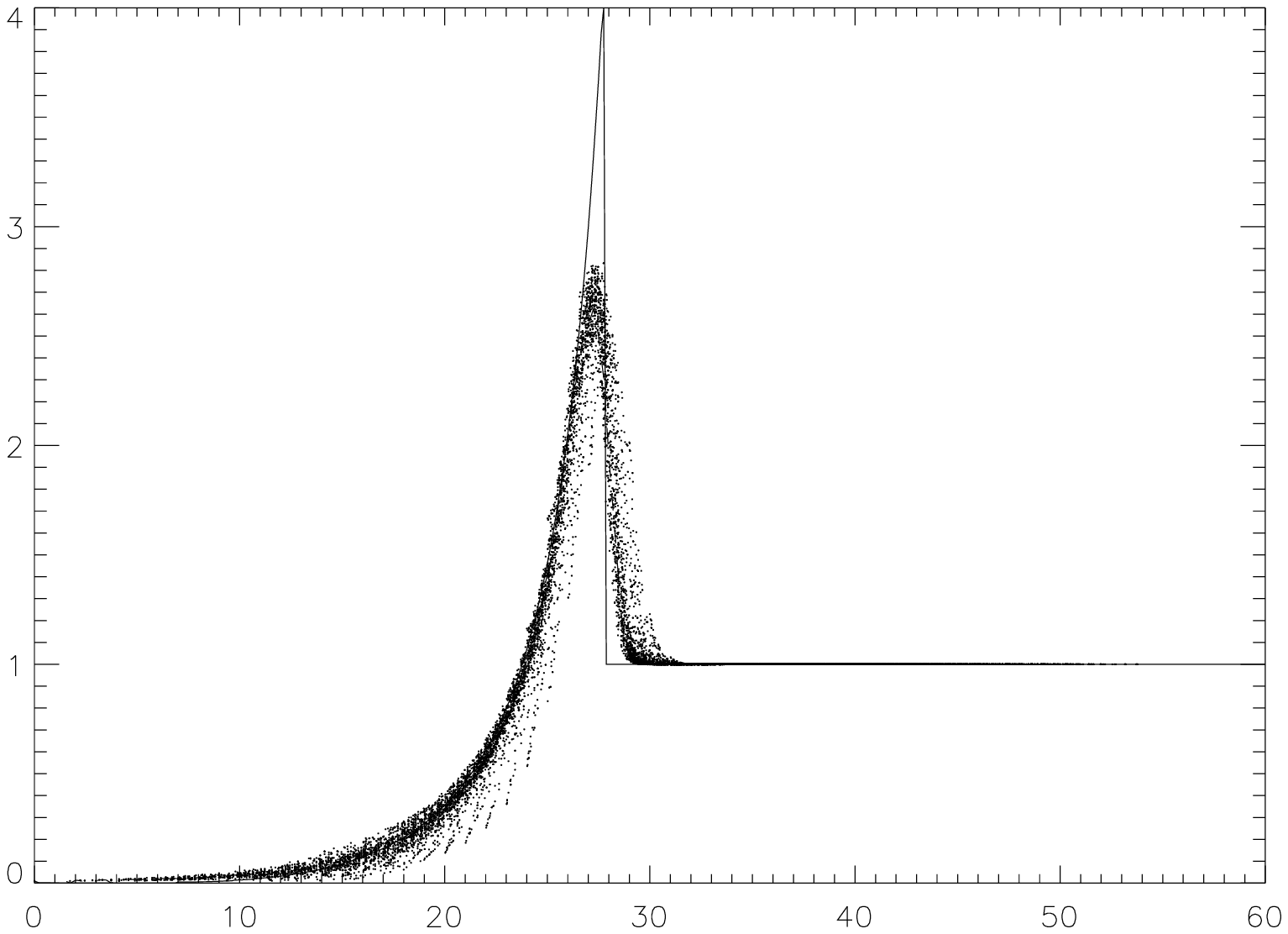}
\includegraphics[width=7.0cm]{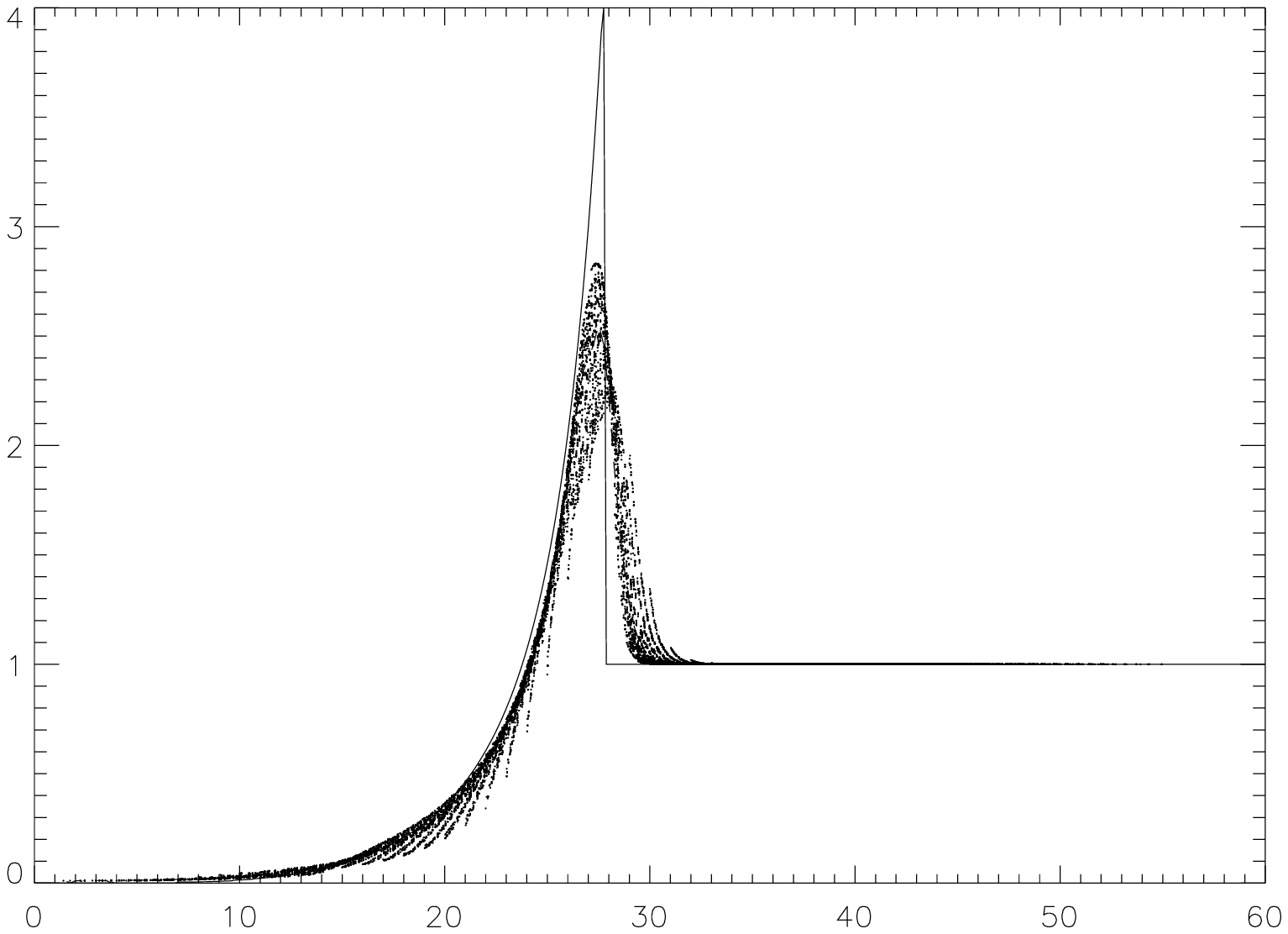}
\includegraphics[width=7.0cm]{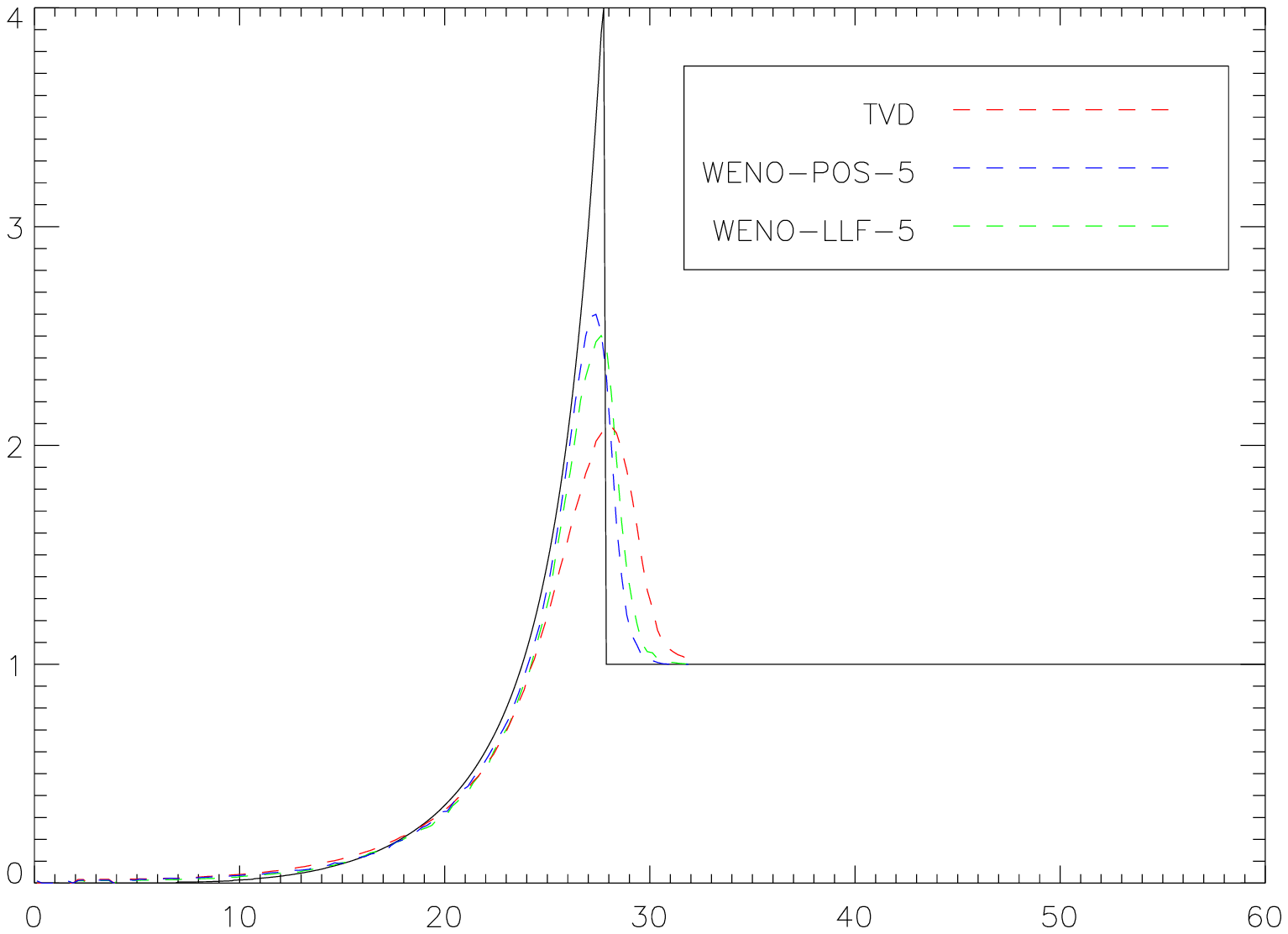}
\caption{Density distribution of $20000$ randomly selected grids 
in the three-dimensional Sedov blast-wave 
test with resolution $256^3$. Top left: TVD. Top right: WENO-POS-5. 
Bottom left: WENO-LLF-5. The bottom 
right panel shows the mean density as a function of radius in 
the three runs with respect to the analytical result
(solid line).}
\end{figure}

Fig. A2. displays the density of randomly selected $20000$ grid cells in 
 three-dimensional Sedov blast-wave test with resolution
$256^3$. The mean density as a function of radius is 
also shown in Fig. A2.  The explosion front 
are captured by all the three schemes. However, it can be seen
clearly that the fifth order WENO schemes resolve the 
explosion front more sharply, although moderate scattering appears 
at the central low density region. The 
density peaks in all the three codes are under-developed with respect 
to the analytic result, but the WENO 
schemes have smaller errors. The local flux splitting in WENO-LLF-5 
shows more dispersion near the peak of the wave than the 
positive-preserving version. 

\section{Resolution Convergence}

We check the resolution convergence by comparing results in 
$256^3$ and $512^3$ samples. Only 
a few selected quantities are presented here for the sake of brevity. 
Fig. A3 shows the power spectra of 
vorticity and velocity. The pattern and 
difference between the WENO and TVD schemes are not changed by different 
resolution. The dissipation scale has been significantly extended by 
improved resolution. 

\begin{figure}[htb]
\centering
\includegraphics[width=12.0cm]{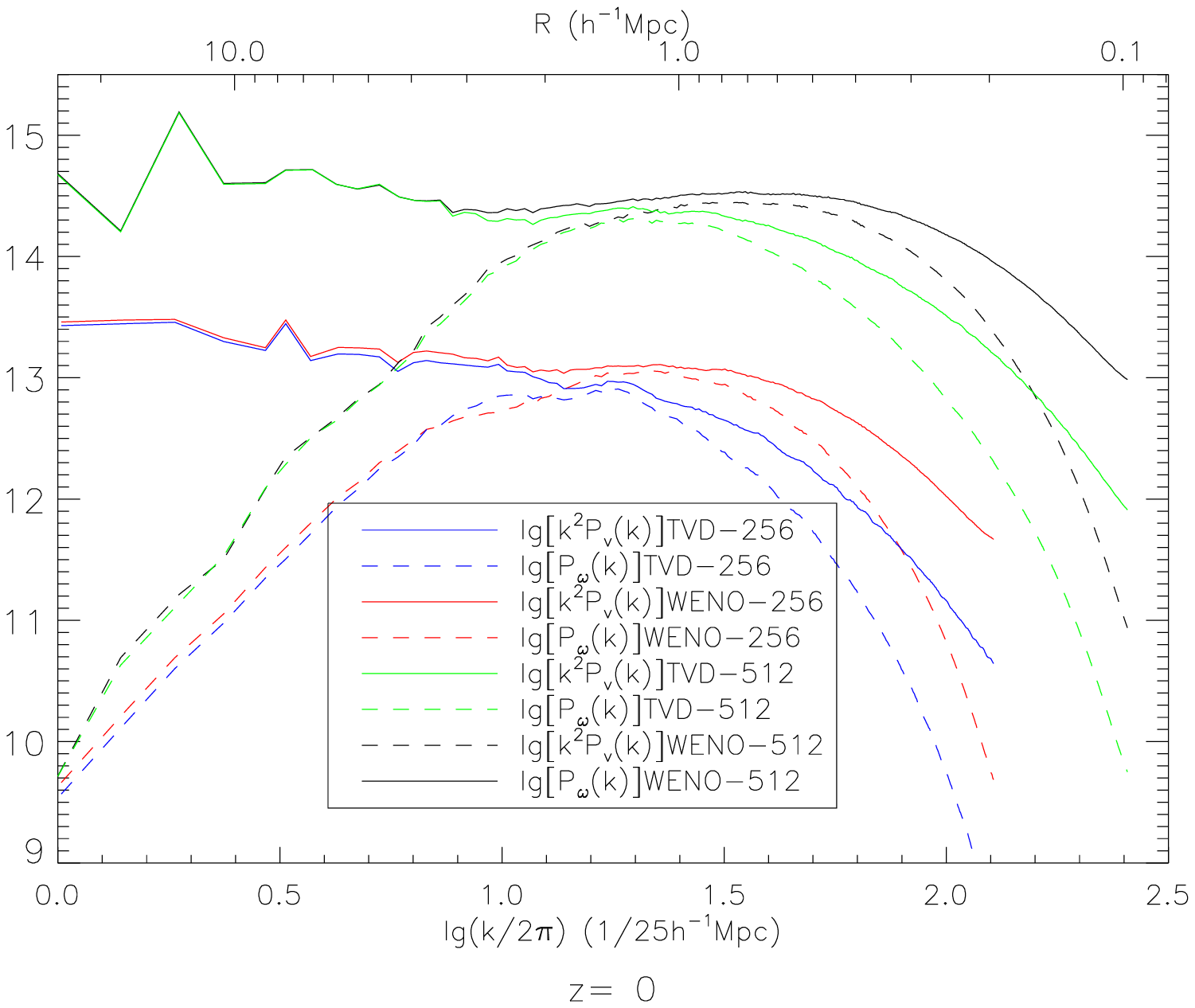}
\caption{Power spectra of vorticity and velocity in $256^3$ and 
$512^3$ samples. The results of $256^3$ 
samples have been shifted down by one dex for the sake of clarity.}
\end{figure}

Fig. A4 reports the kinetic turbulent energy ratio in the $256^3$ 
run. The general trends are similar to results 
in $512^3$. The magnitude of the ratio is systematically lower than 
the $512^3$ result, which results from a higher numerical 
viscosity for a lower resolution. Meanwhile, the kinetic 
turbulent energy ratio in the WENO-256 run is comparable to 
the TVD-512 run. 

\begin{figure}[htb]
\centering
\includegraphics[width=8.0cm]{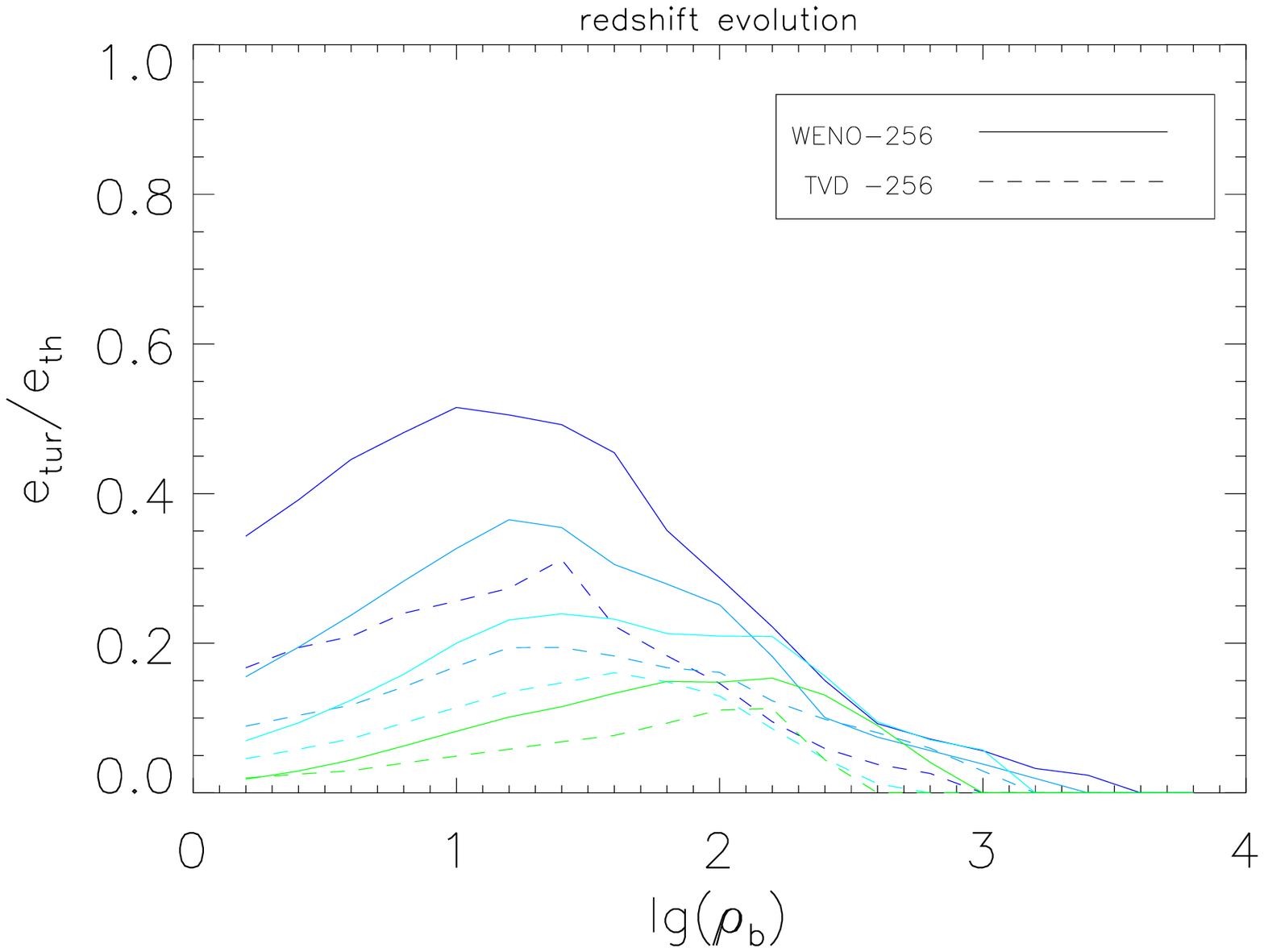}
\includegraphics[width=8.0cm]{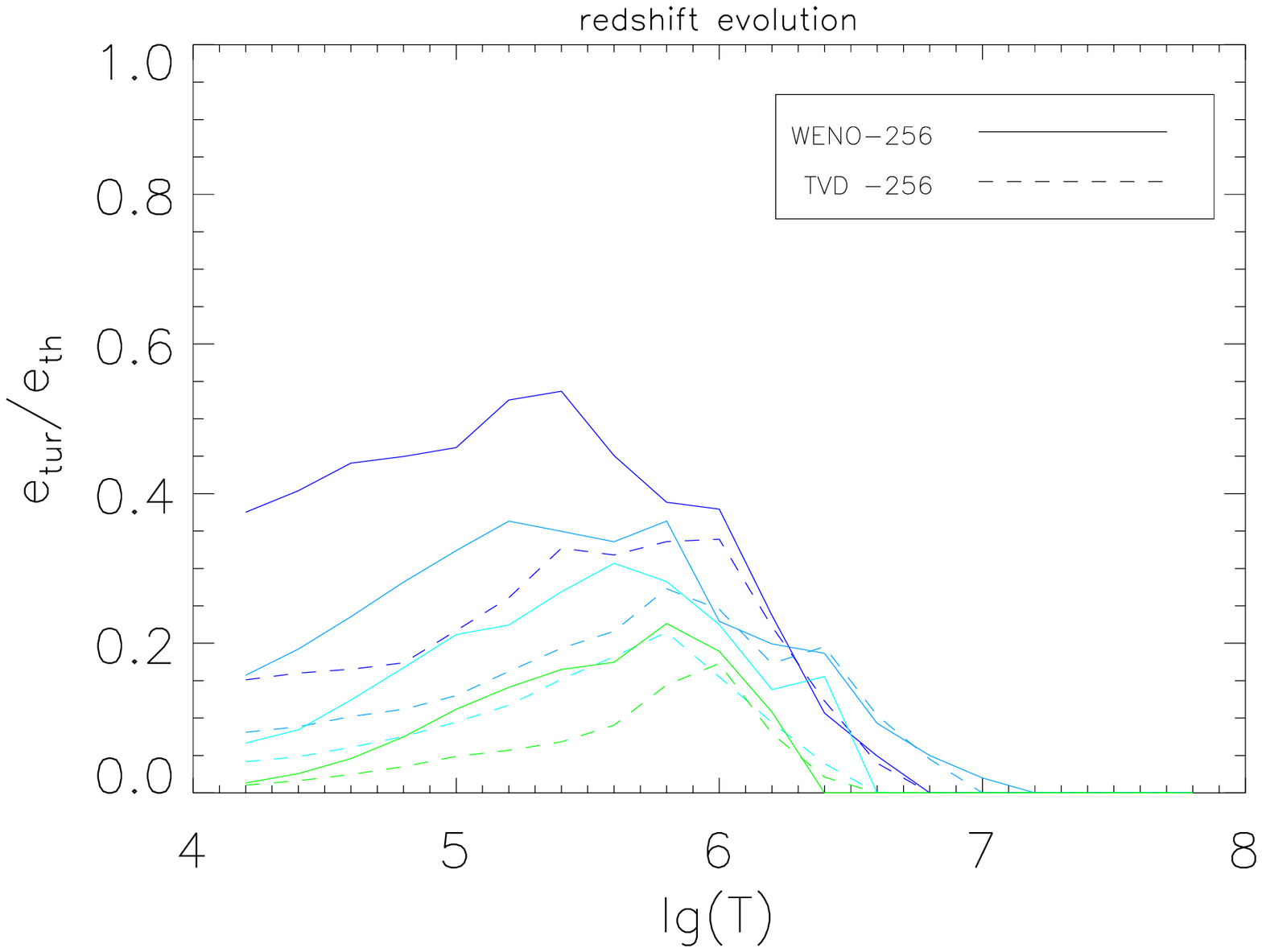}
\caption{The ratio of kinetic turbulent energy to thermal energy as 
function of gas density (left) and 
temperature (right). Solid (dash) lines are the result in WENO-256
(TVD-256).}
\end{figure}

\begin{figure}[htb]
\centering
\includegraphics[width=8.0cm]{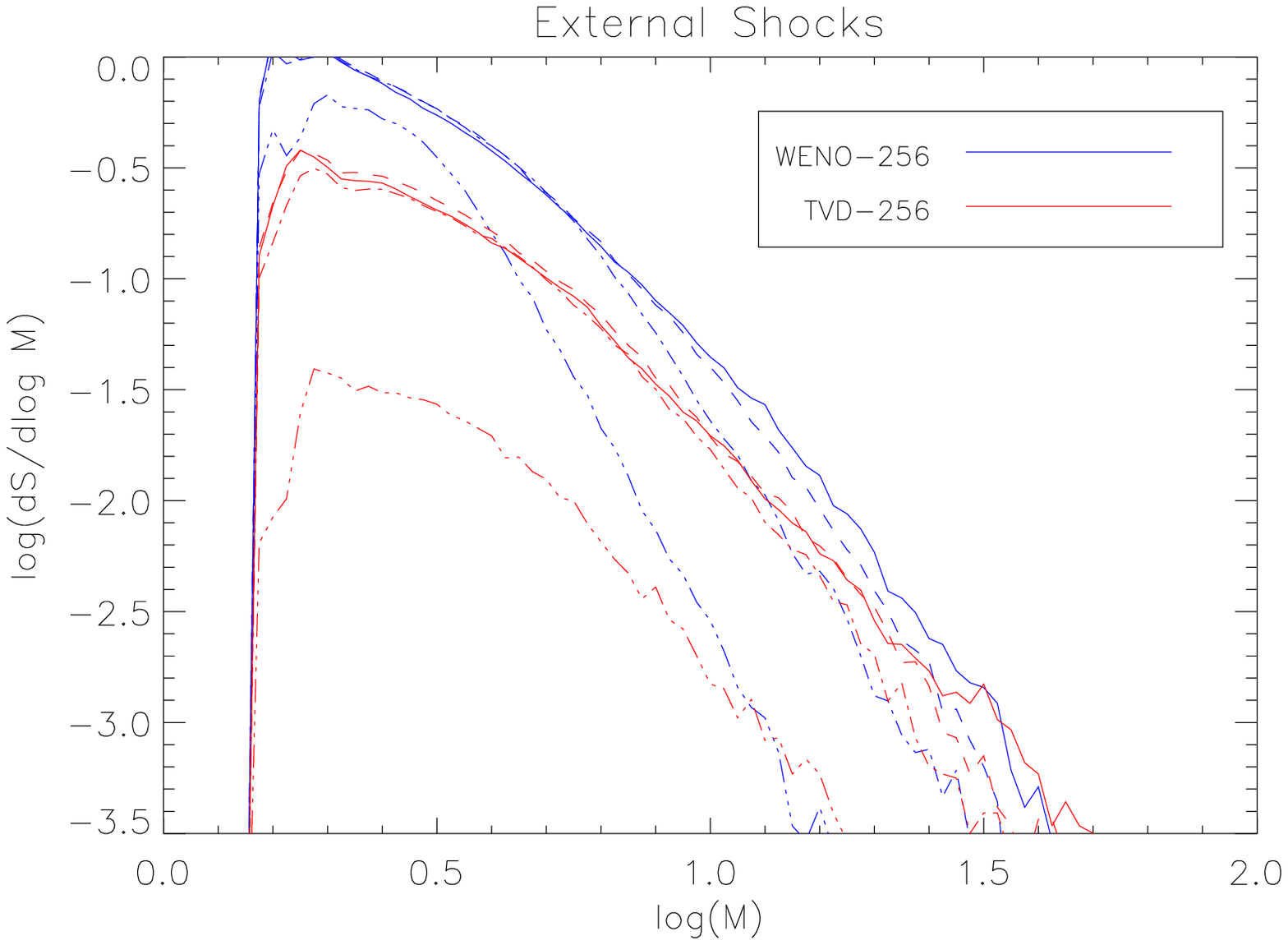}
\includegraphics[width=8.0cm]{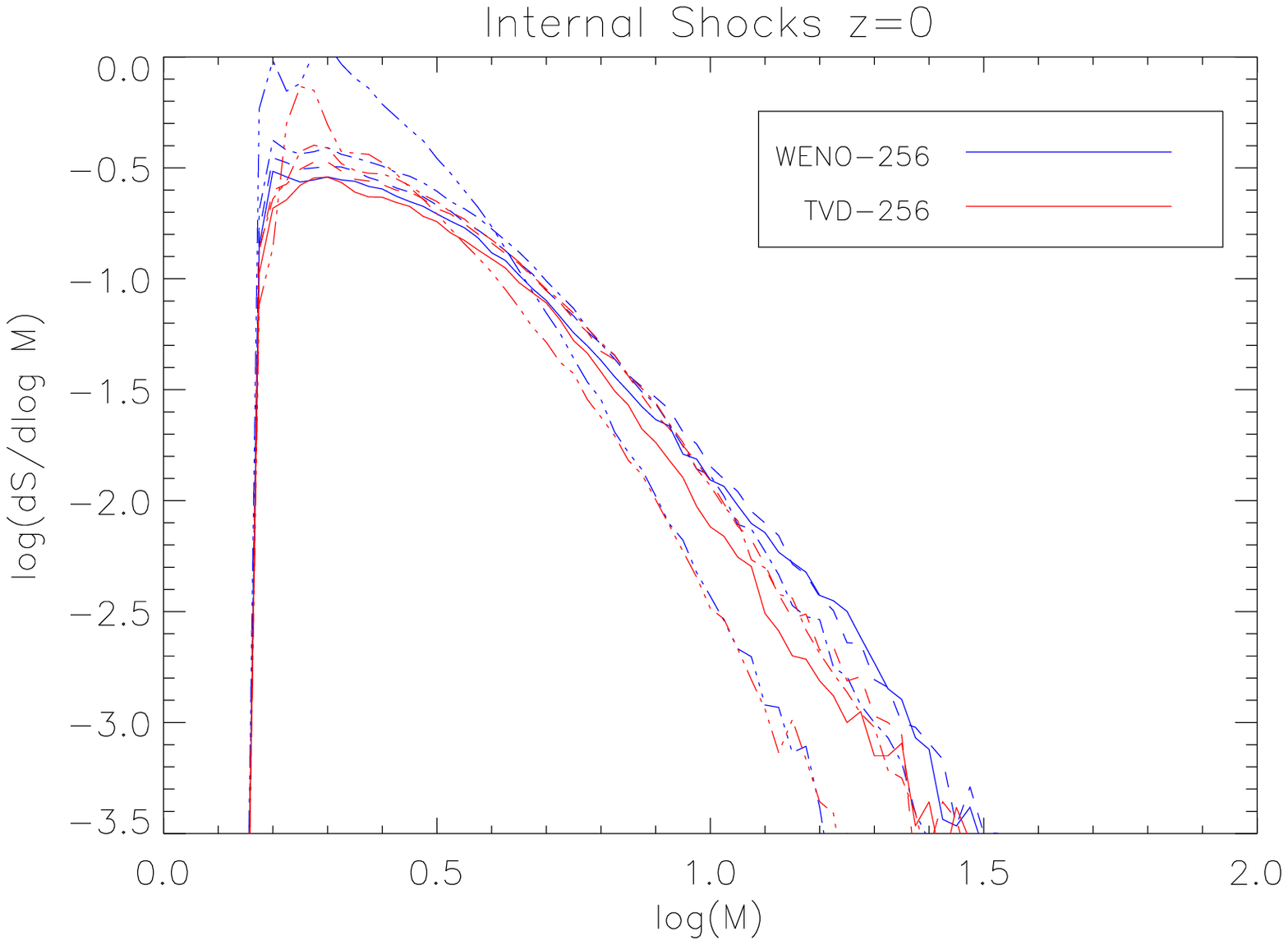}
\caption{Surface area of external and internal shocks in the $256^3$ simulations. 
Blue lines: WENO-256. Red lines: 
TVD-256. Solid (dash, dash-dotted, dash-triple-dotted) lines are 
the results at $z=0,0.5,1.0,2.0$.}
\end{figure}

Fig. A5 shows the distribution of external and internal 
shocks in the $256^3$ samples. The difference is 
very similar to the $512^3$ simulations, suggesting that 
the different levels of vorticity and turbulence between
two codes at a given resolution may be mainly 
caused by the shocks outside of clusters.

\end{document}